\documentclass[12pt]{article}
\usepackage{amsmath,amssymb,amsfonts}
\usepackage{color,graphicx,cite,soul}
\usepackage{array,booktabs,longtable}
\usepackage{caption,microtype,url}

\newcommand\Code[1]{\ensuremath{\texttt{#1}}}
\newcommand\Var[1]{\ensuremath{\mathit{#1}}}
\newcommand\Vi{\Var{i}}
\newcommand\Vj{\Var{j}}

\newcommand\tb{\tan\beta}
\newcommand\TB{t_\beta}

\newcommand\LP{\left(}
\newcommand\RP{\right)}
\newcommand\LB{\left[}
\newcommand\RB{\right]}
\newcommand\LV{\left\{}
\newcommand\RV{\right\}}

\renewcommand\Re{\mathop{\mathrm{Re}}}

\newcommand\ReDiag{\mathop{%
  \raise .5pt\hbox{[}%
  \widetilde{\mathrm{Re}}%
  \raise .5pt\hbox{]}}}
\newcommand\ReOffDiag{\mathop{%
  \raise .5pt\hbox{$\llbracket$}%
  \widetilde{\mathrm{Re}}%
  \raise .5pt\hbox{$\rrbracket$}}}
\newcommand\SE[1]{\Sigma_{#1}}

\newcommand\DRbar{\ensuremath{\smash{\overline{\mathrm{DR}}}}}
\newcommand\MSbar{\ensuremath{\overline{\mathrm{MS}}}}
\newcommand\matr[1]{\mathbf{#1}}

\newcommand\unity{\mathrm{1\mskip-4.25mu l}}

\newcommand\cMt{{\cal M}_{\text{tree}}}
\newcommand\cMl{{\cal M}_{\text{1-loop}}}

\newcommand\SW{s_\mathrm{w}}
\newcommand\CW{c_\mathrm{w}}
\newcommand\MW{M_W}
\newcommand\MZ{M_Z}

\newcommand\MHp{M_{H^\pm}}

\newcommand\mb{m_b}

\newcommand\At{A_t}

\newcommand\Sf{{\tilde f}}

\newcommand\Se{\mathrm{\tilde e}}
\newcommand\Fe{\mathrm{e}}

\newcommand\Fu{\mathrm{u}}

\newcommand\Sd{\mathrm{\tilde d}}
\newcommand\Fd{\mathrm{d}}

\newcommand\cind{c^{\phantom{\prime}}}
\newcommand\cpri{c^\prime}
\newcommand\nind{n^{\phantom{\prime}}}
\newcommand\npri{n^\prime}
\newcommand\sind{s^{\phantom{\prime}}}
\newcommand\spri{s^\prime}
\newcommand\dZm[1]{\delta\matr{Z}_{#1}}
\newcommand\dbZm[1]{\delta\matr{\breve Z}_{#1}}

\newcommand\dTB{\delta\TB}
\newcommand\ino[1]{\tilde\chi_{#1}}

\newcommand\chapm[1]{\ino{#1}^\pm}
\newcommand\champ[1]{\ino{#1}^\mp}
\newcommand\chap[1]{\ino{#1}^+}
\newcommand\cham[1]{\ino{#1}^-}
\newcommand\cha{\chapm}
\newcommand\mcha[1]{m_{\chapm{#1}}}
\newcommand\mchap[1]{m_{\ino{#1}^+}}
\newcommand\mcham[1]{m_{\ino{#1}^-}}
\newcommand\neu[1]{\ino{#1}^0}
\newcommand\mneu[1]{m_{\neu{#1}}}

\newcommand\refeq[1]{Eq.~(\ref{#1})}
\newcommand\refeqs[1]{Eqs.~(\ref{#1})}
\newcommand\refta[1]{Tab.~\ref{#1}}
\newcommand\refse[1]{Sect.~\ref{#1}}

\newcommand\citere[1]{Ref.~\cite{#1}}
\newcommand\citeres[1]{Refs.~\cite{#1}}

\newcommand\uscore{\symbol{95}}
\newcommand\eg{e.g.\ }
\newcommand\ie{i.e.\ }
\newcommand\wrt{w.r.t.\ }

\newcommand{\CP}{{\cal CP}}
\newcommand{\os}{\mathrm{os}}

\newcommand{\onel}{one-loop}
\newcommand{\tev}{\,\, \mathrm{TeV}}
\newcommand{\gev}{\,\, \mathrm{GeV}}
\newcommand{\mev}{\,\, \mathrm{MeV}}
\newcommand{\Hpm}{H^\pm}

\newcommand\hChaDecay{h_i \to \cham{c} \chap{\cpri}}
\newcommand\hchaechae{h_i \to \champ1 \chapm1}
\newcommand\hchaechaz{h_i \to \champ1 \chapm2}
\newcommand\hchazchaz{h_i \to \champ2 \chapm2}
\newcommand\hNeuDecay{h_i \to \neu{n} \neu{\npri}}
\newcommand\hneueneue{h_i \to \neu1 \neu1}
\newcommand\hneuzneuz{h_i \to \neu2 \neu2}
\newcommand\hneudneud{h_i \to \neu3 \neu3}
\newcommand\hneuvneuv{h_i \to \neu4 \neu4}
\newcommand\hneueneuz{h_i \to \neu1 \neu2}
\newcommand\hneueneud{h_i \to \neu1 \neu3}
\newcommand\hneueneuv{h_i \to \neu1 \neu4}
\newcommand\hneuzneud{h_i \to \neu2 \neu3}
\newcommand\hneuzneuv{h_i \to \neu2 \neu4}
\newcommand\hneudneuv{h_i \to \neu3 \neu4}

\newcommand\Hpdecay{H^+ \to \neu{n} \chap{c}}

\newcommand\HpmDecay{H^\pm \to \neu{n} \chapm{c}}

\newcommand\FA{\texttt{FeynArts}}
\newcommand\FC{\texttt{FormCalc}}
\newcommand\LT{\texttt{LoopTools}}
\newcommand\FH{\texttt{FeynHiggs}}
\newcommand\FT{\texttt{FeynTools}}
\newcommand\HFOLD{\texttt{HFOLD}}
\newcommand\HDECAY{\texttt{HDECAY}}

\newcommand\mh[1]{m_{h_{#1}}}

\newcommand\mstop[1]{m_{\tilde{t}_{#1}}}
\newcommand\msbot[1]{m_{\tilde{b}_{#1}}}

\newcommand\mstau[1]{m_{\tilde{\tau}_{#1}}}
\newcommand\mtausneu{m_{\tilde{\nu}_{\tau}}}

\newcommand{\Sce}{S1}
\newcommand{\Scz}{S2}
\newcommand{\Scd}{S3}
\newcommand{\Scv}{S4}
\newcommand{\Scf}{S5}

\def\order#1{\ensuremath{{\cal O}(#1)}}
\def\reffi#1{\mbox{Fig.~\ref{#1}}}
\def\reffis#1{\mbox{Figs.~\ref{#1}}}
\def\als{\alpha_s}

\def\Ga{\Gamma}
\def\ga{\gamma}
\def\de{\delta}
\def\la{\lambda}

\def\phimu{\varphi_{\mu}}
\def\phiMe{\varphi_{M_1}}
\def\phiMz{\varphi_{M_2}}

\definecolor{Orange}{named}{Orange}
\definecolor{Purple}{named}{Purple}
\definecolor{Lightblue}{cmyk}{0.9,0.1,0.1,0.3}
\definecolor{dgelborange}{cmyk}{0.,0.3,0.5, 0.}
\definecolor{Lila}{rgb}{0.5,0.,1}

\graphicspath{{figs/}}

\evensidemargin \oddsidemargin
\allowdisplaybreaks
\begin{document}
\thispagestyle{empty}

\def\thefootnote{\fnsymbol{footnote}}

\begin{flushright}
\mbox{}
%CERN--PH--TH/2011--291\\
%FR--PHENO--2011--021\\
%arXiv:yymm.nnnn [hep-ph]
\end{flushright}

\vspace{0.5cm}

\begin{center}

{\large\sc {\bf Higgs Decays into Charginos and Neutralinos}} 

\vspace{0.4cm}

{\large\sc {\bf in the Complex MSSM: A Full One-Loop Analysis}}

\vspace{1cm}

{\sc
S.~Heinemeyer$^{1}$%
\footnote{email: Sven.Heinemeyer@cern.ch}%
~and C.~Schappacher$^{2}$%
\footnote{email: schappacher@kabelbw.de}%
\footnote{former address}%
}

\vspace*{.7cm}

{\sl
$^1$Instituto de F\'isica de Cantabria (CSIC-UC), Santander,  Spain

\vspace*{0.1cm}

$^2$Institut f\"ur Theoretische Physik,
Karlsruhe Institute of Technology, \\
D--76128 Karlsruhe, Germany

}

\end{center}

\vspace*{0.1cm}

\begin{abstract}
\noindent
For the search for additional Higgs bosons in the Minimal
Supersymmetric Standard Model (MSSM) as well as for future precision
analyses in the Higgs sector a precise knowledge of their decay
properties is mandatory.
We evaluate all two-body decay modes of the Higgs bosons into 
charginos and neutralinos in the MSSM with complex parameters (cMSSM). 
The evaluation is based on a full one-loop calculation of all decay 
channels, also including hard QED radiation. 
We restricted ourselves to a version of our renormalization scheme which 
is valid for $|M_1| < |M_2|, |\mu|$ and $M_2 \neq \mu$ to simplify
the analysis, even though we are able to switch to other parameter regions.
The dependence of the Higgs boson predictions on the relevant cMSSM parameters 
is analyzed numerically.  We find sizable contributions to many partial decay 
widths.  They are roughly of 10\% of the tree-level results, 
but can go up to 20\% or higher.  The full one-loop contributions 
are important for the correct interpretation of heavy Higgs boson search 
results at the LHC and, if kinematically allowed, at a future linear 
$e^+e^-$ collider.  It is planned to implement the evaluation of the 
branching ratios of the Higgs bosons into the Fortran code \FH, 
together with an automated choice of the renormalization scheme valid for 
the full cMSSM parameter space.
\end{abstract}
%\pacs{}

\def\thefootnote{\arabic{footnote}}
\setcounter{page}{0}
\setcounter{footnote}{0}

\newpage

%%%%%%%%%%%%%%%%%%%%%%%%%%%%%%%%%%%%%%%%%%%%%%%%%%%%%%%%%%%%%%%%%%%%%%%%%%%%%%%
%%%%%%%%%%%%%%%%%%%%%%%%%%%%%%%%%%%%%%%%%%%%%%%%%%%%%%%%%%%%%%%%%%%%%%%%%%%%%%%

\section{Introduction}
\label{sec:intro}

One of the most important tasks at the LHC is to search for physics effects 
beyond the Standard Model (SM), where the Minimal Supersymmetric Standard 
Model (MSSM)~\cite{mssm,HaK85,GuH86} is one of the leading candidates. 
Supersymmetry (SUSY) predicts two scalar partners for all SM fermions as well
as fermionic partners to all SM bosons.
Another important task of the LHC is the investigation of the mechanism
of electroweak symmetry breaking. 
The most frequently studied realizations are the Higgs mechanism within the 
SM and within the MSSM.  Contrary to the case of the SM, in the MSSM two 
Higgs doublets are required.
This results in five physical Higgs bosons instead of the single Higgs
boson in the SM.  In lowest order these are the light and heavy 
$\CP$-even Higgs bosons, $h$ and $H$, the $\CP$-odd Higgs boson, 
$A$, and two charged Higgs bosons, $H^\pm$. Within the MSSM with complex
parameters (cMSSM), taking higher-order corrections into account, the
three neutral Higgs bosons mix and result in the states 
$h_i$ ($i = 1,2,3$)~\cite{mhiggsCPXgen,Demir,mhiggsCPXRG1,mhiggsCPXFD1}.
The Higgs sector of the cMSSM is described at the tree-level by two
parameters: 
the mass of the charged Higgs boson, $\MHp$, and the ratio of the two
vacuum expectation values, $\tb \equiv \TB = v_2/v_1$.
Often the lightest Higgs boson, $h_1$ is identified with the particle 
discovered at the LHC~\cite{ATLASdiscovery,CMSdiscovery} with a mass 
around $\sim 125\gev$.
If the mass of the charged Higgs boson is assumed to be larger than 
$\sim 200\gev$ the four additional Higgs bosons are roughly mass
degenerate, $\MHp \approx \mh2 \approx \mh3$ and referred to as the
``heavy Higgs bosons''. 
Discovering one or more of those additional Higgs bosons would be an
unambiguous sign of physics beyond the SM and could yield important 
information about their possible supersymmetric origin.

If SUSY is realized in nature and the charged Higgs-boson mass is 
$\MHp \lesssim 1.5\tev$, then the heavy Higgs bosons could be detectable 
at the LHC~\cite{ATLAS-HA,CMS-HA} (including its high luminosity upgrade, 
HL-LHC, see \citere{holzner} and references therein) and/or at a future 
linear $e^+e^-$ collider such as the ILC~\cite{ILC-TDR,teslatdr,ilc} or 
CLIC~\cite{CLIC}. (Results on the combination of LHC and ILC results can 
be found in \citere{lhcilc}.)
The discovery potential at the HL-LHC goes up to \order{1\tev} for
large $\tb$ values and somewhat lower at low $\tb$ values. 
At an $e^+e^-$ linear collider the heavy Higgs
bosons are pair produced, and the reach is limited by the center of mass
energy, $\MHp \lesssim \sqrt{s}/2$, roughly independent of $\tb$.
Details about the discovery process(es) depend strongly
on the cMSSM parameters (and will not be further discussed in this paper). 

In the case of a discovery of additional Higgs bosons a subsequent
precision measurement of their properties will be crucial determine
their nature and the underlying (SUSY) parameters. 
In order to yield a sufficient accuracy, one-loop corrections to the 
various Higgs-boson decay modes have to be considered.
Decays to SM fermions have been evaluated at the full one-loop level 
in the cMSSM in \citere{hff}, see also \citere{hff0} as well as 
\citeres{deltab,db2l} for higher-order SUSY corrections.
Decays to (lighter) Higgs bosons have been evaluated at the full
one-loop level in the cMSSM in \citere{hff}, see also \citeres{hhh,hAA}.
Decays to SM gauge bosons (see also \citere{hVV-WH}) can be evaluated 
to a very high precision using the full SM one-loop 
result~\cite{prophecy4f} combined with the appropriate effective 
couplings~\cite{mhcMSSMlong}.
The full one-loop corrections in the cMSSM listed here together with
resummed SUSY corrections have been implemented into the code 
\FH~\cite{feynhiggs,mhiggslong,mhiggsAEC,mhcMSSMlong,Mh-logresum}.
Corrections at and beyond the one-loop level in the MSSM with real
parameters (rMSSM) are implemented into the code 
\HDECAY~\cite{hdecay,hdecay2}.
Both codes were combined by the LHC Higgs Cross Section Working Group to
obtain the most precise evaluation for rMSSM Higgs boson decays to SM
particles and decays to lighter Higgs bosons~\cite{YR3}.

The (heavy) MSSM Higgs bosons can (if kinematically allowed) also decay 
to SUSY particles, \ie to charginos, neutralinos and scalar fermions.
In \citere{benchmark4} it was demonstrated that the SUSY particle modes
can dominate the decay of the heavy Higgs bosons.
The lightest neutral Higgs boson, on the other hand, can have a
substantial branching ratio into the lightest neutralino, 
$h_1 \to \neu1\neu1$, where the $\neu1$ constitutes the Dark Matter
candidate in the MSSM~\cite{EHNOS}. Bounds on $\mneu1$ often assume an
underlying SUSY grand unified theory, based on a simple Lie
group. Dropping these assumptions hardly any bound on $\mneu1$ can be
placed directly (see, e.g., \citere{masslessx} and references therein),
and the decay $h_1 \to \neu1\neu1$ is kinematically possible. In order
to determine the Dark Matter properties a precision measurement of this
process at the LHC or a future $e^+e^-$ collider will be necessary.

Higher-order contributions to MSSM Higgs boson decays to scalar fermions 
have been evaluated in various analyses over the last decade.  For
calculations in the rMSSM, see \citeres{Asfsf_1L,Phisqsq_1L,Phisqsq_als_2} 
and references therein. 
More recently, the results of \citere{Asfsf_1L} were made public in the 
code \HFOLD~\cite{hfold}, using a pure \DRbar\ renormalization for the 
calculation.
In \citere{Phisqsq_als_3} the \order{\als} corrections to Higgs boson 
decays to scalar quarks were re-analyzed and included into the code \HDECAY.
Within the cMSSM a full one-loop calculation of Higgs boson decays to 
scalar fermions has recently been published in \citere{HiggsDecaySferm} and
will be included into the code \FH. 
These results were obtained in a renormalization
scheme~\cite{MSSMCT,SbotRen,Stop2decay,mhcMSSMlong,LHCxC,LHCxN,LHCxNprod},
which has been shown to yield stable results over nearly the full cMSSM
parameters space.
In this work we take another step in the direction of completion 
of the calculation of \textit{all} two-body decays at the one-loop level 
in the cMSSM in this stable and reliable renormalization scheme: we
calculate all two-body decay modes of the Higgs bosons to charginos and 
neutralinos in the cMSSM.  More specifically, we calculate the full 
one-loop corrections to the partial decay widths
\begin{alignat}{3}
\label{eq:hchacha}
&\Ga(\hChaDecay) &\qquad & (i = 1,2,3;\, c,\cpri = 1,2)\,, \\
\label{eq:hneuneu}
&\Ga(\hNeuDecay) &\qquad & (i = 1,2,3;\, n,\npri = 1,2,3,4)\,, \\
\label{eq:Hpmneucha}
&\Ga(\HpmDecay)  &\qquad & (n = 1,2,3,4;\, c = 1,2)\,,
\end{alignat}
where $\chapm{c}$ ($\neu{n}$) denotes the charginos 
(neutralinos). 
While we have calculated the decay of all Higgs bosons, in the numerical 
evaluation below, we will concentrate on the heavy Higgs bosons, $h_{2,3}$ 
and $\Hpm$, but also show results for $h_1 \to \neu1\neu1$.

The evaluation of the channels \refeqs{eq:hchacha} -- (\ref{eq:Hpmneucha}) 
is based on a full one-loop calculation, \ie including electroweak (EW) 
corrections, as well as soft and hard QED radiation. For ``mixed'' decay 
modes, we evaluate in addition the two ``$\CP$-versions'' $(c \ne \cpri)$ 
of \refeq{eq:hchacha} and the two ``$\CP$-versions'' of \refeq{eq:Hpmneucha}, 
which give different results for non-zero complex phases.
We restricted ourselves to a version of our renormalization scheme which 
is valid for $|M_1| < |M_2|, |\mu|$ and $M_2 \neq \mu$ (where $M_1$ and 
$M_2$ denote the soft SUSY-breaking parameter of the $U(1)$ and $SU(2)$
gauginos, and $\mu$ is the Higgs mixing parameter) to simplify the
analysis, even though we are able to switch to other parameter regions, 
see the discussion in \citeres{LHCxC,LHCxN,LHCxNprod} 
(see also \citere{onshellCNmasses}).

Higher-order contributions to MSSM Higgs boson decays to charginos and 
neutralinos have been evaluated in various analyses over the last decade. 
In \citere{AHn2n2c1c1} the leading Yukawa corrections to 
$A/H \to \neu2\neu2, \chap1\cham1$ in the rMSSM have been evaluated,
employing an on-shell (OS) scheme 
(referring to \citere{denner}, but without providing further details).
Next, in \citere{AHc12c12} the full one-loop corrections
to $A/H \to \chap{\cind}\cham{\cpri}$ ($c,\cpri = 1,2$) have been 
presented in the rMSSM (again without details about the OS like
scheme).  An effective Lagrangian approach for heavy neutral Higgs boson
decays in the rMSSM was published in \citere{AHnnccLeff}.  The full one-loop
corrections to all heavy Higgs decays to charginos and neutralinos in
the rMSSM in the \DRbar\ scheme was published in the code 
\HFOLD~\cite{hfold}.  More recently also evaluations of Higgs boson
decays to charginos and neutralinos in the cMSSM became available. 
In \citere{hHAcc-cMSSM} the decays $h_i \to \chap{\cind}\cham{\cpri}$, 
($i = 1,2,3;\, c,\cpri = 1,2$) were presented, together with a short
discussion of different renormalization schemes (see \refse{sec:chaneu}) 
and brief analysis of the dependence on the phases of $\mu$, $M_1$ and the 
trilinear Higgs stop coupling, $\At$. 
The decays $h_i \to \neu{\nind}\neu{\npri}$ 
($i = 1,2,3;\, n,\npri = 1,2,3,4$) were calculated in \citere{dissAF},
where the numerical analysis concentrated on $h_{2,3} \to \neu2\neu2$
and the dependence on the phase of $\At$. The latter two references
are close to the calculations presented in this paper. Small differences
in the renormalization in the chargino/neutralino sector exist (see
\refse{sec:chaneu} and \citere{LHCxN}), where we use consistently the
scheme detailed in \citere{MSSMCT} for \textit{all} two-body decays
\textit{simultaneously}. In our numerical analysis we focus on the
one parameter 
with a possible complex phase entering at the tree-level, $M_1$, see the
discussion in \refse{sec:calc}. A short numerical comparison with the
literature, in particular with \citeres{hfold,hHAcc-cMSSM,dissAF}, 
will be given in \refse{sec:comparisons}.

\medskip

In this paper we present a full one-loop calculation 
for all two-body decay channels of the Higgs bosons into charginos
and neutralinos in the cMSSM, taking into account soft and hard QED 
radiation.  In \refse{sec:renorm} we briefly review the 
relevant sectors of the cMSSM.  Details about the calculation can be 
found in \refse{sec:calc}, and the numerical results for all decay 
channels are presented in \refse{sec:numeval} 
(including comments on comparisons with results from other groups).  
The conclusions can be found in \refse{sec:conclusions}.  
It is planned to implement the evaluation of the branching ratios of the
heavy Higgs bosons into the Fortran code 
\FH~\cite{feynhiggs,mhiggslong,mhiggsAEC,mhcMSSMlong,Mh-logresum}, 
together with an automated choice of the renormalization scheme valid
for the full cMSSM parameter space.

%%%%%%%%%%%%%%%%%%%%%%%%%%%%%%%%%%%%%%%%%%%%%%%%%%%%%%%%%%%%%%%%%%%%%%%%%%%%%%%
%%%%%%%%%%%%%%%%%%%%%%%%%%%%%%%%%%%%%%%%%%%%%%%%%%%%%%%%%%%%%%%%%%%%%%%%%%%%%%%

\section{The complex MSSM}
\label{sec:renorm}

The channels (\ref{eq:hchacha}) -- (\ref{eq:Hpmneucha}) are calculated 
at the one-loop level, including soft and hard QED radiation.  
This requires the simultaneous renormalization of several sectors of 
the cMSSM: the Higgs and gauge boson sector as well as the 
chargino/neutralino sector.  In the following subsections we very 
briefly review these sectors and their renormalization.

%%%%%%%%%%%%%%%%%%%%%%%%%%%%%%%%%%%%%%%%%%%%%%%%%%%%%%%%%%%%%%%%%%%%%%%%%%%%%%%

\subsection{The Higgs- and gauge-boson sector}
\label{sec:higgs}

The Higgs- and gauge-boson sector follow strictly \citere{MSSMCT} and 
references therein (see especially \citere{mhcMSSMlong}).
This defines in particular the counterterm $\de\tb \equiv \dTB$,
as well as the counterterms for the $Z$~boson mass, $\de\MZ^2$, and
for the sine of the weak mixing angle, $\de\SW$
(with $\SW = \sqrt{1 - \CW^2} = \sqrt{1 - \MW^2/\MZ^2}$, 
where $\MW$ denotes the $W$~boson mass).

%%%%%%%%%%%%%%%%%%%%%%%%%%%%%%%%%%%%%%%%%%%%%%%%%%%%%%%%%%%%%%%%%%%%%%%%%%%%%%%

\subsection{The chargino/neutralino sector}
\label{sec:chaneu}

The chargino/neutralino sector is also described in detail in 
\citere{MSSMCT} and references therein, see in particular
\citeres{LHCxC,LHCxN,LHCxNprod}.  In this paper we use the so called 
``CCN'' scheme, \ie OS conditions for two charginos and one 
neutralino, which we chose to be the lightest one. 
Renormalizing the two charged states OS, \ie ensuring that
they have the same mass at the tree- and at the loop-level is 
(in general) crucial for the cancellation of the IR divergencies.
In the notation of \citere{MSSMCT} we used:
\begin{align*}
\Code{\$InoScheme = CCN[1]} \qquad
\text{fixed CCN scheme with on-shell } \neu1\,.
\end{align*}
\noindent
This defines in particular the counterterm $\de\mu$, where $\mu$
denotes the Higgs mixing parameter.
This scheme yields numerically stable results for 
$|M_1| < |M_2|, |\mu|$ and $M_2 \neq \mu$, \ie the lightest neutralino 
is bino-like and defines the counterterm for
$M_1$~\cite{LHCxC,LHCxN,LHCxNprod,onshellCNmasses}. In the numerical 
analysis this mass pattern holds. Switching to a different mass pattern,
e.g.\ with $|M_2| < |M_1|$ and/or $M_2 \sim \mu$ requires to switch to a
different renormalization scheme~\cite{MSSMCT,onshellCNmasses}. While
these schemes are implemented into the \FA/\FC\ framework~\cite{MSSMCT},
so far no automated choice of the renormalization scheme has been
devised. For simplicity we stick to the CCN[1] scheme with a matching
choice of SUSY parameters, see \refse{sec:paraset}.

Since both chargino masses $\mcha{1,2}$ and the lightest neutralino mass 
$\mneu{1}$ have been chosen as independent parameters, the one-loop 
masses of the heavier neutralinos $\neu{n}$ ($n$ = 2,3,4) are obtained 
from the tree-level ones via the shifts \cite{dissAF}
\begin{alignat}{1}
\Delta \mneu{n} 
&= -\Re \bigg\{\mneu{n} \LP \SE{\neu{n}}^L(\mneu{n}^2) 
         + \frac{1}{2} \LB 
           \dZm{\neu{}}^L + \dbZm{\neu{}}^L
         + \dZm{\neu{}}^R + \dbZm{\neu{}}^R 
                       \RB_{nn} \RP \notag \\
&\hspace{1.6cm} + \SE{\neu{n}}^{SL}(\mneu{n}^2) 
    - \mneu{n} \LB \dZm{\neu{}}^L + \dbZm{\neu{}}^L \RB_{nn}
    - \LB \delta \matr{M}_{\neu{}} \RB_{nn}
        \bigg\}\,,
\label{eq:Deltamneu}
\end{alignat}
where the renormalization constants $\dZm{}$ and $\dbZm{}$ can be
found in \citere{MSSMCT}.
For all externally appearing neutralino masses we use the (shifted) 
``on-shell'' masses:
\begin{align}
\mneu{n}^\os &= \mneu{n} + \Delta \mneu{n}\,.
\label{eq:mneuOS}
\end{align}
In order to yield UV-finite results we use the tree-level values $\mneu{n}$ 
for all internally appearing neutralino masses in loop calculations.

%%%%%%%%%%%%%%%%%%%%%%%%%%%%%%%%%%%%%%%%%%%%%%%%%%%%%%%%%%%%%%%%%%%%%%%%%%%%%%%

\subsection{The fermion/sfermion sector}
\label{sec:sfermion}

To be in accordance with \citere{HiggsDecaySferm}, we use shifted 
(s)fermion masses in the loop corrections.  
As requirement for these shifts one needs the renormalization of the 
fermion/sfermion sector:

\begin{itemize}
\item 
  The renormalization of the fermion sector is described in 
  detail in \citere{MSSMCT} and references therein. 
  For simplification we use the \DRbar\ renormalization for all 
  three generations of down-type quarks \textit{and} leptons, 
  again in the notation of \citere{MSSMCT}:
  \begin{align*}
    \Code{UVMf1[4,\,\uscore] = UVDivergentPart} &\qquad
    \text{\DRbar\ renormalization for $m_d$, $m_s$, $m_b$} \\
    \Code{UVMf1[2,\,\uscore] = UVDivergentPart} &\qquad 
    \text{\DRbar\ renormalization for $m_e$, $m_\mu$, $m_\tau$}
  \end{align*}
\item
  The renormalization of the sfermion sector differs 
  slightly from the one described in \citere{MSSMCT}. 
  For the squark sector we follow \citeres{SbotRen,Stop2decay}
  (which agrees with the renormalization scheme used in 
  \citeres{LHCxC,LHCxN,Gluinodecay,Stau2decay}) 
  and the slepton sector can be found in \citere{HiggsDecaySferm}.
  Concerning our notation we denote as 
  $M_{\tilde{Q}_g, \tilde{L}_g, \tilde{U}_g, \tilde{D}_g, \tilde{E}_g}$,
  the ``diagonal'' soft SUSY-breaking parameters for the $SU(2)$ squark, 
  slepton doublet, the u-, d-type squark singlet, and the e-type 
  slepton singlet, respectively, where $g$ is the generation index. 
  Furthermore we use $A_{\Fu_g,\Fd_g,\Fe_g}$ for the trilinear Higgs-scalar 
  u-, d-, e-type fermion couplings, respectively.
\end{itemize}

%%%%%%%%%%%%%%%%%%%%%%%%%%%%%%%%%%%%%%%%%%%%%%%%%%%%%%%%%%%%%%%%%%%%%%%%%%%%%%%
%%%%%%%%%%%%%%%%%%%%%%%%%%%%%%%%%%%%%%%%%%%%%%%%%%%%%%%%%%%%%%%%%%%%%%%%%%%%%%%

\section{Calculation of loop diagrams}
\label{sec:calc}

\begin{figure}[t!]
\begin{center}
\framebox[15cm]{\includegraphics[width=0.8\textwidth]{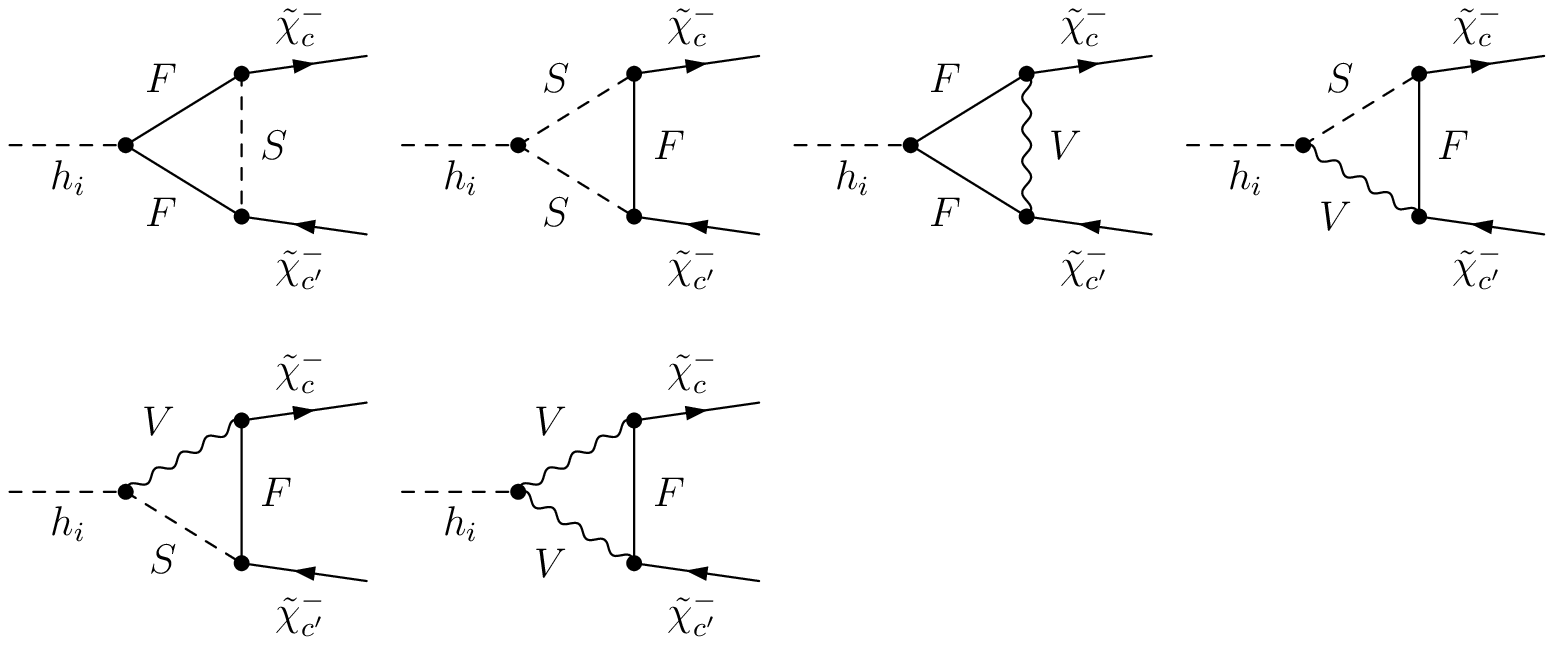}}
%\vspace{1em}
\caption{
  Generic Feynman diagrams for the decay $\hChaDecay$ 
  ($i = 1,2,3;\, c,\cpri = 1,2$).
  $F$ can be a SM fermion, chargino, neutralino; 
  $S$ can be a sfermion or a Higgs/Goldstone boson; 
  $V$ can be a $\ga$, $Z$, $W^\pm$. 
  Not shown are the diagrams with a $h_i$--$Z$ or $h_i$--$G$ 
  transition contribution on the external Higgs boson leg.
}
\label{fig:hchacha}
\vspace{1em}
\end{center}
\end{figure}

\begin{figure}[t!]
\begin{center}
\framebox[15cm]{\includegraphics[width=0.8\textwidth]{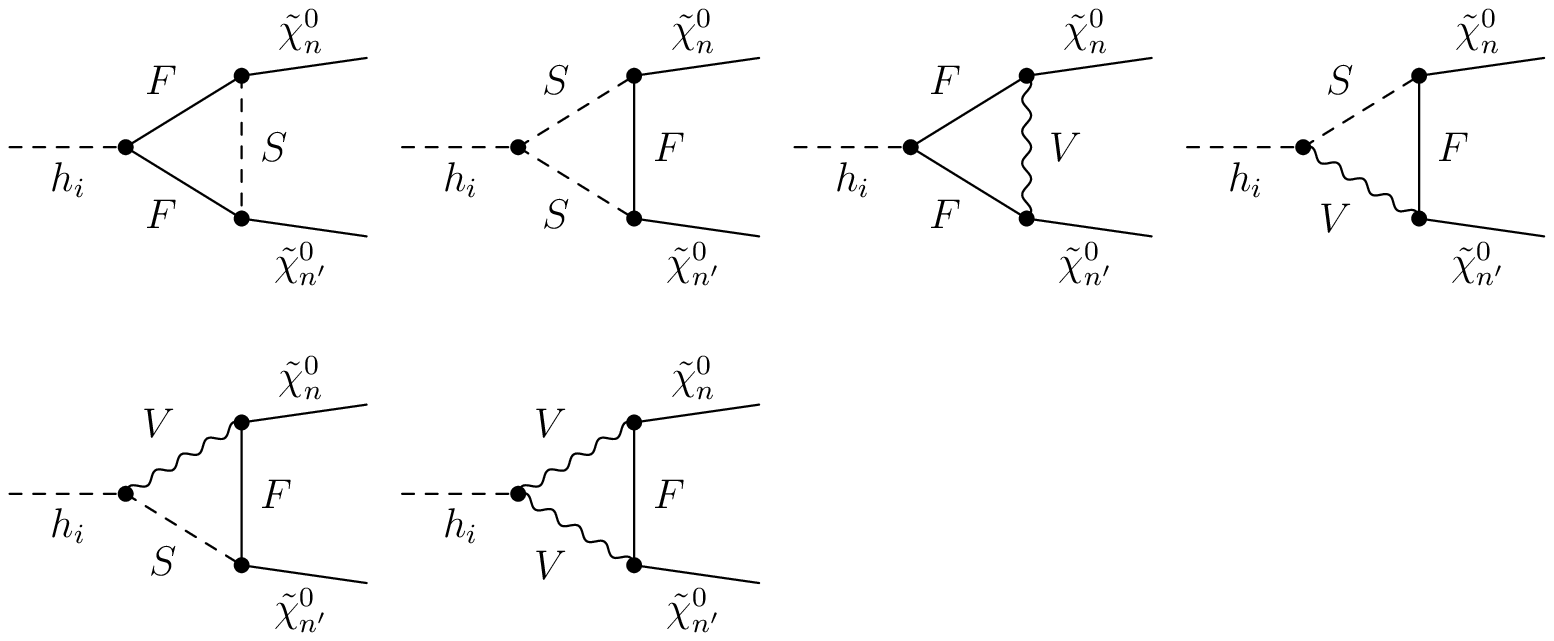}}
\caption{
  Generic Feynman diagrams for the decay $\hNeuDecay$ 
  ($i = 1,2,3;\, n,n^\prime=1,2,3,4$).
  $F$ can be a SM fermion, chargino, neutralino; 
  $S$ can be a sfermion or a Higgs/Goldstone boson; 
  $V$ can be a $Z$ or $W^\pm$. 
  Not shown are the diagrams with a $h_i$--$Z$ or $h_i$--$G$ 
  transition contribution on the external Higgs boson leg.
}
\label{fig:hneuneu}
%\vspace{1em}
\end{center}
\end{figure}

\begin{figure}[t!]
\begin{center}
\framebox[15cm]{\includegraphics[width=0.8\textwidth]{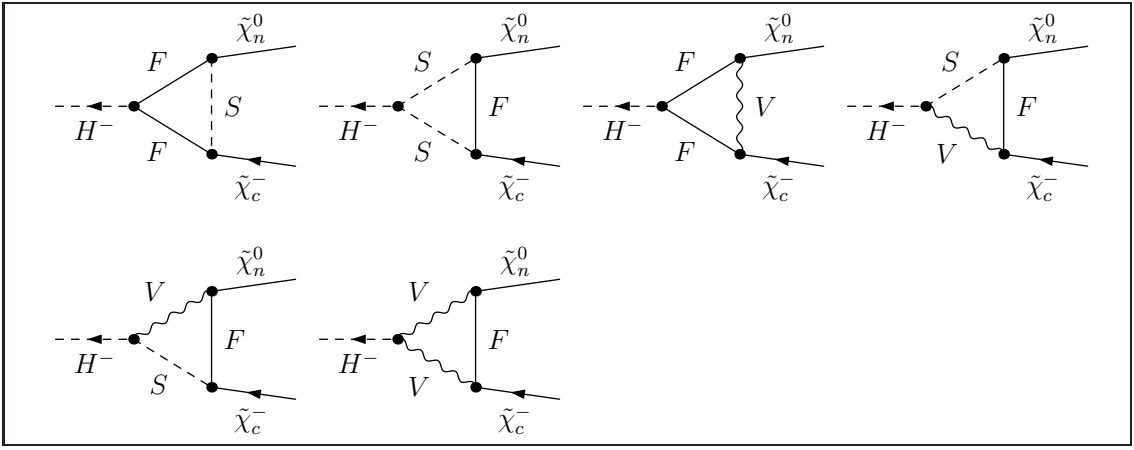}}
\caption{
  Generic Feynman diagrams for the decay $\Hpdecay$
  ($n = 1,2,3,4;\, c = 1,2$).
  (It should be noted that all arrows are inverted in case of a $H^-$ decay.)
  $F$ can be a SM fermion, chargino, neutralino;
  $S$ can be a sfermion or a Higgs/Goldstone boson;
  $V$ can be a $\ga$, $Z$, $W^\pm$. 
  Not shown are the diagrams with a $H^\pm$--$W^\pm$ or $H^\pm$--$G^\pm$ 
  transition contribution on the external Higgs boson leg.
}
\label{fig:Hpneucha}
\vspace{1em}
\end{center}
\end{figure}

In this section we give some details about the calculation of the
higher-order corrections to the partial decay widths of Higgs bosons. 
Sample diagrams for the decays $\hChaDecay$ 
($i = 1,2,3;\, c,\cpri = 1,2$), $\hNeuDecay$ ($i = 1,2,3;\, n,\npri = 1,2,3,4$) 
and $\Hpdecay$ ($n = 1,2,3,4;\, c = 1,2$) are shown in \reffis{fig:hchacha}, 
\ref{fig:hneuneu} and \ref{fig:Hpneucha}, respectively.
Not shown are the diagrams for real (hard and soft) photon radiation. 
They are obtained from the corresponding tree-level diagrams by 
attaching a photon to the electrically charged particles. 
The internal generically depicted particles in \reffis{fig:hchacha},
\ref{fig:hneuneu} and \ref{fig:Hpneucha} are labeled as follows:
$F$ can be a SM fermion $f$, chargino $\cha{c}$, neutralino $\neu{n}$; 
$S$ can be a sfermion $\Sf_s$ or a Higgs (Goldstone) boson $h_i$ $(G)$; 
$V$ can be a photon $\ga$ or a massive SM gauge boson, $Z$ or $W^\pm$. 
For internally appearing Higgs bosons no higher-order
corrections to their masses or couplings are taken into account; 
these corrections would correspond to effects beyond one-loop order.%
\footnote{
  We found that using loop corrected Higgs boson masses 
  in the loops leads to a UV divergent result.
}
For external Higgs bosons, as discussed in \citere{mhcMSSMlong}, the 
appropriate $\hat{Z}$~factors are applied and OS masses 
(including higher-order corrections) are used~\cite{mhcMSSMlong}, 
obtained with 
\FH~\cite{feynhiggs,mhiggslong,mhiggsAEC,mhcMSSMlong,Mh-logresum}.

%%%%%%%%%%%%%%%%%%%%%%%%% F I G U R E %%%%%%%%%%%%%%%%%%%%%%%%%%%%%%%%%%%%%%%%%
%%%%%%%%%%%%%%%%%%%%%%%%% F I G U R E %%%%%%%%%%%%%%%%%%%%%%%%%%%%%%%%%%%%%%%%%

%%%%%%%%%%%%%%%%%%%%%%%%% F I G U R E %%%%%%%%%%%%%%%%%%%%%%%%%%%%%%%%%%%%%%%%%
%%%%%%%%%%%%%%%%%%%%%%%%% F I G U R E %%%%%%%%%%%%%%%%%%%%%%%%%%%%%%%%%%%%%%%%%

%%%%%%%%%%%%%%%%%%%%%%%%% F I G U R E %%%%%%%%%%%%%%%%%%%%%%%%%%%%%%%%%%%%%%%%%
%%%%%%%%%%%%%%%%%%%%%%%%% F I G U R E %%%%%%%%%%%%%%%%%%%%%%%%%%%%%%%%%%%%%%%%%

Also not shown are the diagrams with a Higgs boson--gauge/Goldstone
self-energy contribution on the external Higgs boson leg. 
They appear in the decay $\hChaDecay$, \reffi{fig:hchacha} 
and $\hNeuDecay$, \reffi{fig:hneuneu}, with a $h_i$--$Z/G$ 
transition and in the decay $\HpmDecay$, \reffi{fig:Hpneucha}, 
with a $H^\pm$--$W^\pm$/$G^\pm$ transition.%
\footnote{
  From a technical point of view, the $H^\pm$--$W^\pm$/$G^\pm$ transitions 
  have been absorbed into the respective counterterms, while the 
  $h_i$--$Z/G$ transitions have been calculated explicitly.
}

Furthermore, in general, in \reffis{fig:hchacha} -- \ref{fig:Hpneucha} 
we have  omitted diagrams with self-energy type corrections of external 
(on-shell) particles.  While the contributions from the real parts of 
the loop functions are taken into account via the renormalization 
constants defined by OS renormalization conditions, the 
contributions coming from the imaginary part of the loop functions can 
result in an additional (real) correction if multiplied by complex 
parameters.  In the analytical and numerical evaluation, these diagrams 
have been taken into account via the prescription described in 
\citere{MSSMCT}. 

Within our one-loop calculation we neglect finite width effects that 
can help to cure threshold singularities.  Consequently, in the close 
vicinity of those thresholds our calculation does not give a reliable
result.  Switching to a complex mass scheme \cite{complexmassscheme} 
would be another possibility to cure this problem, but its application 
is beyond the scope of our paper.

The diagrams and corresponding amplitudes have been obtained with \FA\ 
\cite{feynarts}.  The model file, including the MSSM counterterms, 
is largely based on \citere{MSSMCT}, however adjusted to match exactly 
the renormalization prescription described in \refse{sec:renorm}. 
The further evaluation has been performed with \FC\ and \LT\ 
\cite{formcalc}.

\subsubsection*{Ultraviolet divergences}

As regularization scheme for the UV divergences we
have used constrained differential renormalization~\cite{cdr}, 
which has been shown to be equivalent to 
dimensional reduction~\cite{dred} at the \onel\ level~\cite{formcalc}. 
Thus the employed regularization scheme preserves SUSY~\cite{dredDS,dredDS2}
and guarantees that the SUSY relations are kept intact, \eg that the 
gauge couplings of the SM vertices and the Yukawa couplings of the 
corresponding SUSY vertices also coincide to \onel\ order in the SUSY limit. 
Therefore no additional shifts, which might occur when using a different 
regularization scheme, arise.
All UV divergences cancel in the final result.

\subsubsection*{Infrared divergences}

The IR divergences from diagrams with an internal photon have
to cancel with the ones from the corresponding real soft radiation. 
In the case of QED we have included the soft photon contribution
following the description given in \citere{denner}. 
The IR divergences arising from the diagrams involving a $\ga$ 
are regularized by introducing a photon mass parameter, $\la$. 
All IR divergences, \ie all divergences in the limit $\la \to 0$, 
cancel once virtual and real diagrams for one decay channel are added.

\subsubsection*{Tree-level formulas}

For completeness we show here also the formulas that have been
used to calculate the tree-level decay widths:
\begin{align}
\label{HpmTree}
\Gamma^{\rm tree}(\HpmDecay) &= \Big[ \LP 
          |C(H^\pm, \neu{n}, \cha{c})_L|^2
        + |C(H^\pm, \neu{n}, \cha{c})_R|^2 \RP 
          (\MHp^2 - \mneu{n}^2 - \mcha{c}^2) \notag \\
&\qquad - 4\, \Re \LV C(H^\pm, \neu{n}, \cha{c})_L^*\, 
                      C(H^\pm, \neu{n}, \cha{c})_R \RV\,
          \mneu{n}\, \mcha{c} \Big] \times \notag \\
&\mathrel{\phantom{=}} 
          \frac{\la^{1/2}(\MHp^2,\mneu{n}^2,\mcha{c}^2)}
               {16\, \pi\, \MHp^3} \qquad 
          (c = 1,2;\, n = 1,2,3,4)\,, \\
\label{hChaTree}
\Gamma^{\rm tree}(\hChaDecay) &= \Big[ \LP 
          |C(h_i, \cham{\cind}, \chap{\cpri})_L|^2
        + |C(h_i, \cham{\cind}, \chap{\cpri})_R|^2 \RP 
          (\mh{i}^2 - \mcham{\cind}^2 - \mchap{\cpri}^2) \notag \\
&\qquad - 4\, 
%\grey{(-1)^{\de_{i3}}}\, 
\Re \LV C(h_i, \cham{\cind}, \chap{\cpri})_L^*\, 
                      C(h_i, \cham{\cind}, \chap{\cpri})_R \RV\,
          \mcham{\cind}\, \mchap{\cpri} \Big] \times \notag \\
&\mathrel{\phantom{=}}
          \frac{\la^{1/2}(\mh{i}^2,\mcham{\cind}^2,\mchap{\cpri}^2)}
               {16\, \pi\, \mh{i}^3} \qquad 
          (i = 1,2,3;\, c,\cpri = 1,2)\,, \\
\label{hNeuTree}
\Gamma^{\rm tree}(\hNeuDecay) &= \Big[ \LP 
          |C(h_i, \neu{n}, \neu{\npri})_L|^2
        + |C(h_i, \neu{n}, \neu{\npri})_R|^2 \RP 
          (\mh{i}^2 - \mneu{\nind}^2 - \mneu{\npri}^2) \notag \\
&\qquad - 4\, 
%\grey{(-1)^{\de_{i3}}}\, 
\Re \LV C(h_i, \neu{\nind}, \neu{\npri})_L^*\, 
                      C(h_i, \neu{\nind}, \neu{\npri})_R \RV\,
          \mneu{\nind}\, \mneu{\npri} \Big] \times \notag \\
&\mathrel{\phantom{=}} 
          \frac{\la^{1/2}(\mh{i}^2,\mneu{\nind}^2,\mneu{\npri}^2)}
               {16\, \pi\, \mh{i}^3} \qquad 
          (i = 1,2,3;\, n,\npri = 1,2,3,4)\,,
\end{align}
where $\la(x,y,z) = (x - y - z)^2 - 4yz$. 
The couplings $C(a, b, c)$ can be found in the \FA~model 
files, see \citere{feynarts-mf}.
$C(a, b, c)_{L,R}$ denote the part of the coupling which
is proportional to $(\unity \mp \ga_5)/2$.
For the later interpretation of the results in the numerical analysis the 
following should be kept in mind.  In \refeqs{hChaTree}, (\ref{hNeuTree}) 
the couplings of the Higgs to charginos/neutralinos result in a relative 
plus (minus) sign between the two terms (in the first and second line of 
each equation, respectively) for $h_i$ being a $\CP$-odd ($\CP$-even) Higgs, 
leading to an enhancement (suppression) of the decay width.
In case of equal final state masses ($c = \cpri$ or $n = \npri$) one finds 
in \FA\ convention%
\footnote{
  It should be noted that the convention for Feynman rules in
  \citere{HaK85} and in \citere{feynarts-mf} differ by a 
  global factor of $-i$, which would formally lead to 
  $C(h_i, \ino{}, \ino{})_L = +C(h_i, \ino{}, \ino{})^*_R$. 
  However, the physics outcome remains, of course, unchanged.
}%
~$C(h_i, \ino{}, \ino{})_L = -C(h_i, \ino{}, \ino{})^*_R =: 
C(h_i, \ino{}, \ino{})$,
and the general structure of the tree-level decay width simplifies 
for \textit{real} parameters to
\begin{align}
\label{hOddTree}
\Gamma^{\rm tree}(h_i \to \ino{} \ino{}) &= 
\frac{|C(h_i, \ino{}, \ino{})|^2}{8\, \pi} 
\LB \mh{i}^2 - 4\, m_{\ino{}}^2 \RB^{(1/2)} 
\qquad \text{for $h_i$ $\CP$-odd}\,, \\
\label{hEvenTree}
\Gamma^{\rm tree}(h_i \to \ino{} \ino{}) &= 
\frac{|C(h_i, \ino{}, \ino{})|^2}{8\, \pi\, \mh{i}^2} 
\LB \mh{i}^2 - 4\, m_{\ino{}}^2 \RB^{(3/2)} 
\qquad \text{for $h_i$ $\CP$-even}\,.
\end{align}
The latter decay width exhibits a $p$-wave suppression.

%%%%%%%%%%%%%%%%%%%%%%%%%%%%%%%%%%%%%%%%%%%%%%%%%%%%%%%%%%%%%%%%%%%%%%%%%%%%%%%
%%%%%%%%%%%%%%%%%%%%%%%%%%%%%%%%%%%%%%%%%%%%%%%%%%%%%%%%%%%%%%%%%%%%%%%%%%%%%%%

\section{Numerical analysis}
\label{sec:numeval}

In this section we present the comparisons with results from other groups
and our numerical analysis of the light and heavy Higgs boson decay 
channels into charginos and neutralinos in the cMSSM. 
In the various figures below we show the partial decay widths and their 
relative correction at the tree-level (``tree'') and at the one-loop 
level (``full'').

%%%%%%%%%%%%%%%%%%%%%%%%%%%%%%%%%%%%%%%%%%%%%%%%%%%%%%%%%%%%%%%%%%%%%%%%%%%%%

\subsection{Comparisons}
\label{sec:comparisons}

We performed exhaustive comparisons with results from other groups
for Higgs boson decays into charginos and neutralinos.  
Most of these comparisons were restricted to the MSSM with real 
parameters.

\begin{itemize}

\item
%hep-ph/0201132: A/H -> neu2 neu2, cha1 cha1, leading Yukawa corrections,
%                "full OS scheme" (a la SM, keine Details) in rMSSM
A comparison with \citere{AHn2n2c1c1} (in the rMSSM) gave an overall
qualitative agreement for the decays $H/A \to \chap1 \cham1$, as was 
to be expected, because that work took into account only the 
leading Yukawa corrections and used a different renormalization 
scheme.  On the other hand, we omit a comparison with the results for 
$H/A \to \neu2 \neu2$ of \citere{AHn2n2c1c1}, because in their set-up 
neutralino masses were used as input parameters, which is rather 
difficult to adapt to our numerical analysis set-up.

\item
%hep-ph/0405187: A/H -> cha1,2 cha1,2 full 1L in rMSSM,
%                chargino Renormierung "OS", keine Details.
For the comparison with \citere{AHc12c12} we calculated the decays 
$A/H \to \chap1 \cham1$ at $\order{\alpha(\MZ)}$, using their input 
parameters as far as possible. We found good (qualitative) agreement 
with \citere{AHc12c12} (where the calculation was restricted to the 
rMSSM).  We successfully reproduced their Figs.~2, 4, 5 and~6, where 
only a small difference remains due to the different renormalization 
schemes, see also \refta{tab:Ibrahim} with differences below 5\%.

\item
%arXiv:0704.1913/arXiv:0803.4134: betrachtet H/A -> cha cha/neu neu
%                                 "effective 1L Lagrangian" im rMSSM
We performed a numerical comparison with \citere{AHnnccLeff} for the 
decay $H/A \to \chap{1} \cham{1}$ at $\order{\alpha(\MZ)}$ (in the rMSSM),
see \refta{tab:Ibrahim}, where the columns for \citere{AHc12c12}
and \citere{AHnnccLeff} where taken over from Tab.~1 of the first
article in \citere{AHnnccLeff}.  Their set-up differs from ours in the 
renormalization of the chargino/neutralino sector, leading to different 
loop corrections. 
Furthermore they used an ``effective one-loop Lagrangian''.
Nevertheless, using their input parameters as far as possible, we found 
differences below the 10\%~level.

%%%%%%%%%%%%%%%%%%%%% T A B L E %%%%%%%%%%%%%%%%%%%%%%%%%%%%%%%%%%%%%%%%%%%%%%
\begin{table}[t!]
\caption{\label{tab:Ibrahim}
  Comparison of the 1-loop corrected partial decay widths 
  (in GeV) with \citere{AHnnccLeff}.
}
\centering
\begin{tabular}{llrrr}
\toprule
Process & Mass & \citere{AHc12c12} & \citere{AHnnccLeff} & \FT \\
\midrule
$A^0 \to \chap1 \cham1$ & $M_A = 700$ & 0.85 & 0.80 & 0.83 \\
$A^0 \to \chap1 \cham1$ & $M_A = 800$ & 1.00 & 0.91 & 0.96 \\
$H^0 \to \chap1 \cham1$ & $M_H = 800$ & 0.63 & 0.58 & 0.64 \\
$H^0 \to \chap1 \cham1$ & $M_H = 900$ & 0.73 & 0.70 & 0.75 \\
\bottomrule
\end{tabular}
\vspace{1em}
\end{table}
%%%%%%%%%%%%%%%%%%%%% T A B L E %%%%%%%%%%%%%%%%%%%%%%%%%%%%%%%%%%%%%%%%%%%%%%

\item
% HFOLD, arXiv:1012.5025 (im rMSSM, DRbar):
A numerical comparison with the program \HFOLD~\citere{hfold}
at the benchmark point SPS1a` (proposed in the SPA project~\cite{spa}) 
can be found in \refta{tab:hfold}.
Only for this point sufficient details about the chargino/neutralino 
masses was available for a numerical comparison. 
In \refta{tab:hfold} we show the full one-loop results 
of \HFOLD, using \DRbar\ masses for the internal and external 
particles, corresponding to the full \DRbar\ renormalization used in 
the code (where the renormalization scale was set to $1\tev$~\cite{spa}). 
Our results, labeled \FT, are evaluated using our 
renormalization scheme, but inserting the \HFOLD\ \DRbar\ masses.
In the tree-level results we find more than 10 digits agreement and
in the full results we find agreement of 3\% -- 15\% (7\% on average).

\HFOLD\ also offers to switch to (the recommended) OS masses 
for the external particles.  In this case, we are including in our 
calculation $\mneu{2}^\os$ as described in \refse{sec:chaneu},
\refeq{eq:mneuOS}, but using the same OS Higgs boson masses as in
\HFOLD. With it the agreement between the two calculations is 
11\% on average (5\% -- 22\%).

%%%%%%%%%%%%%%%%%%%%% T A B L E %%%%%%%%%%%%%%%%%%%%%%%%%%%%%%%%%%%%%%%%%%%%%%
\begin{table}[t!]
\caption{\label{tab:hfold}
  Comparison of the 1-loop corrected partial decay widths 
  (in $10^{-1}\gev$) with \HFOLD.
}
\centering
\begin{tabular}{lrrcrr}
\toprule
& \multicolumn{2}{c}{OS masses} & & \multicolumn{2}{c}{\DRbar\ masses} \\ 
\cmidrule{2-3}\cmidrule{5-6} 
Process & \HFOLD & \FT & & \HFOLD & \FT \\
\midrule
$H^0 \to \neu1 \neu1$   & 0.1381 & 0.1648 & & 0.1046 & 0.1229 \\
$H^0 \to \neu1 \neu2$   & 0.4584 & 0.4908 & & 0.2690 & 0.2828 \\
$H^0 \to \neu2 \neu2$   & 0.2061 & 0.2259 & & 0.0117 & 0.0111 \\
$H^0 \to \chap1 \cham1$ & 0.5262 & 0.5672 & & 0.0345 & 0.0332 \\
$A^0 \to \neu1 \neu1$   & 0.2044 & 0.2404 & & 0.1704 & 0.2016 \\
$A^0 \to \neu1 \neu2$   & 0.9693 & 1.0248 & & 0.7334 & 0.7750 \\
$A^0 \to \neu2 \neu2$   & 1.1652 & 1.0747 & & 0.3966 & 0.3791 \\
$A^0 \to \chap1 \cham1$ & 2.8604 & 2.6454 & & 1.0236 & 0.9928 \\
$H^+ \to \neu1 \chap1$  & 1.2981 & 1.4307 & & 0.9333 & 0.9996 \\
$H^+ \to \neu2 \chap1$  & 0.0063 & 0.0081 & & 0.0026 & 0.0030 \\
\bottomrule
\end{tabular}
\vspace{1em}
\end{table}
%%%%%%%%%%%%%%%%%%%%% T A B L E %%%%%%%%%%%%%%%%%%%%%%%%%%%%%%%%%%%%%%%%%%%%%%

\item
%arXiv:1211.3134: h_a -> cha1,2 cha1,2, full 1L in cMSSM,
%                 kurze Betrachtung von CCN, CNN, NNN,
%                 kurze Analyze von phiAt, phiM1, phimu.
Decays of $h_{2,3}$ to charginos in the cMSSM at the full one-loop 
level have been numerically compared with \citere{hHAcc-cMSSM} using their 
latest \FA/\FC\ model file implementation.  We found overall agreement 
better than 1\% in the loop corrections for real and complex parameters.%
\footnote{
  It should be noted that the original code used for \citere{hHAcc-cMSSM} 
  is no longer available~\cite{aoife}, where we found some numerical 
  differences with the results shown in \citere{hHAcc-cMSSM} in the case 
  of complex parameters.
}

\item
%PhD von Alison Fowler: h_a -> neu neu full 1L in cMSSM
%                       Numerische Analyse von h_2,3 -> neu2 neu2 (phiAt).
$h_{2,3}$ boson decays into $\neu2 \neu2$ in the cMSSM have been analyzed 
in \citere{dissAF}.  Again we had to use here the latest \FA/\FC\ model 
file implementation of \citere{hHAcc-cMSSM} (which bases mainly on code 
from \citere{dissAF}) for the same reasons as described in the
previous item.  In comparison with that model file \cite{hHAcc-cMSSM} we 
found overall agreement better than 2\% in the loop corrections for real 
and complex parameters.

\end{itemize}

%%%%%%%%%%%%%%%%%%%%%%%%%%%%%%%%%%%%%%%%%%%%%%%%%%%%%%%%%%%%%%%%%%%%%%%%%%%%%

\subsection{Parameter settings}
\label{sec:paraset}

The renormalization scale $\mu_R$ has been set to the mass of the 
decaying Higgs boson.  The SM parameters are chosen as follows; 
see also \cite{pdg}:
\begin{itemize}

\item Fermion masses (on-shell masses, if not indicated differently):
\begin{align}
m_e    &= 0.510998928\mev\,, & m_{\nu_e}    &= 0\mev\,, \notag \\
m_\mu  &= 105.65837515\mev\,, & m_{\nu_{\mu}} &= 0\mev\,, \notag \\
m_\tau &= 1776.82\mev\,,      & m_{\nu_{\tau}} &= 0\mev\,, \notag \\
m_u &= 68.7\mev\,,           & m_d         &= 68.7\mev\,, \notag \\ 
m_c &= 1.275\gev\,,          & m_s         &= 95.0\mev\,, \notag \\
m_t &= 173.21\gev\,,         & m_b         &= 4.18\gev\,.
\end{align}
According to \citere{pdg}, $m_s$ is an estimate of a so-called 
"current quark mass" in the \MSbar\ scheme at the scale 
$\mu \approx 2\gev$.  $m_c \equiv m_c(m_c)$ and $m_b \equiv m_b(m_b)$ 
are the "running" masses in the \MSbar\ scheme.%
\footnote{
  It should be noted, that in the analysis below, we use the \DRbar\ 
  mass $m_b^{\DRbar}$ from Eq.(19) of \citere{HiggsDecaySferm}.
}
$m_u$ and $m_d$ are effective parameters, calculated through the hadronic
contributions to
\begin{align}
\Delta\alpha_{\text{had}}^{(5)}(M_Z) &= 
      \frac{\alpha}{\pi}\sum_{f = u,c,d,s,b}
      Q_f^2 \Bigl(\ln\frac{M_Z^2}{m_f^2} - \frac 53\Bigr) \sim 0.027723\,.
\end{align}

\item Gauge boson masses\index{gaugebosonmasses}:
\begin{align}
M_Z = 91.1876\gev\,, \qquad M_W = 80.385\gev\,.
\end{align}

\item Coupling constant\index{couplingconstants}:
\begin{align}
\alpha(0) = 1/137.0359895\,.
\end{align}
\end{itemize}

The Higgs sector quantities (masses, mixings, etc.) have been
evaluated using \FH\ (version 2.10.2)
\cite{feynhiggs,mhiggslong,mhiggsAEC,mhcMSSMlong,Mh-logresum}.

%%%%%%%%%%%%%%%%%%%%% T A B L E %%%%%%%%%%%%%%%%%%%%%%%%%%%%%%%%%%%%%%%%%%%%%%
\begin{table}[t!]
\caption{\label{tab:para}
  MSSM parameters for the initial numerical investigation; all parameters 
  (except of $\TB$) are in GeV (calculated masses are rounded to 1 MeV). 
  In our analysis $M_{\tilde Q_3}$, $M_{\tilde U_3}$, $M_{\tilde D_3}$,
  $M_{\tilde L_3}$ and $M_{\tilde E_3}$ are chosen such that the values of 
  $\mstop1$, $\mstop2$, $\msbot2$, $\mtausneu$ and $\mstau2$ are realized. 
  For the sfermion sector the shifts in $M_{\tilde Q, \tilde D}(\Sd_g)$ and 
  $M_{\tilde L, \tilde E}(\Se_g)$ as defined in \citere{HiggsDecaySferm} are 
  taken into account.  The values for the trilinear sfermion Higgs
  couplings, $A_{\Fu_g,\Fd_g,\Fe_g}$ ($g = 1,2,3$; identical for all~$g$)
  are chosen such that charge- and/or color-breaking minima are 
  avoided~\cite{ccb}.  It should be noted that for the first and second 
  generation of sfermions we chose instead $M_{\tilde L, \tilde E} = 1500\gev$ 
  and $M_{\tilde Q, \tilde U, \tilde D} = 2000\gev$.  For the neutralino sector 
  the shifts in \refeq{eq:Deltamneu} are taken into account.
}
\centering
\begin{tabular}{lrrrrrrrrrrrrr}
\toprule
Scen. & $\TB$ & $\mu$ & $A_{\Fu_g}$ & $A_{\Fd_g}$ & $A_{\Fe_g}$ & $|M_1|$ &
$M_2$ & $M_3$ & $\mstop1$ & $\mstop2$ & $\msbot2$ & $\mtausneu$ & $\mstau2$ \\ 
\midrule
SX &  10 & 500 & 1200 & 600 & 1000 & 300 & 600 & 1500 
   & 394 & 771 &  582 & 280 &  309 \\
\bottomrule
\end{tabular}

\vspace{1em}

\begin{tabular}{lrrrrr}
\toprule
Scen.     &    \Sce &    \Scz &    \Scd &     \Scv &     \Scf \\
$\MHp$    &     700 &     900 &    1000 &     1200 &     1400 \\
\midrule
$\mh1$    & 123.487 & 123.509 & 123.517 &  123.529 &  123.539 \\
$\mh2$    & 694.483 & 895.594 & 996.769 & 1195.095 & 1397.300 \\
$\mh3$    & 695.425 & 896.931 & 996.818 & 1197.407 & 1398.600 \\
$\mstau1$ & 282.705 & 282.573 & 282.517 &  282.420 &  282.336 \\
$\msbot1$ & 513.733 & 513.621 & 513.578 &  513.509 &  513.455 \\
$\mneu1$  & 295.269 & 295.269 & 295.269 &  295.269 &  295.269 \\
$\mneu2$  & 476.772 & 476.763 & 476.759 &  476.753 &  476.748 \\
$\mneu3$  & 496.992 & 496.988 & 496.986 &  496.983 &  496.980 \\
$\mneu4$  & 632.326 & 632.324 & 632.324 &  632.323 &  632.322 \\
$\mcha1$  & 472.534 & 472.534 & 472.534 &  472.534 &  472.534 \\
$\mcha2$  & 632.167 & 632.167 & 632.167 &  632.167 &  632.167 \\
\bottomrule
\end{tabular}
\end{table}
%%%%%%%%%%%%%%%%%%%%% T A B L E %%%%%%%%%%%%%%%%%%%%%%%%%%%%%%%%%%%%%%%%%%%%%%

We emphasize again that the analytical calculation has been done for 
\textit{all} decays into charginos/neutralinos.  Results are shown for 
some representative numerical examples.  The parameters are chosen 
according to the scenarios, SX (X = 1,2,...,5), shown in \refta{tab:para},
unless otherwise noted. 
The scenarios are defined such that a maximum number of decay modes are 
open simultaneously to permit an analysis of all channels, \ie not picking 
specific parameters for each decay.  For the same reason we do not demand 
that the lightest Higgs boson has a mass around $\sim 125\gev$, although 
for most of the parameter space this is given.  
For the light Higgs we will show the variation with 
$\MHp$, $|\mu|$, $M_1$ and $\phiMe$
(where the latter denotes the phase of the gaugino mass parameter $M_1$), 
whereas for the heavy Higgs bosons we will analyze the variation of 
$\MHp$ and $\phiMe$.

The numerical results shown in the next subsections are of course 
dependent on choice of the SUSY parameters. Nevertheless, they give an 
idea of the relevance of the full one-loop corrections.  Channels 
(and their respective one-loop corrections) that may look unobservable 
due to the smallness of their decay width in the plots shown below, 
could become important if other channels are kinematically forbidden.

%%%%%%%%%%%%%%%%%%%%%%%%%%%%%%%%%%%%%%%%%%%%%%%%%%%%%%%%%%%%%%%%%%%%%%%%%%%%%%

\subsection{Full one-loop results for varying \boldmath{$\MHp$},
  \boldmath{$M_1$}, and \boldmath{$\phiMe$}}
\label{sec:full1L}

The results shown in this and the following subsections consist of 
``tree'', which denotes the tree-level value and of ``full'', which 
is the partial decay width including \textit{all} one-loop corrections 
as described in \refse{sec:calc}. 
We restrict ourselves to the analysis of the decay widths themselves, 
since the one-loop effects on the branching ratios are strongly 
parameter dependent, as discussed in the previous subsection.

When performing an analysis involving complex parameters it should be 
noted that the results for physical observables are affected only by 
certain combinations of the complex phases of the parameters $\mu$, 
the trilinear couplings $A_f$ and the gaugino mass parameters 
$M_{1,2,3}$~\cite{MSSMcomplphasen,SUSYphases}.
It is possible, for instance, to rotate the phase $\phiMz$ away.
Experimental constraints on the (combinations of) complex phases 
arise, in particular, from their contributions to electric dipole 
moments of the electron and the neutron (see \citeres{EDMrev2,EDMPilaftsis} 
and references therein), of the deuteron~\cite{EDMRitz} and of heavy 
quarks~\cite{EDMDoink}.
While SM contributions enter only at the three-loop level, due to its
complex phases the MSSM can contribute already at one-loop order.
Large phases in the first two generations of sfermions can only be 
accommodated if these generations are assumed to be very heavy 
\cite{EDMheavy} or large cancellations occur~\cite{EDMmiracle};
see, however, the discussion in \citere{EDMrev1}. 
A review can be found in \citere{EDMrev3}.
Accordingly (using the convention that $\phiMz = 0$, as done in this paper), 
in particular, the phase $\phimu$ is tightly constrained~\cite{plehnix}, 
while the bounds on the phases of the third generation trilinear couplings 
are much weaker.  Setting $\phimu = 0$ and $\varphi_{A_f} = 0$ leaves us with 
$\phiMe$ as the only complex valued parameter. 

Since now the complex gaugino mass parameter $M_1$ can appear in the 
couplings, contributions from absorptive parts of self-energy type 
corrections on external legs can arise.  The corresponding formulas 
for an inclusion of these absorptive contributions via finite wave 
function correction factors can be found in \citeres{MSSMCT,Stop2decay}.

We begin the numerical analysis with partial decay widths of $H^\pm$
evaluated as a function of $\MHp$, starting at $\MHp = 600\gev$ 
up to $\MHp = 1.6\tev$, which roughly coincides with the reach of 
the LHC for high-luminosity running as well as an $e^+e^-$ collider
with a center-of-mass energy up to $\sqrt{s} \sim 3\tev$~\cite{CLIC}.
Then we turn to the $h_i$ ($i = 1,2,3$) decays. 
Finally, it should be noted that we expect from the tree-level 
formulas \refeqs{HpmTree} -- (\ref{hNeuTree}) that the decay widths 
increase (roughly) linearly with the corresponding Higgs boson masses.

%%%%%%%%%%%%%%%%%%%%%%%%%%%%%%%%%%%%%%%%%%%%%%%%%%%%%%%%%%%%%%%%%%%%%%%%%%%%%%

\subsubsection{\boldmath{$H^\pm$} decays into charginos/neutralinos}
\label{Hpdecays}

\begin{figure}[htb!]
\begin{center}
\begin{tabular}{c}
\includegraphics[width=0.49\textwidth,height=7.5cm]{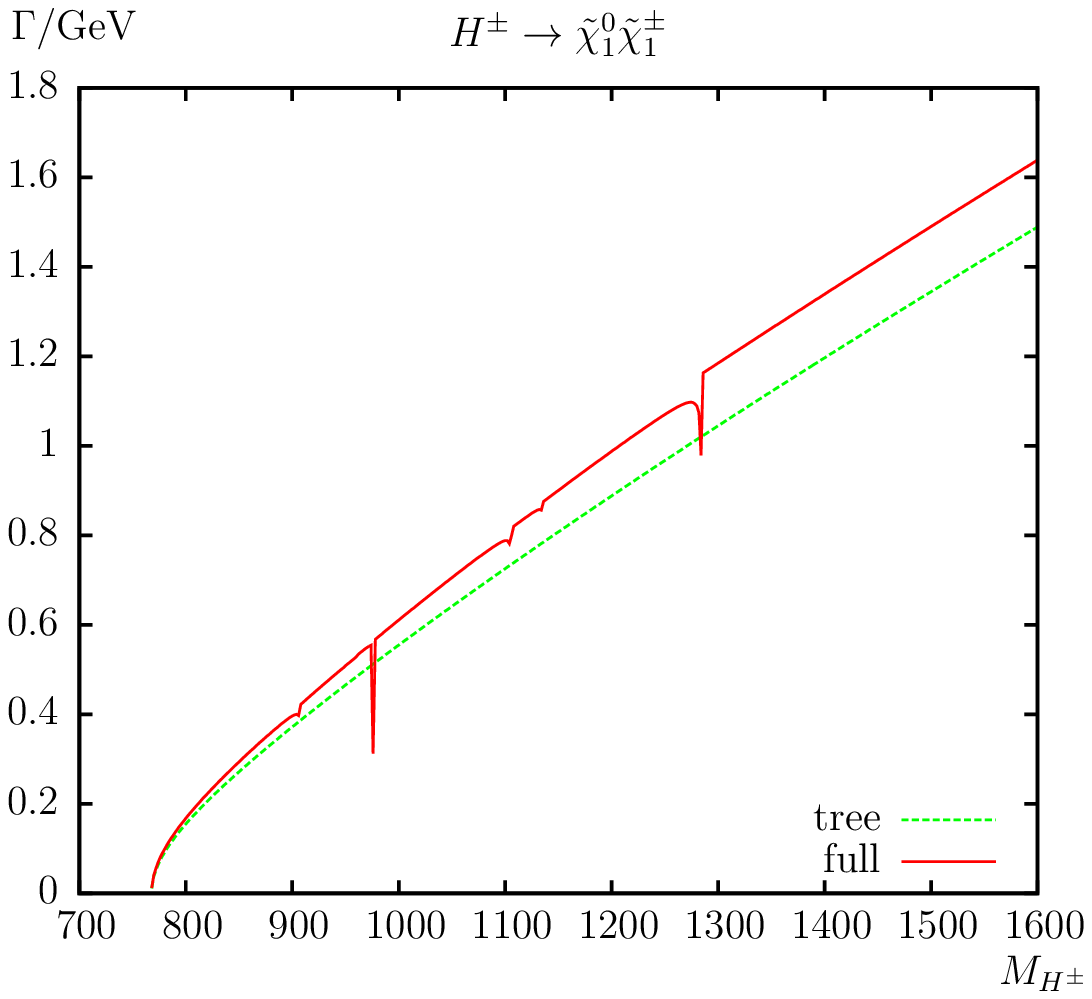}
\hspace{-4mm}
\includegraphics[width=0.49\textwidth,height=7.5cm]{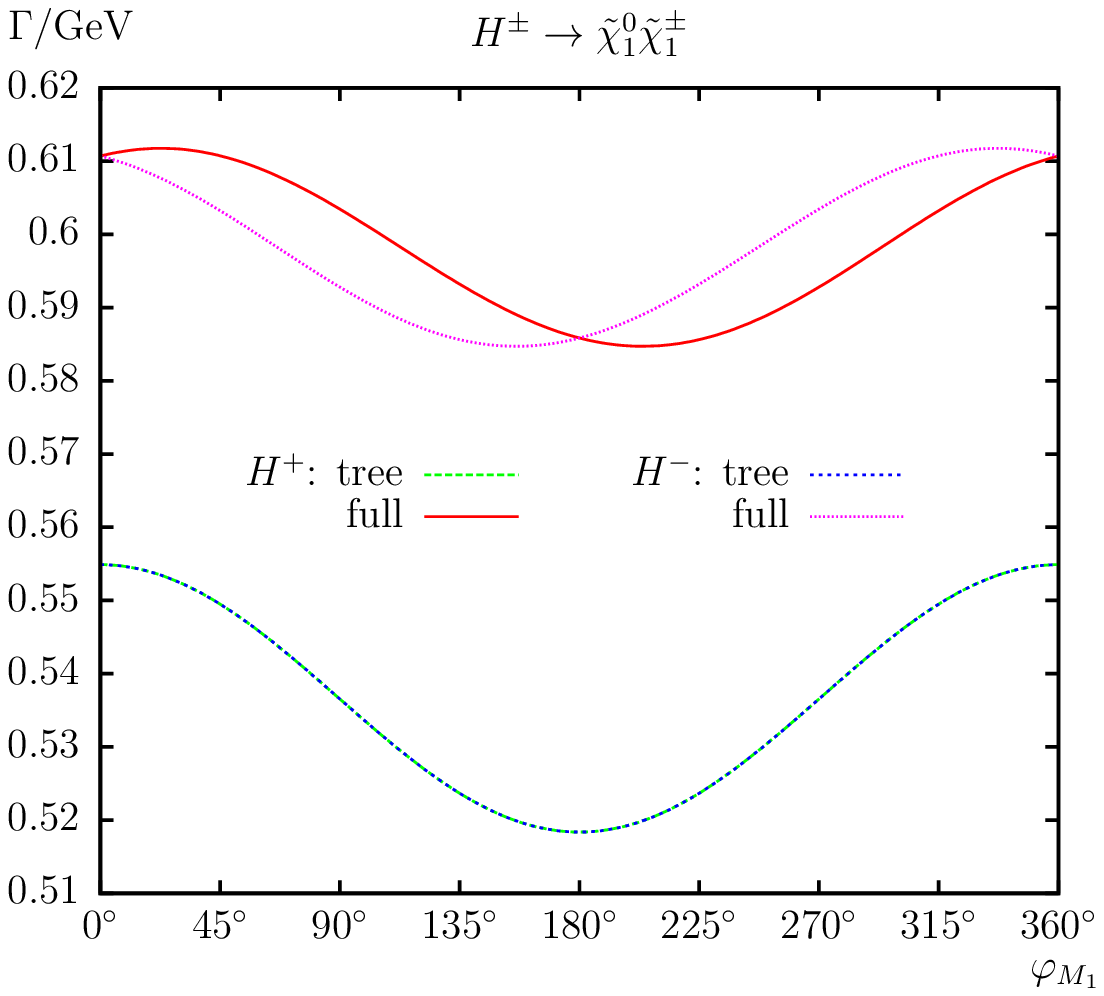}
\end{tabular}
\vspace{1em}
\caption{\label{fig:Hneu1cha1}
  $\Ga(H^\pm \to \neu1 \chapm1)$. 
  Tree-level and full one-loop corrected partial decay widths are shown. 
  The left plot shows the partial decay width with $\MHp$ varied.
  The right plot shows the complex phase $\phiMe$ varied with parameters 
  chosen according to \Scd\ (see \refta{tab:para}).
}
\vspace{6em}
\begin{tabular}{c}
\includegraphics[width=0.49\textwidth,height=7.5cm]{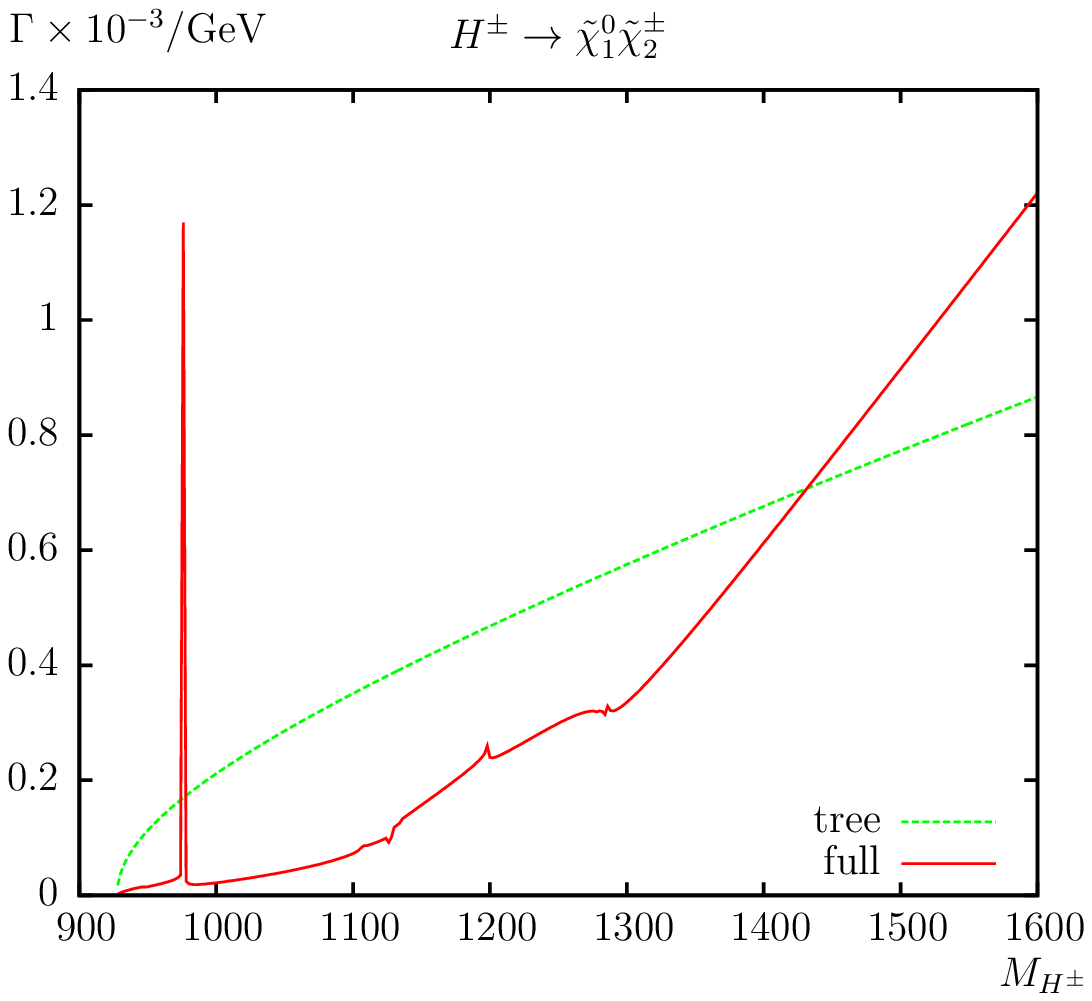}
\hspace{-4mm}
\includegraphics[width=0.49\textwidth,height=7.5cm]{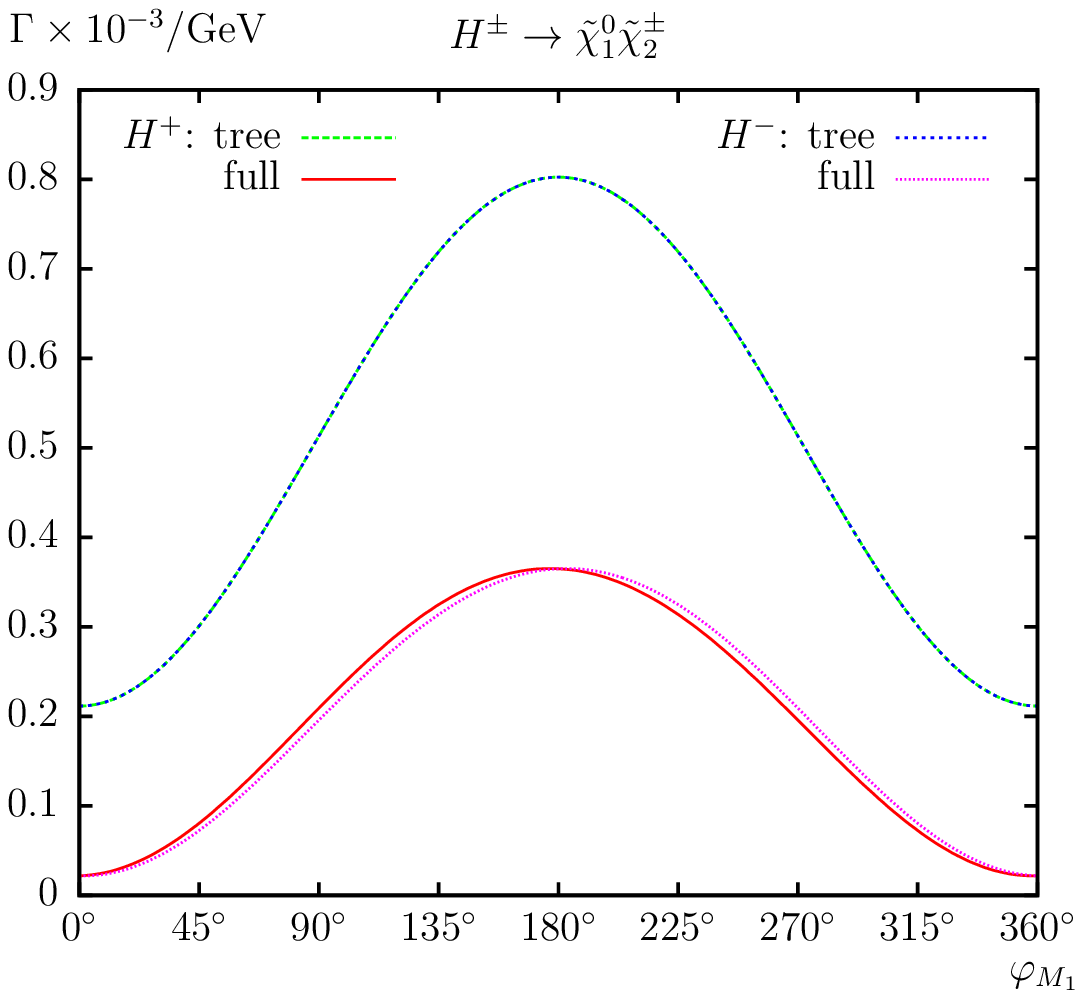}
\end{tabular}
\vspace{1em}
\caption{\label{fig:Hneu1cha2}
  $\Ga(H^\pm \to \neu1 \chapm2)$. 
  Tree-level and full one-loop corrected partial decay widths are shown. 
  The left plot shows the partial decay width with $\MHp$ varied.
  The right plot shows the complex phase $\phiMe$ varied with parameters 
  chosen according to \Scd\ (see \refta{tab:para}).
}
\end{center}
\end{figure}

\begin{figure}[htb!]
\begin{center}
\begin{tabular}{c}
\includegraphics[width=0.49\textwidth,height=7.5cm]{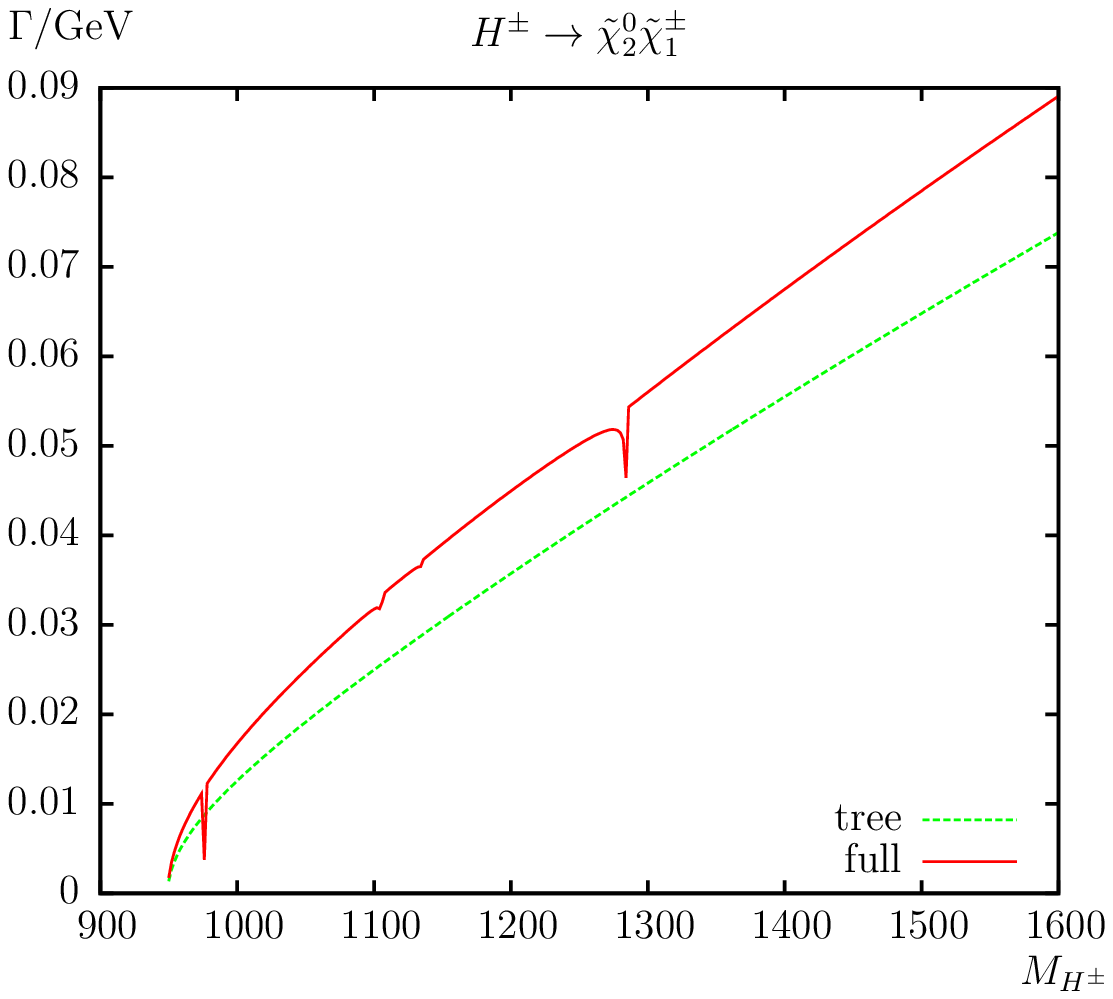}
\hspace{-4mm}
\includegraphics[width=0.49\textwidth,height=7.5cm]{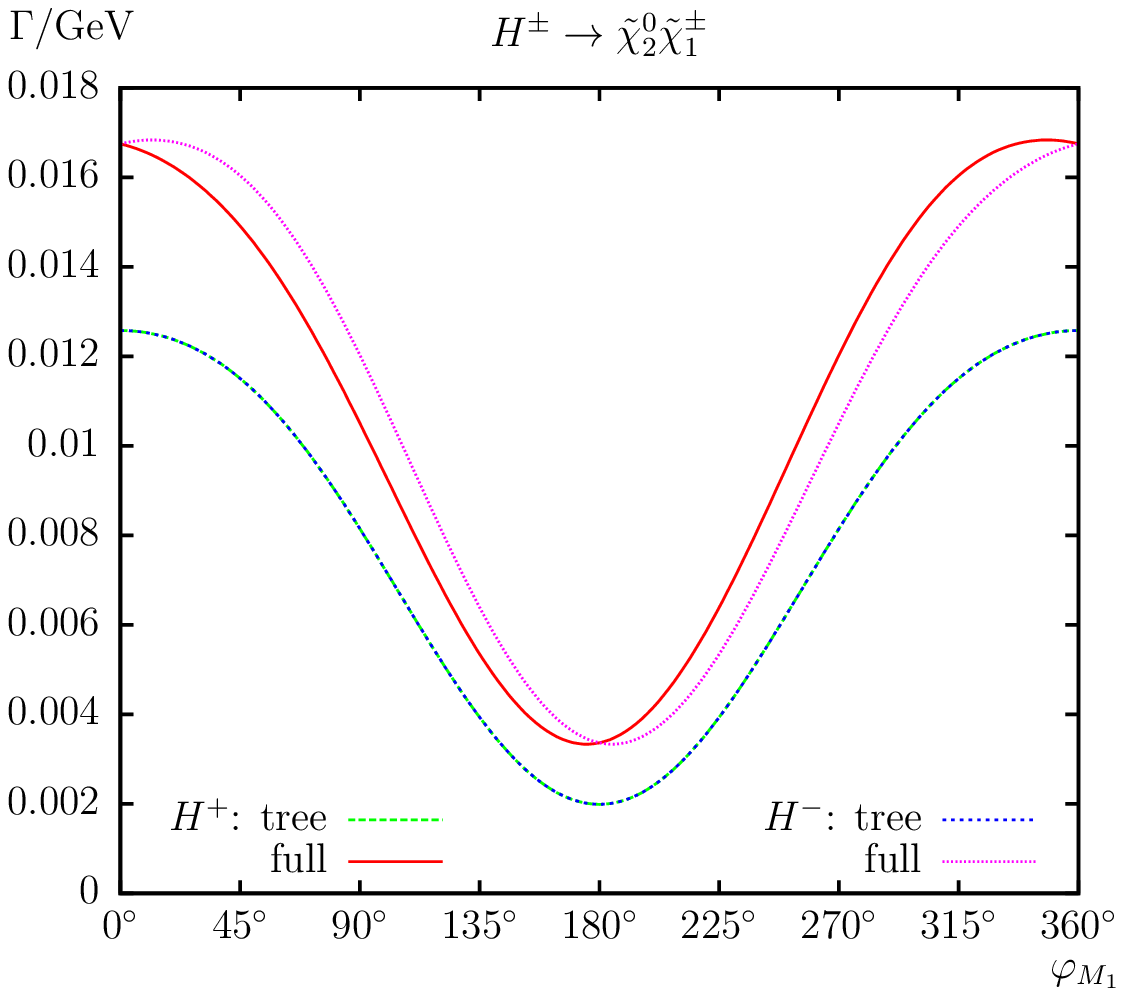}
\end{tabular}
\vspace{1em}
\caption{\label{fig:Hneu2cha1}
  $\Ga(H^\pm \to \neu2 \chapm1)$. 
  Tree-level and full one-loop corrected partial decay widths are shown. 
  The left plot shows the partial decay width with $\MHp$ varied.
  The right plot shows the complex phase $\phiMe$ varied with parameters 
  chosen according to \Scd\ (see \refta{tab:para}).
}
\vspace{6em}
\begin{tabular}{c}
\includegraphics[width=0.49\textwidth,height=7.5cm]{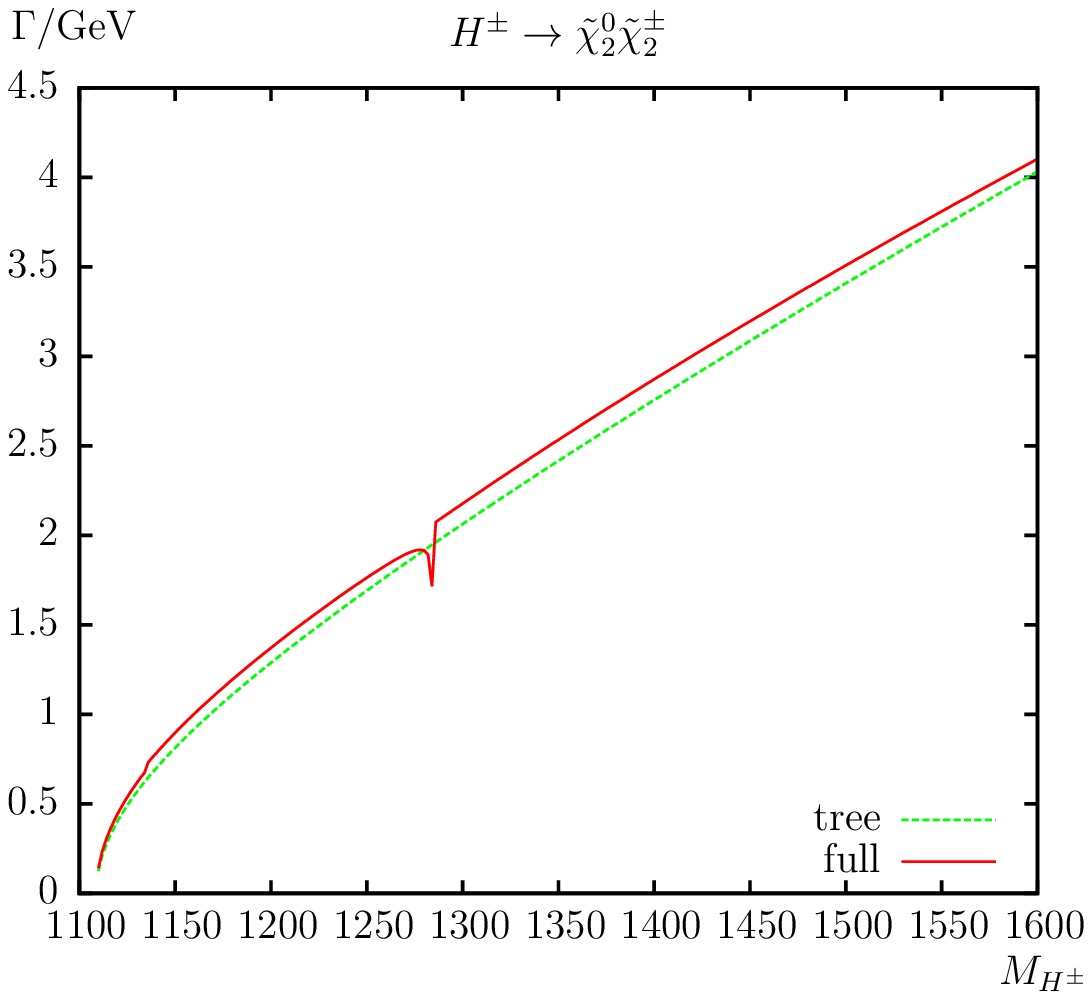}
\hspace{-4mm}
\includegraphics[width=0.49\textwidth,height=7.5cm]{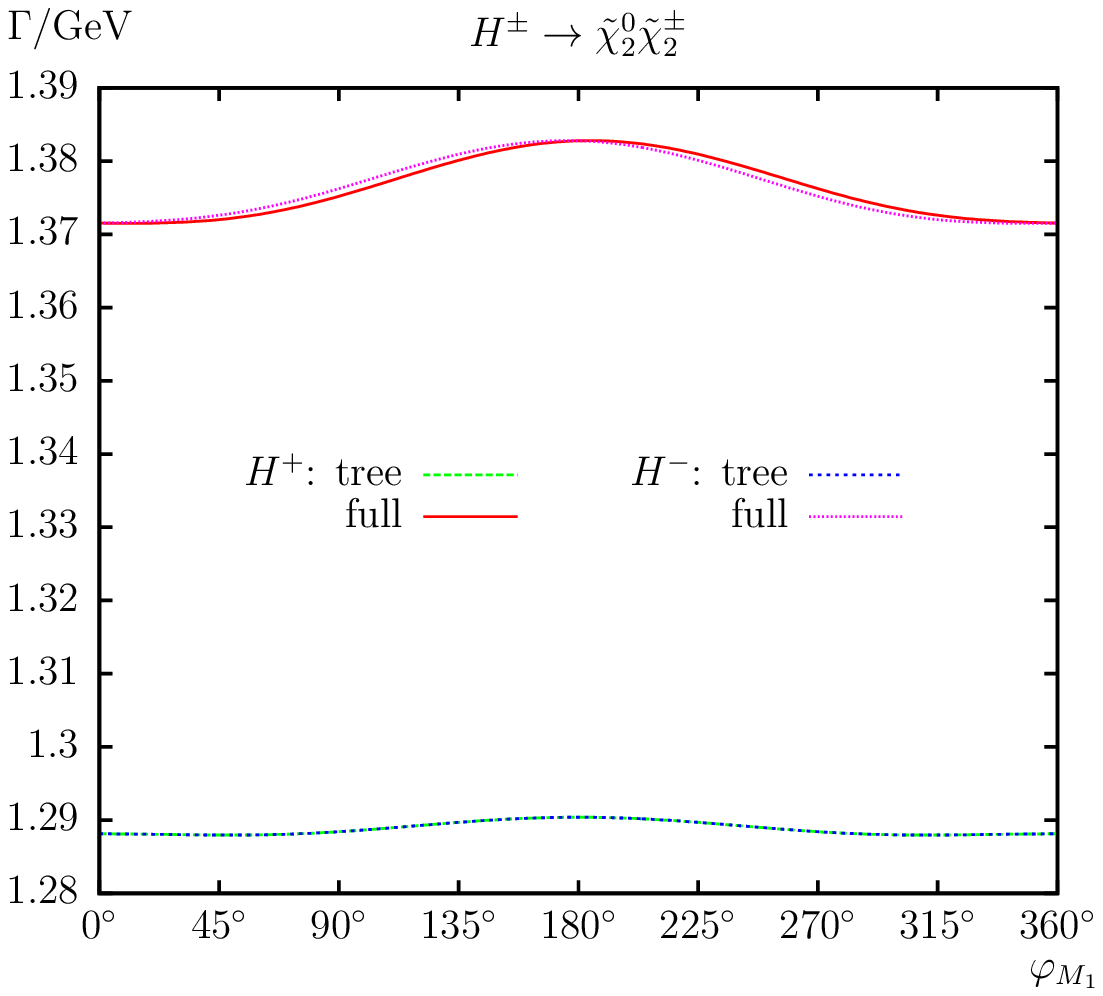}
\end{tabular}
\vspace{1em}
\caption{\label{fig:Hneu2cha2}
  $\Ga(H^\pm \to \neu2 \chapm2)$. 
  Tree-level and full one-loop corrected partial decay widths are shown. 
  The left plot shows the partial decay width with $\MHp$ varied.
  The right plot shows the complex phase $\phiMe$ varied with parameters 
  chosen according to \Scv\ (see \refta{tab:para}).
}
\end{center}
\end{figure}

\begin{figure}[htb!]
\begin{center}
\begin{tabular}{c}
\includegraphics[width=0.49\textwidth,height=7.5cm]{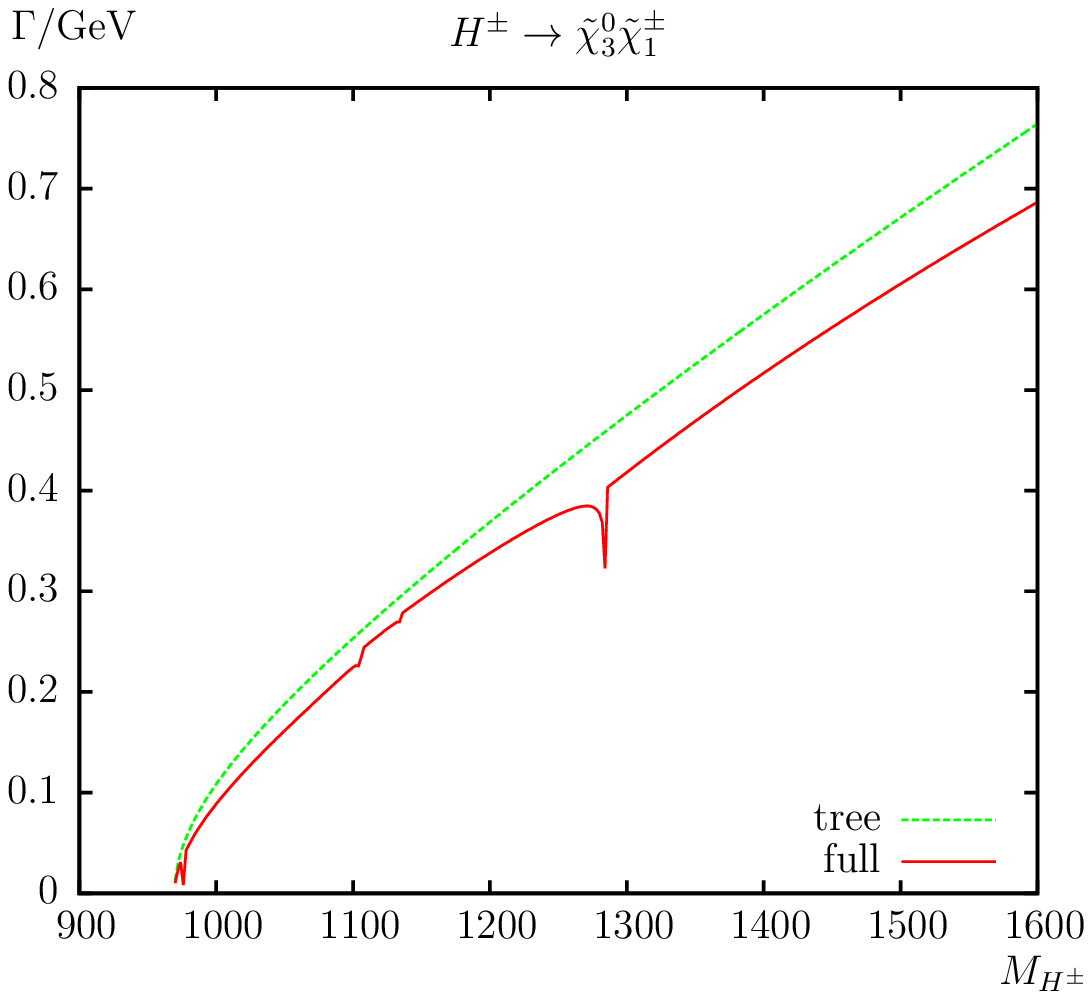}
\hspace{-4mm}
\includegraphics[width=0.49\textwidth,height=7.5cm]{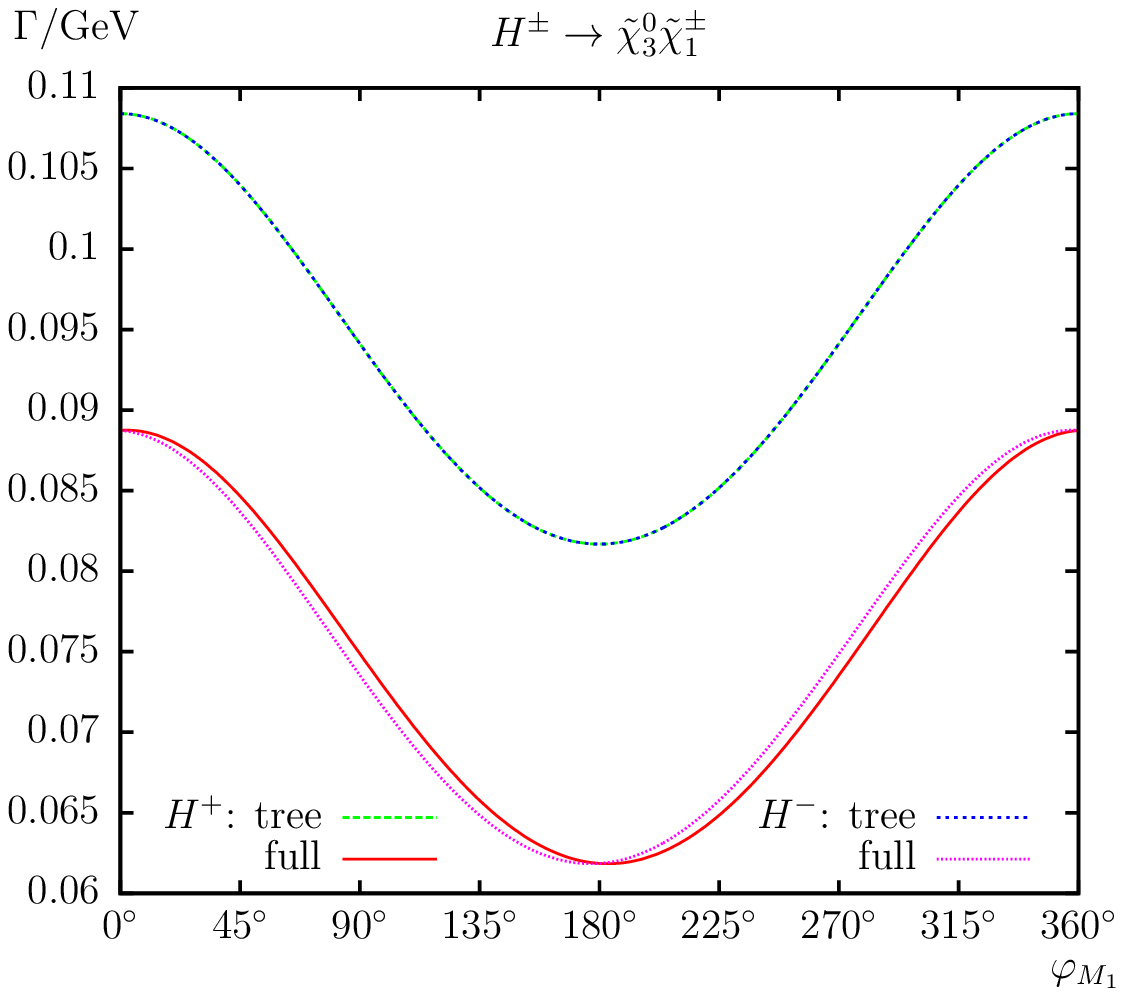}
\end{tabular}
\vspace{1em}
\caption{\label{fig:Hneu3cha1}
  $\Ga(H^\pm \to \neu3 \chapm1)$. 
  Tree-level and full one-loop corrected partial decay widths are shown. 
  The left plot shows the partial decay width with $\MHp$ varied.
  The right plot shows the complex phase $\phiMe$ varied with parameters 
  chosen according to \Scd\ (see \refta{tab:para}).
}
\vspace{6em}
\begin{tabular}{c}
\includegraphics[width=0.49\textwidth,height=7.5cm]{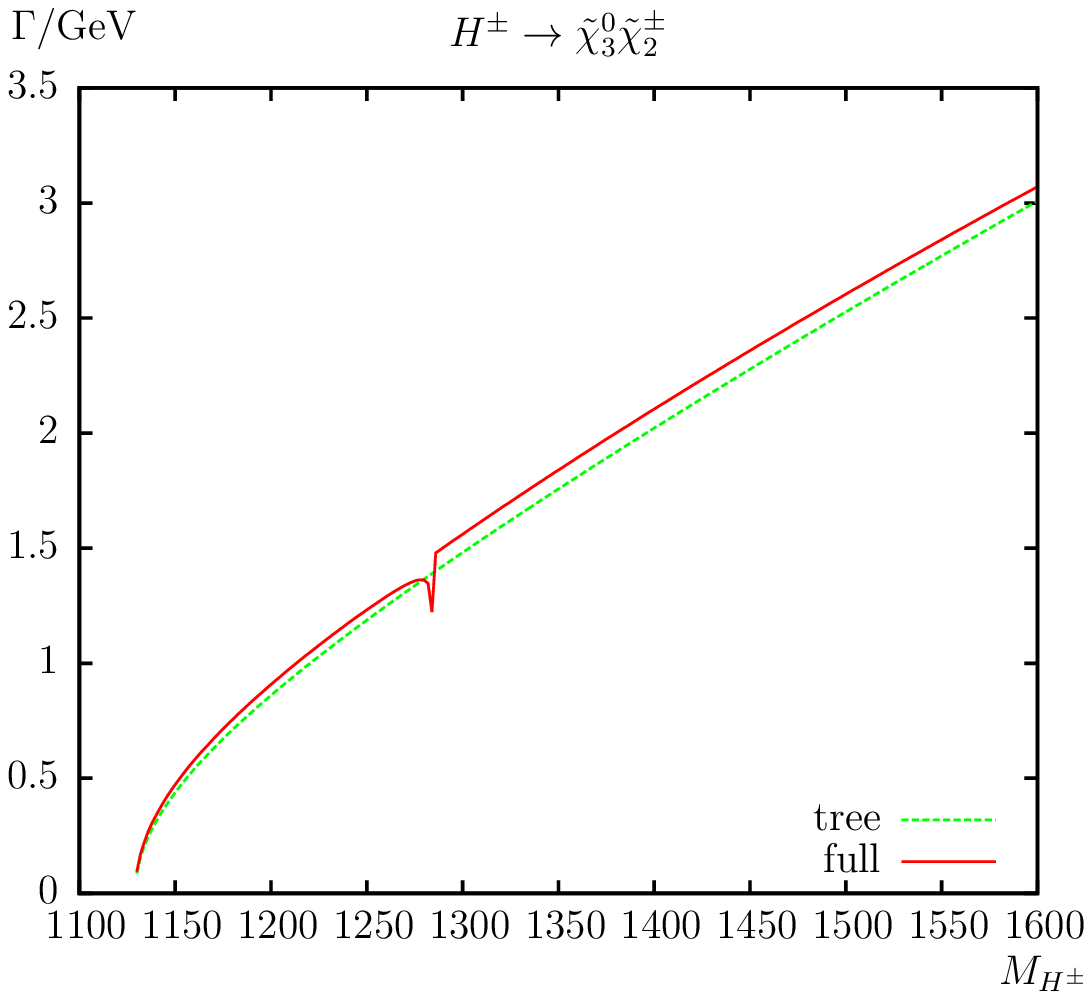}
\hspace{-4mm}
\includegraphics[width=0.49\textwidth,height=7.5cm]{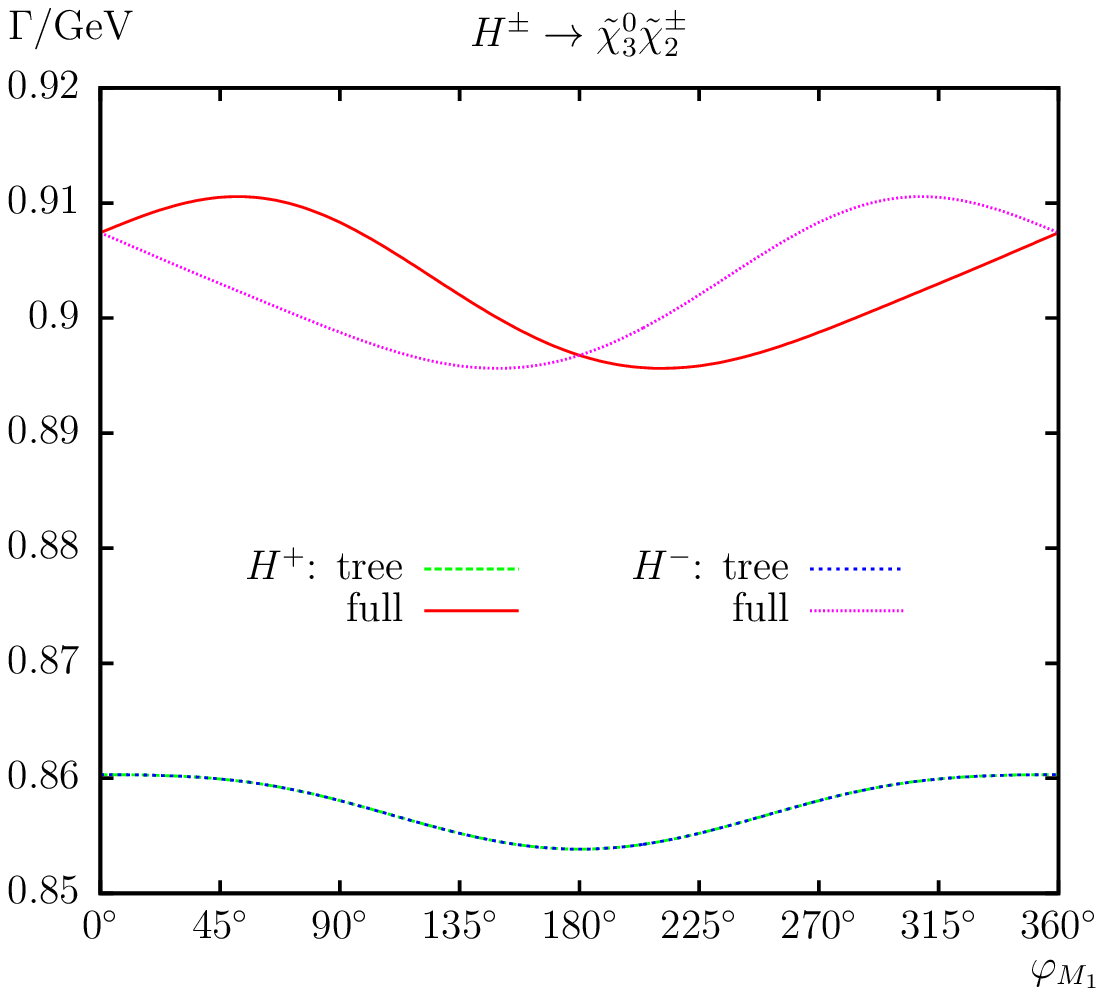}
\end{tabular}
\vspace{1em}
\caption{\label{fig:Hneu3cha2}
  $\Ga(H^\pm \to \neu3 \chapm2)$. 
  Tree-level and full one-loop corrected partial decay widths are shown. 
  The left plot shows the partial decay width with $\MHp$ varied.
  The right plot shows the complex phase $\phiMe$ varied with parameters 
  chosen according to \Scv\ (see \refta{tab:para}).
}
\end{center}
\end{figure}

\begin{figure}[htb!]
\begin{center}
\begin{tabular}{c}
\includegraphics[width=0.49\textwidth,height=7.5cm]{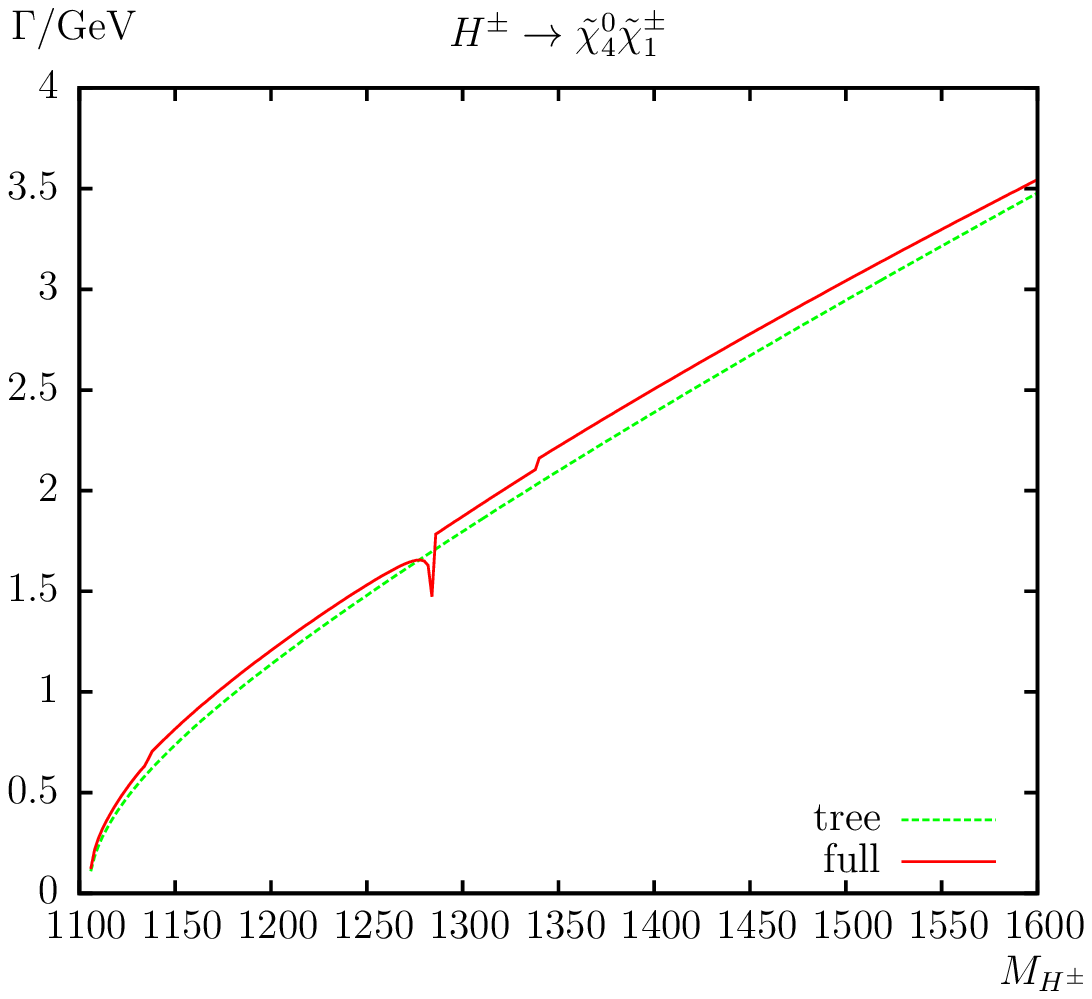}
\hspace{-4mm}
\includegraphics[width=0.49\textwidth,height=7.5cm]{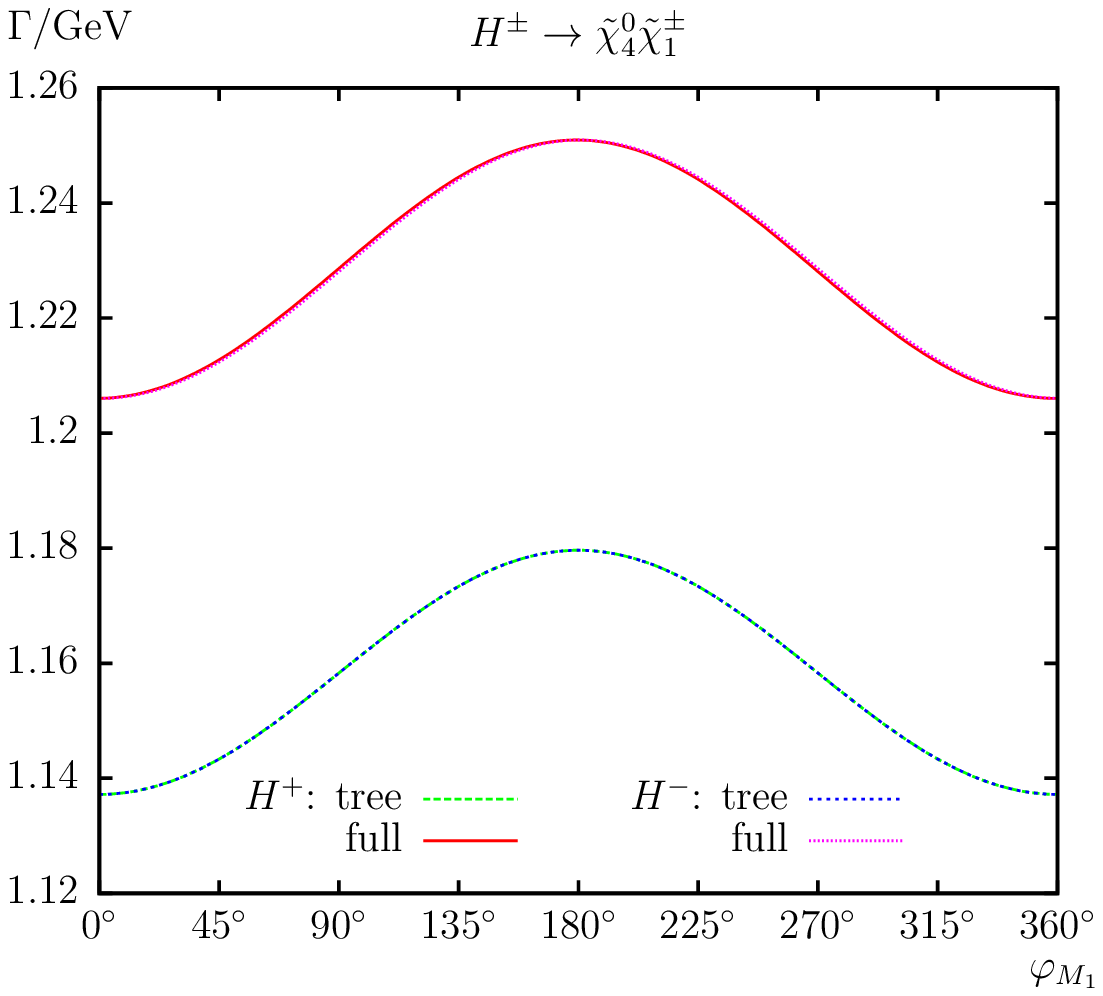}
\end{tabular}
\vspace{1em}
\caption{\label{fig:Hneu4cha1}
  $\Ga(H^\pm \to \neu4 \chapm1)$. 
  Tree-level and full one-loop corrected partial decay widths are shown. 
  The left plot shows the partial decay width with $\MHp$ varied.
  The right plot shows the complex phase $\phiMe$ varied with parameters 
  chosen according to \Scv\ (see \refta{tab:para}).
}
\vspace{6em}
\begin{tabular}{c}
\includegraphics[width=0.49\textwidth,height=7.5cm]{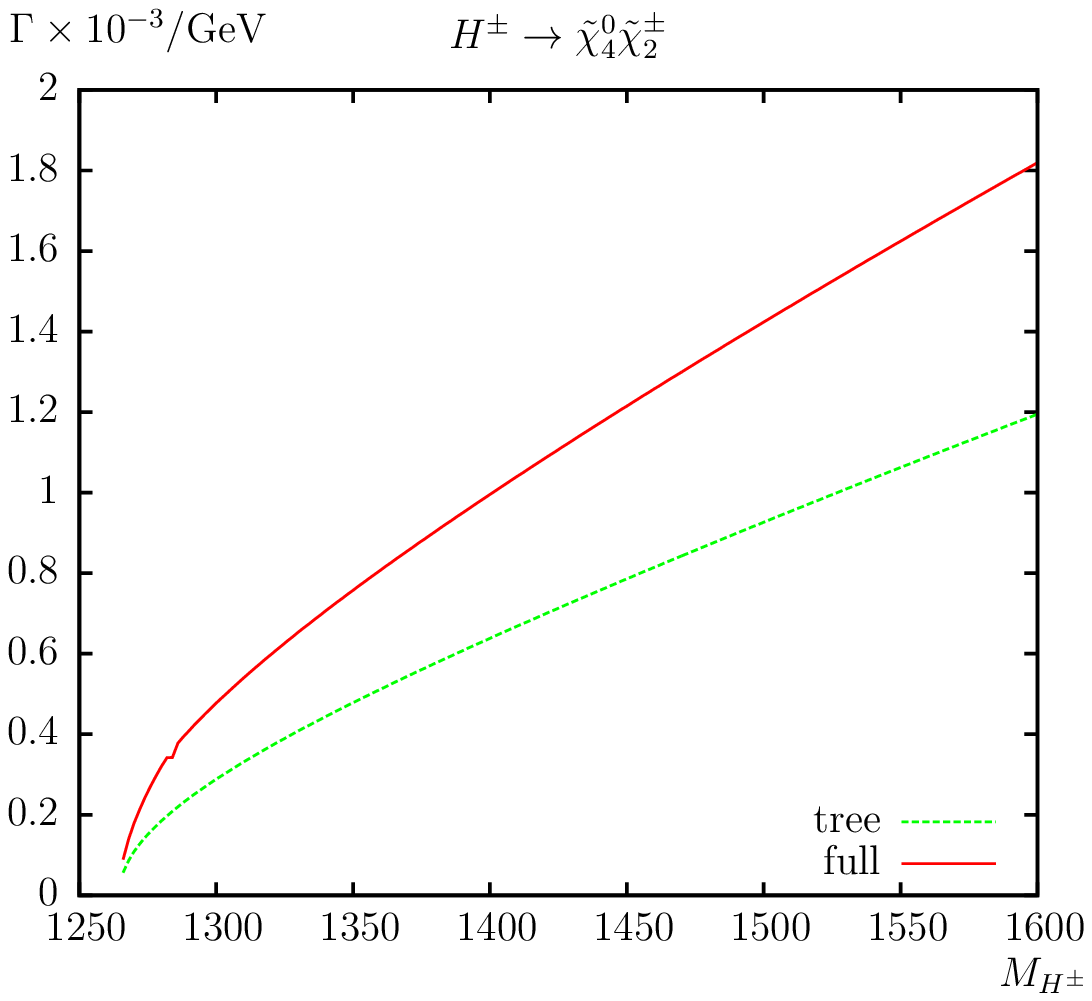}
\hspace{-4mm}
\includegraphics[width=0.49\textwidth,height=7.5cm]{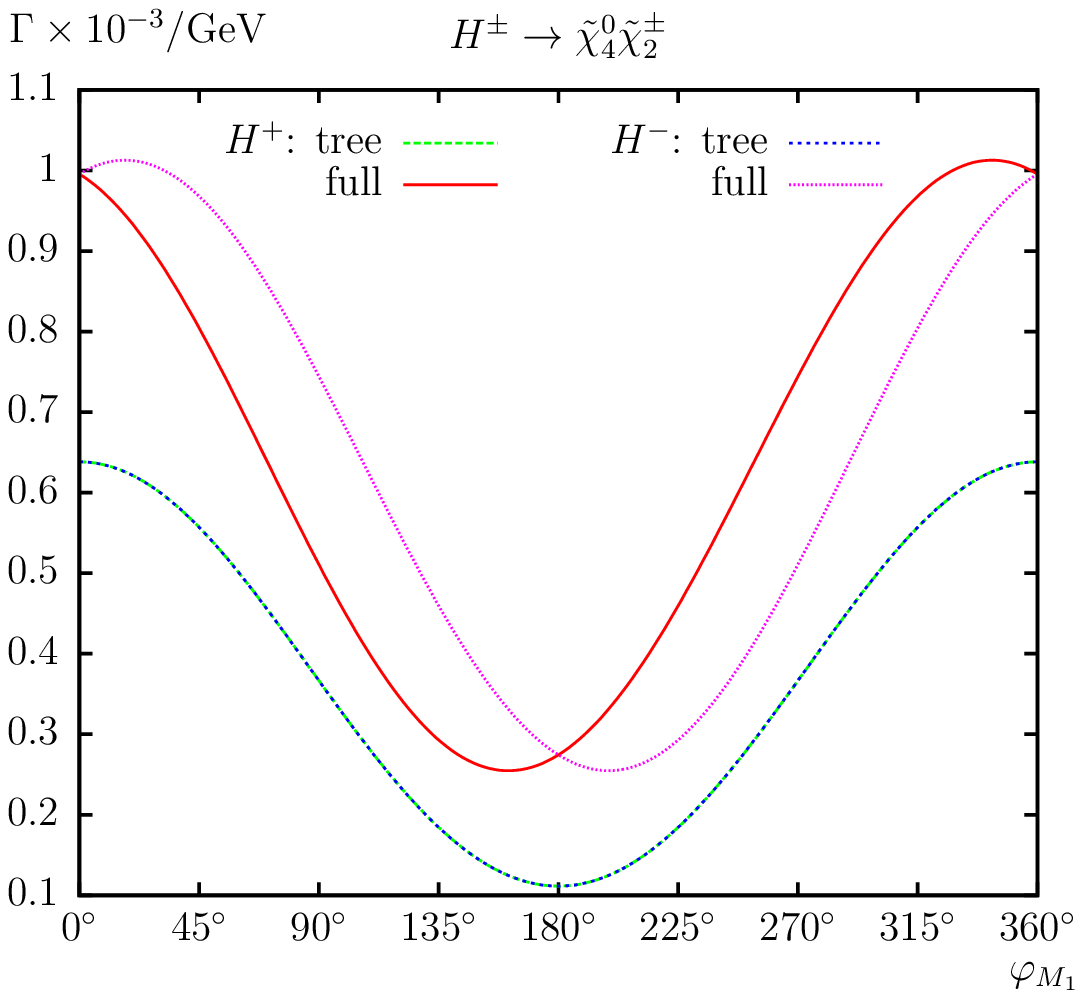}
\end{tabular}
\vspace{1em}
\caption{\label{fig:Hneu4cha2}
  $\Ga(H^\pm \to \neu4 \chapm2)$. 
  Tree-level and full one-loop corrected partial decay widths are shown. 
  The left plot shows the partial decay width with $\MHp$ varied.
  The right plot shows the complex phase $\phiMe$ varied with parameters 
  chosen according to \Scf\ (see \refta{tab:para}).
}
\end{center}
\end{figure}

In \reffis{fig:Hneu1cha1} -- \ref{fig:Hneu4cha2} we show the results 
for the processes $\HpmDecay$ ($n = 1,2,3,4;\, c = 1,2$) as a function 
of $\MHp$ and as a function of the relevant complex phase $\phiMe$. 
These are of particular interest for LHC 
analyses~\cite{stopstophiggs-LHC,Higgsincascades} 
(as emphasized in \refse{sec:intro}). 
The various visible (or hardly visible) dips/thresholds occurring for 
different values of $\MHp$ in the plots are summarized in 
\refta{tab:cthreshold}, labeled TC1 to TC7.

\medskip

We start with the decay $\Hpm \to \neu1 \cha1$. 
In the left plot of \reffi{fig:Hneu1cha1} the first (small) dip 
is the threshold TC1, see \refta{tab:cthreshold}.  
The second (large) dip is an effect due to the threshold TC2.
The third ``apparently single'' dip is in reality two dips coming 
from the thresholds TC3 and TC4. 
The fourth (small) dip is the threshold TC5 and the last 
(large) one is the threshold TC6. 
The size of the corrections of the partial decay widths can be 
especially large very close to the production threshold%
\footnote{
  It should be noted that a calculation very close to the production 
  threshold requires the inclusion of additional (nonrelativistic) 
  contributions, which is beyond the scope of this paper. 
  Consequently, very close to the production threshold our calculation 
  (at the tree- and loop-level) does not provide a very accurate 
  description of the decay width.
}
from which on the considered decay mode is kinematically possible. 
Away from this production threshold relative corrections of $\sim +10\%$ 
are found.

In the right plot of \reffi{fig:Hneu1cha1} we show the results for 
the complex phase $\phiMe$ varied for $\MHp = 1000\gev$.  The full
corrections are up to $\sim +13\%$ at $\phiMe = 180^\circ$.  
At $\phiMe = 90^\circ$ the $H^+$ ($H^-$) full corrections reach 
$\sim +12\%$ ($\sim +10\%$).  Clearly visible is the $\CP$-asymmetry
for the decays of the $H^+$ and $H^-$, which can reach the level of 
several per-cent.

\medskip

In \reffi{fig:Hneu1cha2} we show the results for $\Hpm \to \neu1 \chapm2$. 
The tree-level decay width $\Ga(H^\pm \to \neu1 \chapm2)$ is accidently 
very small for the parameter set chosen, see \refta{tab:para}.  
Because of this smallness, the relative size of the one-loop correction 
becomes larger than the tree-level result, and can even turn the
decay width, $\propto |\cMt|^2 + 2\Re\{\cMt^*\, \cMl^{}\}$, negative. 
Therefore, in this case we added $|\cMl|^2$ to the full 
one-loop result to obtain a positive decay width.
In the left plot the first (large) spike is the threshold 
TC2, see \refta{tab:cthreshold} enhanced through the two-loop
contribution $|\cMl|^2$ (\ie without the explicit two-loop correction 
the spike would be a ``usual dip'').
The second ``apparently single spike'' (hardly visible) is (again) in 
reality the two thresholds TC3 and TC4.
The next (apparently single) ``dip'' is in reality two steps 
(anomalous thresholds, see \citere{tHooft}) traced back to the 
$C$-functions
$C_{0,1,2}(\MHp^2,\mcha2^2,\mneu1^2,\mneu4^2,\mcha1^2,\mh1^2)$ at 
$\MHp \approx 1126\gev$ and
$C_{0,1,2}(\MHp^2,\mcha2^2,\mneu1^2,\mcha2^2,\mneu2^2,\MW^2)$ at 
$\MHp \approx 1129\gev$.%
\footnote{
  In addition both steps are contorted through 
  the higher order contributions $|\cMl|^2$.
}
Not visible (in the plot) is a spike, which is the threshold TC5.
The last spike is the threshold TC6.
Relative corrections of $\sim -90\%$ are found at $\MHp = 1000\gev$
(see \refta{tab:para}), where it should be kept in mind that the 
tree-level is already accidentally very small and thus loop corrections 
can have a relatively large impact. 

In the right plot of \reffi{fig:Hneu1cha2} the results are shown for
\Scd\ as a function of $\phiMe$.  At $\phiMe = 180^\circ$ the full 
corrections reach $\sim -55\%$, again related to the accidentally
small tree-level result.  At $\phiMe = 90^\circ$ the $H^+$ 
($H^-$) full corrections reach $\sim -59\%$ ($\sim -62\%$), showing
a small $\CP$-asymmetry.

\medskip

Next, in \reffi{fig:Hneu2cha1} the results for $\Hpm \to \neu2 \cha1$
are displayed.  In the left plot the results are shown as a function 
of $\MHp$.  The four visible dips here are exactly the same as in 
\reffi{fig:Hneu1cha1} (described above), beginning at $\MHp = 976\gev$.
Relative corrections of $\sim +33\%$ ($\sim +21\%$) are found at 
$\MHp = 1000\gev$ ($\MHp = 1400\gev$), see \refta{tab:para}.

In the right plot the results are displayed as a function of $\phiMe$
in \Scd, \ie for $\MHp = 1000\gev$. 
One can see that the size of the tree-level as well as the corrections 
to the partial decay width vary substantially with the complex phase 
$\phiMe$.  For all $\phiMe$ the full corrections lie between 
$+29\%$ and $+70\%$.%
\footnote{
  It should be noted that the loop corrections can reach 
  $+70\%$ of the tree results because at $\phiMe = 180^\circ$ 
  the tree-level decay width is accidently small, see the 
  right plot of \reffi{fig:Hneu2cha1}. 
}
At $\phiMe = 90^\circ$ the $H^+$ ($H^-$) full one-loop corrections 
reach $\sim +29\%$ ($\sim +48\%$), \ie the $\CP$-asymmetry is
rather large with $\sim \pm 19\%$.

\medskip

The decay $\Hpm \to \neu2 \cha2$ is shown in \reffi{fig:Hneu2cha2}.
In the left plot the results are shown as a function of $\MHp$. 
The first (hardly visible) dip is (again) the threshold 
TC5, see \refta{tab:cthreshold} and the second (large) one is 
the threshold TC6.
The decay width turns out to be relatively large at \order{1\gev}.
Relative corrections of $\sim +6\%$ ($\sim +4\%$) are found at 
$\MHp = 1200\gev$ ($\MHp = 1400\gev$), see \refta{tab:para}.

In the right plot of \reffi{fig:Hneu2cha2} the results are displayed 
as a function of $\phiMe$ in \Scv, \ie for $\MHp = 1200\gev$.  
The full corrections at $\phiMe = 180^\circ$ reach $\sim +7\%$.
On the other hand it can be seen that the variation with $\phiMe$ 
and the $\CP$-asymmetries ($\sim \pm 0.1\%$) are rather small.

\medskip

Next, in \reffi{fig:Hneu3cha1} the results for $\Hpm \to \neu3 \cha1$
are displayed.  In the left plot the results are shown as a function 
of $\MHp$.  Here the four visible dips are the same as in 
\reffi{fig:Hneu1cha1}, beginning at $\MHp = 976\gev$.
Relative corrections of $\sim -18\%$ ($\sim -10\%$) are found at 
$\MHp = 1000\gev$ ($\MHp = 1400\gev$), see \refta{tab:para}.

In the right plot the results are displayed as a function of $\phiMe$
in \Scd.  One can see that again the tree-level results as well as
the size of the corrections to the partial decay width vary substantially 
with the complex phase $\phiMe$.  The full corrections can reach 
$\sim -24\%$ and the $\CP$-asymmetry is found to be small at the level 
of $\sim \pm 2\%$.

\medskip

In \reffi{fig:Hneu3cha2} we show the results for $\Hpm \to \neu3 \chapm2$.
In the left plot (with $\MHp$ varied) the dip is (again) the threshold 
TC6, see \refta{tab:cthreshold}.  The decay width is found to be of 
the same order as for $\Hpm \to \neu2\cha2$.  One-loop corrections of 
$\sim +5\%$ ($\sim +4\%$) are found at $\MHp = 1200\gev$ in \Scv\ 
($\MHp = 1400\gev$ in \Scf), see \refta{tab:para}. 

In the right plot of \reffi{fig:Hneu3cha2} the results are shown for
\Scv\ as a function of $\phiMe$.  At $\phiMe = 180^\circ$ the full 
corrections reach $\sim +5\%$.  At $\phiMe = 90^\circ$ the $H^+$ 
($H^-$) full corrections reach $\sim +6\%$ ($\sim +5\%$), \ie the 
$\CP$-asymmetries are at the level of $\sim \pm 1\%$.

\medskip

We finish the charged Higgs-boson analysis with the decays
involving the heaviest neutralino in \reffis{fig:Hneu4cha1} 
and \ref{fig:Hneu4cha2}, showing the results for 
$\Ga(\Hpm \to \neu4\cha1)$ and $\Ga(\Hpm \to \neu4\cha2)$, respectively.

In the left plot of \reffi{fig:Hneu4cha1} the first dip 
(not visible in the plot) is the threshold TC4, see \refta{tab:cthreshold}. 
The second (small) dip is (again) the threshold TC5 and the third 
(large) dip is the threshold TC6.
The first step (not visible in the plot) at $\MHp \approx 1136\gev$ 
is the anomalous threshold of the $C$-functions
$C_{0,1,2}(\MHp^2,\mcha1^2,\mneu4^2,\mneu3^2,\mcha2^2,\MZ^2)$.
The second anomalous threshold at $\MHp \approx 1340\gev$ is 
caused by  
$C_{0,1,2}(\mneu4^2,\MHp^2,\mcha1^2,m_b^2,\msbot1^2,\mstop2^2)$.
The last dip (also not visible) is the threshold TC7.
The decay width is again found at \order{1\gev} with relative 
corrections of $\sim +6\%$ in \Scv\ (see \refta{tab:para}).

In the right plot of \reffi{fig:Hneu4cha1} we show the complex phase
$\phiMe$ varied at $\MHp = 1200\gev$.  The full corrections are up to 
$\sim +6\%$ at $\phiMe = 180^\circ$.  Here the asymmetries are extremely 
small and hardly visible.

\medskip

Finally, we discuss the decay $\Hpm \to \neu4\cha2$ in
\reffi{fig:Hneu4cha2}.
The overall size of this decay width (with real phases) 
is (accidentally) very small around $1 \times 10^{-3}\gev$.
Consequently, the loop corrections, can be larger than the tree-level 
result.  In the left plot the results are shown as a function of $\MHp$. 
The (small) dip is the threshold TC6, see \refta{tab:cthreshold}.
Relative corrections of $\sim +56\%$ are found at $\MHp = 1400\gev$ 
(see \refta{tab:para}).

In the right plot the results are displayed as a function of $\phiMe$
in \Scf.  One can see that the size of the corrections to the partial 
decay width vary substantially with the complex phase $\phiMe$. 
For all $\phiMe$ the full corrections deviate between $+40\%$ 
and $+146\%$.  (The latter value is reached at $\phiMe = 180^\circ$
where the tree is extremely small $\sim 1 \times 10^{-4}\gev$.)
At $\phiMe = 90^\circ$ the $H^+$ ($H^-$) full one-loop corrections 
reach $\sim +40\%$ ($\sim +103\%$), \ie the $\CP$-asymmetries are 
very large with $\sim \pm 60\%$.

\medskip

Overall, for the charged Higgs boson decays to a chargino/neutralino 
pair we observe, as expected, an increasing decay width $\propto \MHp$.%
\footnote{
  An exception are the loop corrections in the left plot of 
  \reffi{fig:Hneu1cha2}, because there we added $|\cMl|^2$.
}
The full one-loop corrections reach a level of $10\%$ for decay widths 
of \order{1\gev}.  The variation with $\phiMe$ is found largest for very
small decay widths, but can reach the level of $10-50\%$ for widths 
at or below the $1\gev$~level. The $\CP$-asymmetries exceed the level 
of a few per-cent only for very small decay widths.

%%%%%%%%%%%%%%%%%%%%% T A B L E %%%%%%%%%%%%%%%%%%%%%%%%%%%%%%%%%%%%%%%%%%%%%%
\begin{table}[t]
\caption{\label{tab:cthreshold} Thresholds in charged Higgs boson decays.}
\centering
\begin{tabular}{lrr}
\toprule
TC1: & $\MHp \approx \phantom{0}907\gev\quad$ & $\mstop1 + \msbot1 = \MHp$ \\
TC2: & $\MHp    =    \phantom{0}976\gev\quad$ & $\mstop1 + \msbot2 = \MHp$ \\
TC3: & $\MHp \approx \phantom{}1105\gev\quad$ & $\mcha1  + \mneu4  = \MHp$ \\
TC4: & $\MHp \approx \phantom{}1108\gev\quad$ & $\mcha2  + \mneu2  = \MHp$ \\
TC5: & $\MHp \approx \phantom{}1135\gev\quad$ & $\mcha2  + \mneu3  = \MHp$ \\
TC6: & $\MHp \approx \phantom{}1284\gev\quad$ & $\mstop2 + \msbot1 = \MHp$ \\
TC7: & $\MHp \approx \phantom{}1353\gev\quad$ & $\mstop2 + \msbot2 = \MHp$ \\
\bottomrule
\end{tabular}
\end{table}
%%%%%%%%%%%%%%%%%%%%% T A B L E %%%%%%%%%%%%%%%%%%%%%%%%%%%%%%%%%%%%%%%%%%%%%%

\clearpage
\newpage

%%%%%%%%%%%%%%%%%%%%%%%%%% F I G U R E %%%%%%%%%%%%%%%%%%%%%%%%%%%%%%%%%%%%%%%%%
%%%%%%%%%%%%%%%%%%%%%%%%%% F I G U R E %%%%%%%%%%%%%%%%%%%%%%%%%%%%%%%%%%%%%%%%%

%\newpage

%%%%%%%%%%%%%%%%%%%%%%%%%% F I G U R E %%%%%%%%%%%%%%%%%%%%%%%%%%%%%%%%%%%%%%%%%
%%%%%%%%%%%%%%%%%%%%%%%%%% F I G U R E %%%%%%%%%%%%%%%%%%%%%%%%%%%%%%%%%%%%%%%%%

%\newpage

%%%%%%%%%%%%%%%%%%%%%%%%%% F I G U R E %%%%%%%%%%%%%%%%%%%%%%%%%%%%%%%%%%%%%%%%%
%%%%%%%%%%%%%%%%%%%%%%%%%% F I G U R E %%%%%%%%%%%%%%%%%%%%%%%%%%%%%%%%%%%%%%%%%

%\newpage

%%%%%%%%%%%%%%%%%%%%%%%%%% F I G U R E %%%%%%%%%%%%%%%%%%%%%%%%%%%%%%%%%%%%%%%%%
%%%%%%%%%%%%%%%%%%%%%%%%%% F I G U R E %%%%%%%%%%%%%%%%%%%%%%%%%%%%%%%%%%%%%%%

\clearpage
\newpage

%%%%%%%%%%%%%%%%%%%%%%%%%%%%%%%%%%%%%%%%%%%%%%%%%%%%%%%%%%%%%%%%%%%%%%%%%%%%%%

\subsubsection{\boldmath{$h_i$} decays into charginos and neutralinos}
\label{hndecays}

We now turn to the decay modes $\hChaDecay$ ($i = 2,3;\, c,\cpri = 1,2$)
and $\hNeuDecay$ ($i = 1,2,3;\, n,\npri = 1,2,3,4$). 
Results are shown in the \reffis{fig:hcha1cha1} -- \ref{fig:hneu3neu4}.

Before discussing every figure in detail, it should be noted that 
there is a subtlety concerning the mixture of the $h_i$ bosons.
Depending on the input parameters, the higher-order corrections to the
three neutral Higgs boson masses can vary substantially. 
The mass ordering $\mh1 < \mh2 < \mh3$ (as performed automatically by
\FH), even in the case of real parameters, can yield a heavy $\CP$-even
Higgs mass higher \textit{or} lower than the (heavy) $\CP$-odd Higgs mass. 
Such a transition in the mass ordering (or ``mass crossing'') is
accompanied by an abrupt change in the Higgs mixing matrix 
$\matr{\hat Z}$.%
\footnote{In our case the $Z$-factor matrix is given by
  $\hat{Z}_{ij} \equiv \Code{ZHiggs[\Vi,\,\Vj]}$, see \citere{MSSMCT} 
  (and \citere{mhcMSSMlong}), which contributes at tree-level.
  Furthermore $\matr{\hat{Z}}$ is calculated by \FH\ which uses 
  $\mb(\mb)$ and tree-level sfermion masses instead of the shifted 
  masses, causing a slight displacement in the threshold position.
}
For our input parameters (see \refta{tab:para}) there are two (possible) 
crossings.  The first (called ``MC1'' below) appears at 
$\MHp \approx 1006\gev$.  Before the crossing we find 
$h_2 \sim H$ ($h_3 \sim A$), whereas after the crossing it changes to 
$h_2 \sim A$ ($h_3 \sim H$).  The second crossing (called ``MC2'') is 
found at $\MHp \approx 1532\gev$, \ie the changing of the mixture from 
$h_2 \sim A$ ($h_3 \sim H$) to $h_2 \sim H$ ($h_3 \sim A$).
Very close to the mass crossings the $\matr{\hat Z}$~matrix can 
yield small numerical instabilities.  As an example, for 
$1532\gev \lesssim \MHp \lesssim 1536\gev$ the $\matr{\hat Z}$~matrix
causes structures appearing similar to ``usual'' dips from thresholds
(see also the discussion in \citere{HiggsDecaySferm}).
All the dips/thresholds (some are hardly visible) appearing in the 
figures below are listed in \refta{tab:nthreshold}, 
labeled as TN1 to TN13.

%%%%%%%%%%%%%%%%%%%%%%%%%%%%%%%%%%%%%%%%%%%%%%%%%%%%%%%%%%%%%%%%%%%%%%%%%%%%%%

\subsubsection*{\boldmath{$h_i$} decays into charginos}

\begin{figure}[t!]
\begin{center}
\begin{tabular}{c}
\includegraphics[width=0.49\textwidth,height=7.5cm]{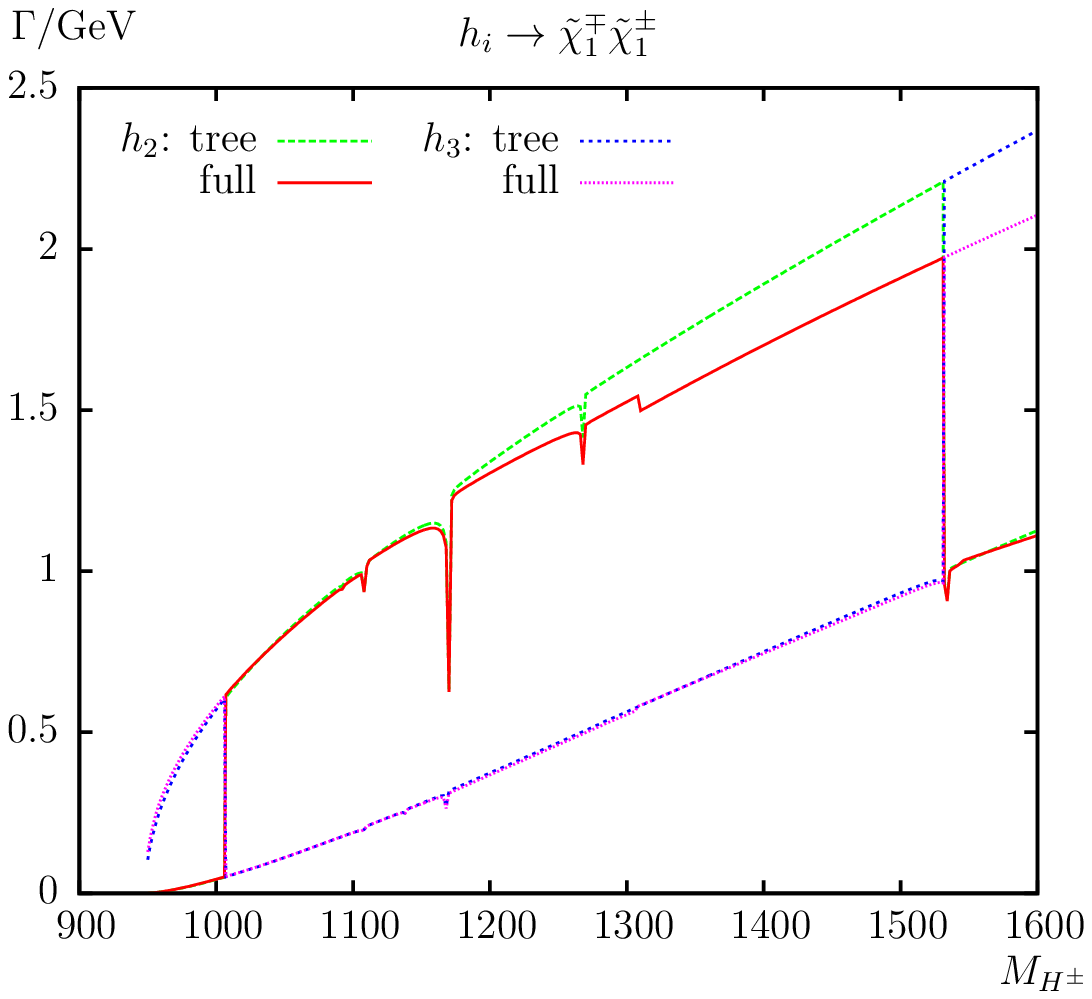}
\hspace{-4mm}
\includegraphics[width=0.49\textwidth,height=7.5cm]{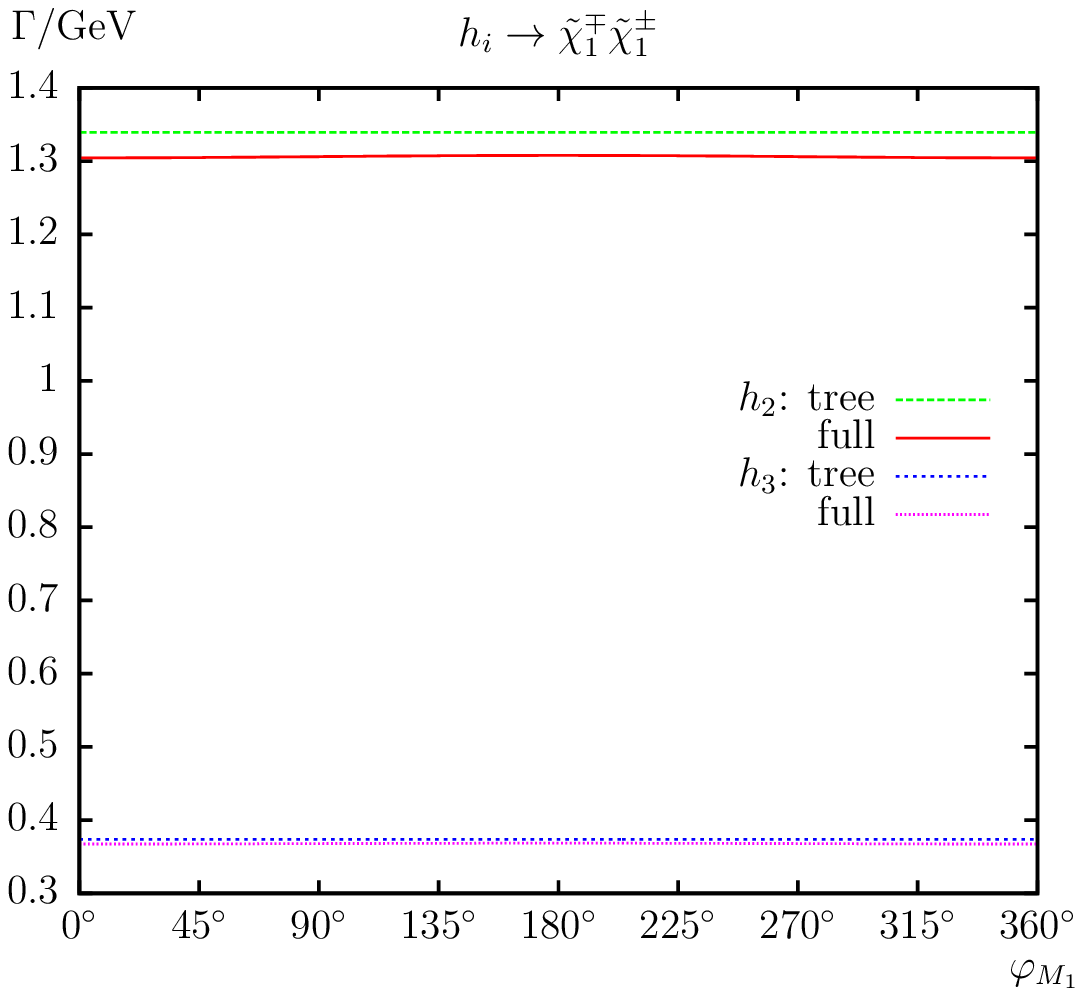}
\end{tabular}
\vspace{1em}
\caption{\label{fig:hcha1cha1}
  $\Ga(\hchaechae)$. 
  Tree-level and full one-loop corrected partial decay widths are shown. 
  The left plot shows the partial decay width with $\MHp$ varied. 
  The right plot shows the complex phase $\phiMe$ varied with parameters 
  chosen according to \Scv\ (see \refta{tab:para}).
}
\end{center}
\end{figure}

In this subsection we analyze the decays of the heavy neutral Higgs
bosons into charginos.  We start with the decay $\hchaechae$ (i = 2,3) 
as shown in \reffi{fig:hcha1cha1}. 
The left plot shows the results as a function of $\MHp$, whereas in the 
right plot we present the decay widths as a function of $\phiMe$ in \Scv. 
We show separately the results for the $h_2$ and $h_3$ decay widths.
In the left plot of \reffi{fig:hcha1cha1} the first ``apparently single'' 
dip in the $h_2$ decay (upper lines) is in reality coming from the 
thresholds TN6 and TN7, see \refta{tab:nthreshold}.
The second (large) dip is the threshold TN10. 
The last ``apparently single'' dip is in reality coming from the 
thresholds TN11 and TN12.  
The ``step'' (anomalous threshold) at $\MHp \approx 1310\gev$ 
could be traced back to the $C$-functions  
$C_{0,1,2}(\mcha1^2,\mh2^2,\mcha1^2,m_b^2,\mstop{\sind}^2,\mstop{\spri}^2)$
with $\sind \ne \spri$.
Away from the production threshold relative corrections of $\sim -3\%$ 
are found in \Scv\ (see \refta{tab:para}) for the $h_2$ decay.  
The loop corrections increase with increasing $\MHp$ and reach
$\sim -10\%$ in \Scf.
In case of the $h_3$ decay the dips are the same as for $h_2$ and the 
relative corrections are only $\sim -2\%$ in \Scv\ (see \refta{tab:para}). 
The two mass crossings MC1 and MC2 are clearly visible at 
$\MHp \approx 1006\gev$ and $\MHp \approx 1532\gev$ as described above, 
where $h_2$ and $h_3$ change their role.  Between MC1 and MC2 we find 
$\Ga(h_2 \to \champ1\chapm1) > \Ga(h_3 \to \champ1\chapm1)$,
outside it is vice versa, as can be clearly observed in the left plot 
of \reffi{fig:hcha1cha1}.
The suppression of the $\CP$-even decay (lower lines) vs.\ the 
$\CP$-odd decay (upper lines) is clearly visible, where at threshold 
the behavior follows \refeqs{hOddTree}, (\ref{hEvenTree}).
After the threshold the decays grow roughly linear with the Higgs 
boson masses.

We now turn to the phase dependence of the decay width shown in \Scv, 
\ie for $\MHp = 1200\gev$, where the right plot in \reffi{fig:hcha1cha1} 
shows the dependence of $\Ga(\hchaechae)$ on $\phiMe$.  Since $M_1$ does 
not appear in the chargino mass matrix, the effects of varying its phase 
enter only via loop corrections and are extremely small.
The relative corrections in \Scv\ are the same as in the left plot.

%%%%%%%%%%%%%%%%%%%%%%%%%% F I G U R E %%%%%%%%%%%%%%%%%%%%%%%%%%%%%%%%%%%%%%%%%
%%%%%%%%%%%%%%%%%%%%%%%%%% F I G U R E %%%%%%%%%%%%%%%%%%%%%%%%%%%%%%%%%%%%%%%%%

\medskip

\begin{figure}[htb!]
\begin{center}
\begin{tabular}{c}
\includegraphics[width=0.49\textwidth,height=7.5cm]{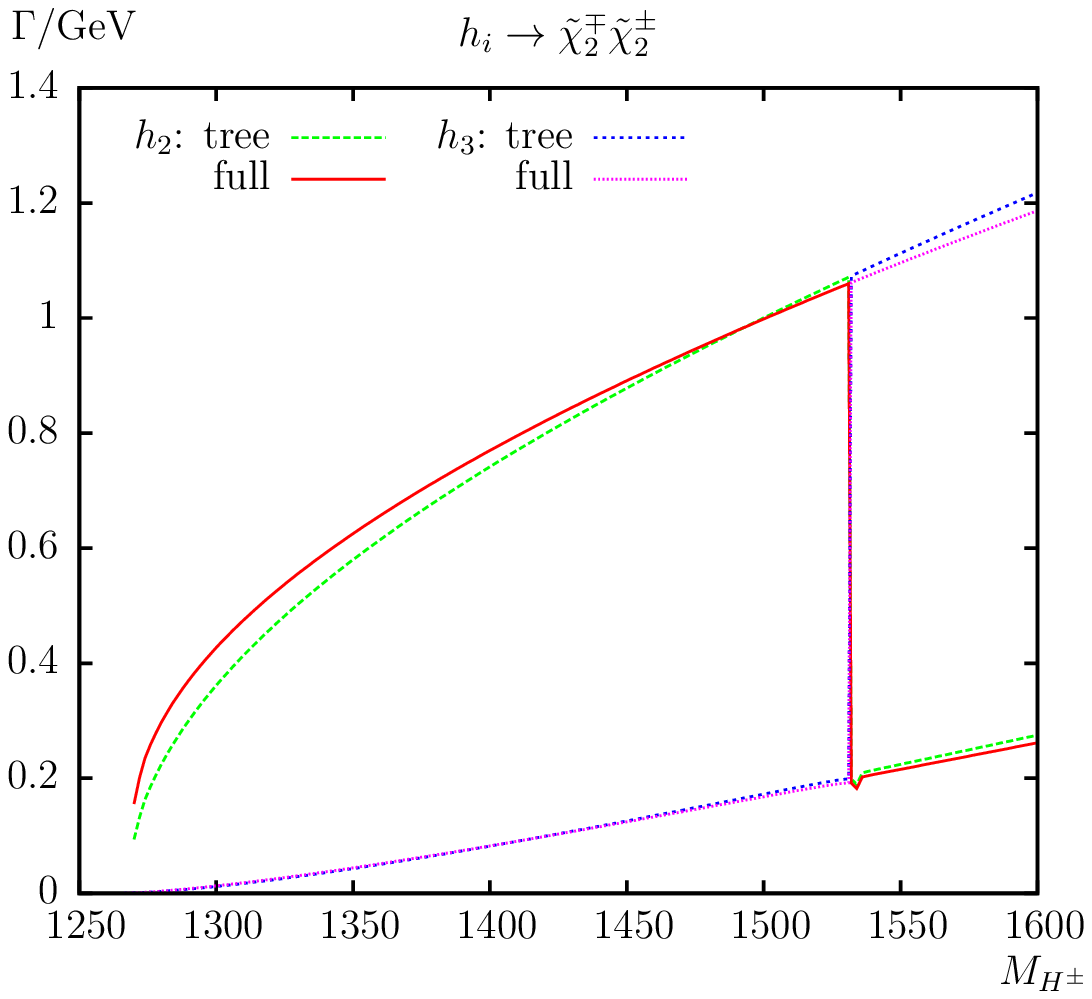}
\hspace{-4mm}
\includegraphics[width=0.49\textwidth,height=7.5cm]{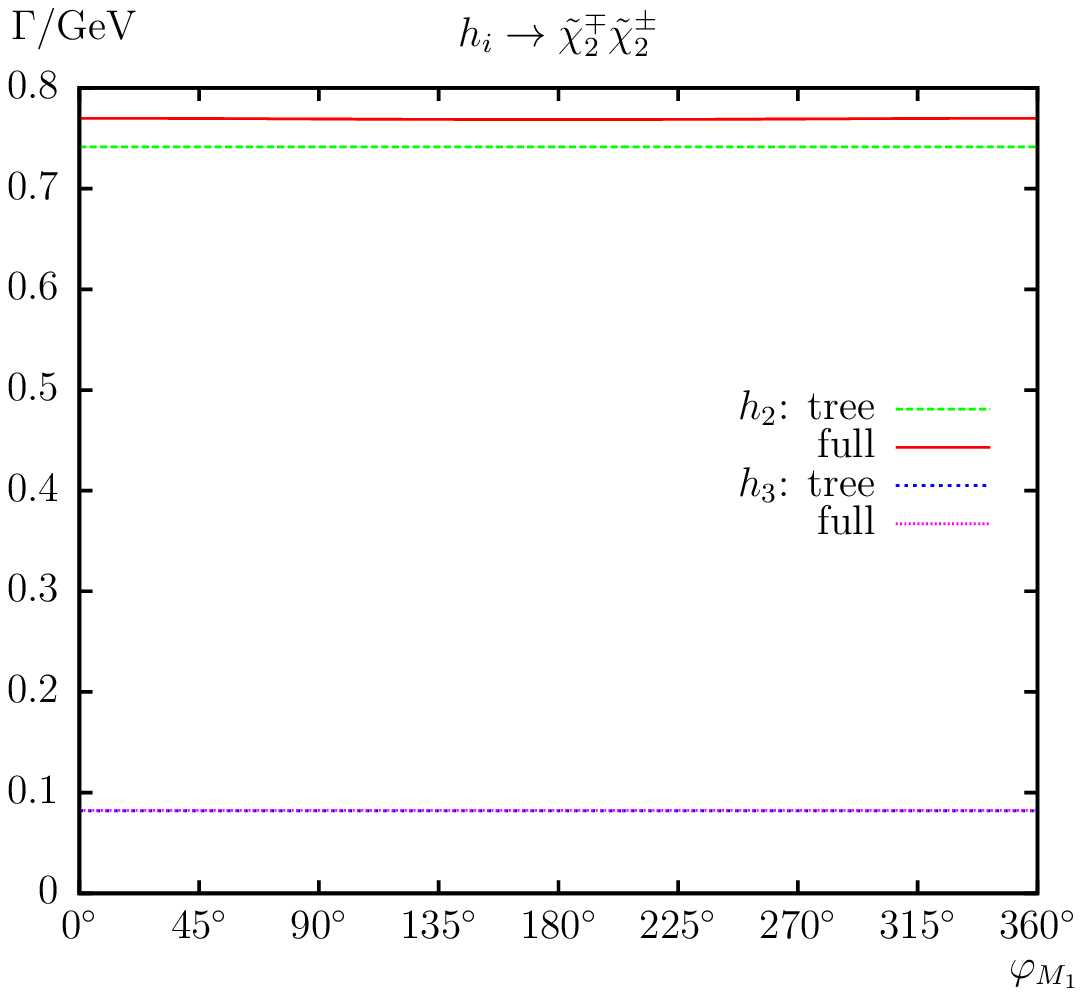}
\end{tabular}
\vspace{1em}
\caption{\label{fig:hcha2cha2}
  $\Ga(\hchazchaz)$. 
  Tree-level and full one-loop corrected partial decay widths are shown. 
  The left plot shows the partial decay width with $\MHp$ varied. 
  The right plot shows the complex phase $\phiMe$ varied with parameters 
  chosen according to \Scf\ (see \refta{tab:para}).
}
\vspace{6em}
\begin{tabular}{c}
\includegraphics[width=0.49\textwidth,height=7.5cm]{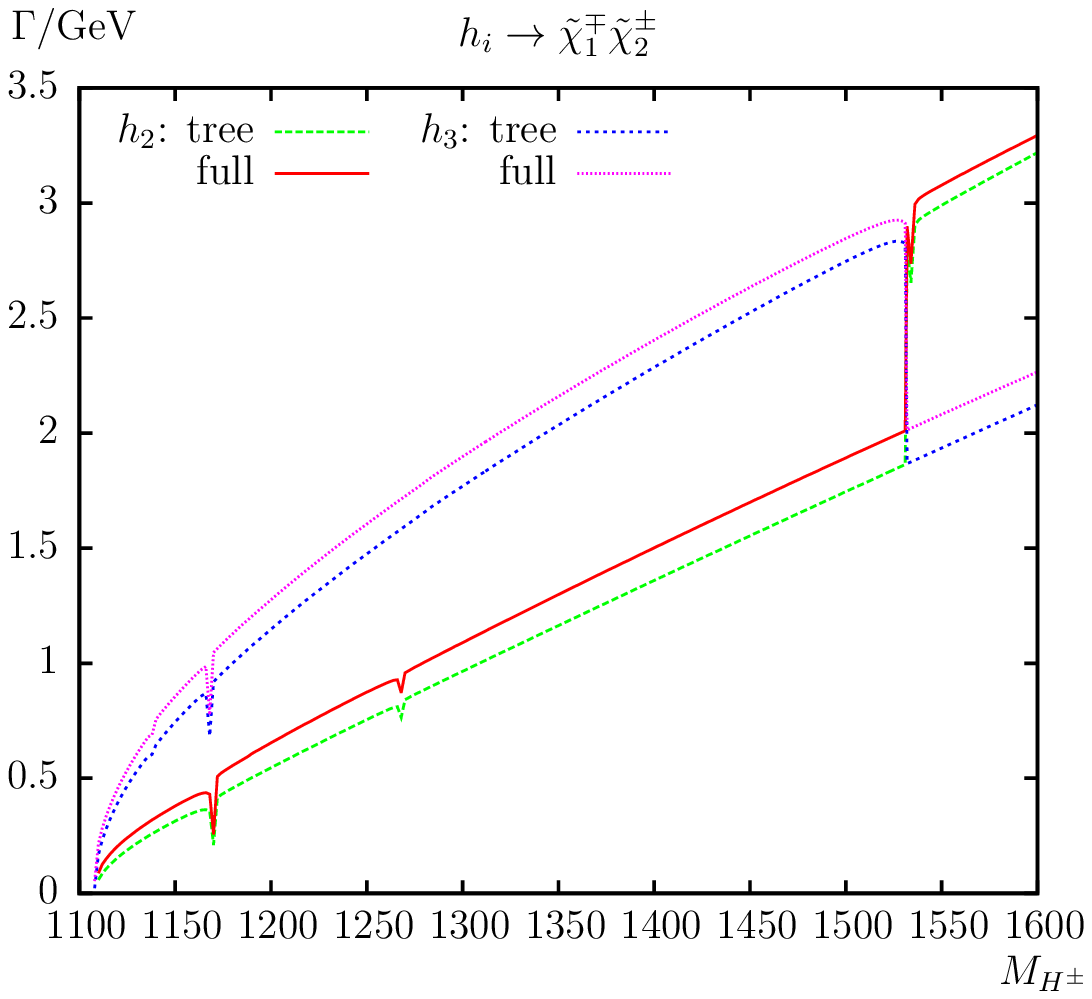}
\hspace{-4mm}
\includegraphics[width=0.49\textwidth,height=7.5cm]{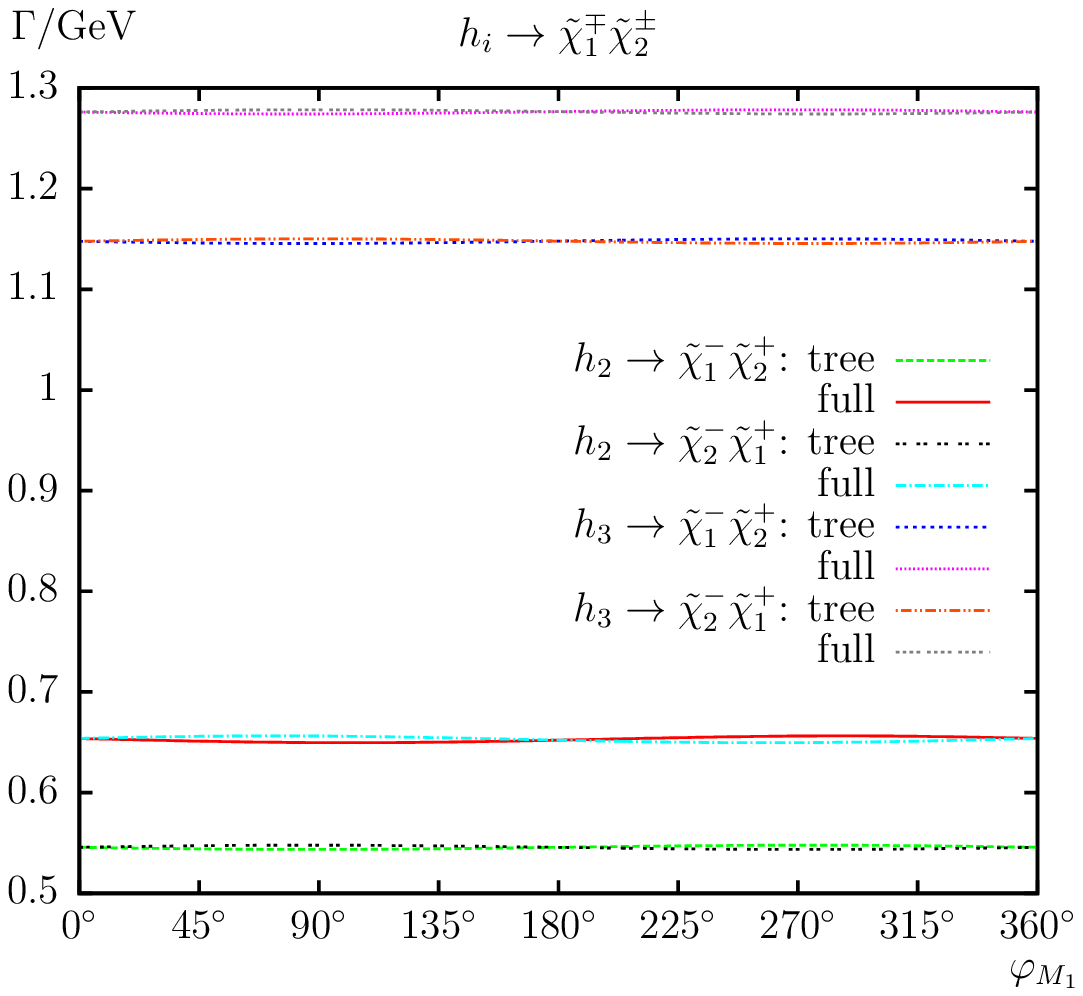}
\end{tabular}
\vspace{1em}
\caption{\label{fig:hcha1cha2}
  $\Ga(\hchaechaz)$.
  Tree-level and full one-loop corrected partial decay widths are shown. 
  The left plot shows the partial decay width with $\MHp$ varied. 
  The right plot shows the complex phase $\phiMe$ varied with parameters 
  chosen according to \Scv\ (see \refta{tab:para}).
}
\end{center}
\end{figure}

The results for $\Ga(\hchazchaz)$, as shown in the left plot of 
\reffi{fig:hcha2cha2} are smaller by roughly a factor of 2 \wrt
$\Ga(\hchaechae)$, largely related to the kinematic suppression.
At $\MHp = 1400\gev$ the full one-loop corrections to the $h_2$ 
decay reach $\sim +4\%$.  For the decay of the $h_3$ at 
$\MHp = 1400\gev$ we find full corrections at the level of less 
than $+1\%$.  As in the upper left plot one can observe the MC2 
with an ``interchange'' of $h_2$ and $h_3$.
The same suppression of the $\CP$-even vs.\ the $\CP$-odd decay, 
as in \reffi{fig:hcha1cha2} is clearly visible.

In the right plot of \reffi{fig:hcha2cha2} we show the complex phase
$\phiMe$ varied at $\MHp = 1400\gev$.  The variation with $\phiMe$ is
extremely small (for the same reasons as explained above), 
therefore the full relative corrections in \Scf\ are the same as in 
the left plot, see above.

\medskip

The results for the ``mixed'' decay, $\Ga(\hchaechaz)$, are shown in 
\reffi{fig:hcha1cha2}, where in the left (right) plot we show the 
dependence on $\MHp$ ($\phiMe$). In the left plot the first dip in 
the $h_2$ decay (lower lines) is the threshold TN10, 
see \refta{tab:nthreshold}.  
The remaining dip (at $\MHp \approx 1268\gev$) is caused by the two 
thresholds TN11 and TN12.  
At $\MHp = 1200\gev$ the full one-loop corrections to the $h_2$ decay 
reach $\sim +20\%$.  Now we turn to the corresponing $h_3$ decay.
The first dip (hardly visible in the upper lines) is the threshold TN8.
The second dip is the same as for the $h_2$ decay, see above.
For the decay of $h_3$ at $\MHp = 1200\gev$ we find full corrections 
at the level of $+10\%$.  As in \reffi{fig:hcha1cha1} one can observe 
the MC2 with an ``interchange'' of $h_2$ and $h_3$.

In the right plot of \reffi{fig:hcha1cha2} one can see that the variation 
with $\phiMe$ is again very small with tiny $\CP$-asymmetries and the same
corrections as in the left plot (for the same reasons as explained above).

\medskip

Overall, for the neutral Higgs decays to a chargino pair we observe, 
again as expected, an increasing decay width $\propto \MHp$, as $\mh{2,3}$ 
increase nearly linearly with our input parameter $\MHp$.
The full one-loop corrections reach a level of $10\%$ for decay widths 
being of \order{1\gev}, and they can reach up to $20\%$ in the ``mixed'' 
decay mode.  The variation with $\phiMe$ is found to be negligible, as 
expected, since $M_1$ enters only via the loop corrections.

%%%%%%%%%%%%%%%%%%%%% T A B L E %%%%%%%%%%%%%%%%%%%%%%%%%%%%%%%%%%%%%%%%%%%%%%
\begin{table}[t!]
\caption{\label{tab:nthreshold} Thresholds in neutral Higgs boson decays.}
\centering
\begin{tabular}{lrr}
\toprule
TN1: & $\MHp \approx\phantom{0}805\gev\quad$ & 
$\mneu1 + \mneu3 = \mh2 \approx\phantom{0}799\gev$ \\
TN2: & $\MHp \approx\phantom{0}948\gev\quad$ & 
$\mcha1 + \mcha1  = \mh3 \approx\phantom{0}945\gev$ \\
TN3: & $\MHp \approx\phantom{0}954\gev\quad$ & 
$\mneu2 + \mneu2  = \mh3 \approx\phantom{0}951\gev$ \\
TN4: & $\MHp \approx\phantom{}1092\gev\quad$ & 
$\msbot1 + \msbot2 = \mh2 \approx\phantom{}1086\gev$ \\
TN5: & $\MHp \approx\phantom{}1107\gev\quad$ & 
$\mcha1 + \mcha2  = \mh3 \approx\phantom{}1105\gev$ \\
TN6: & $\MHp \approx\phantom{}1108\gev\quad$ & 
$\mcha1 + \mcha2  = \mh2 \approx\phantom{}1105\gev$ \\
TN7: & $\MHp \approx\phantom{}1112\gev\quad$ & 
$\mneu2 + \mneu4  = \mh2 \approx\phantom{}1108\gev$ \\
TN8: & $\MHp \approx\phantom{}1138\gev\quad$ & 
$\mneu3 + \mneu4  = \mh3 \approx\phantom{}1135\gev$ \\
TN9: & $\MHp \approx\phantom{}1168\gev\quad$ & 
$\mstop1 + \mstop2 = \mh3 = \phantom{}1165\gev$ \\
TN10: & $\MHp \approx\phantom{}1171\gev\quad$ & 
$\mstop1 + \mstop2 = \mh2 = \phantom{}1165\gev$ \\
TN11: & $\MHp \approx\phantom{}1268\gev\quad$ & 
$\mcha2 + \mcha2 = \mh2 \approx\phantom{}1264\gev$ \\
TN12: & $\MHp \approx\phantom{}1268\gev\quad$ & 
$\mneu4 + \mneu4 = \mh2 \approx\phantom{}1265\gev$ \\
TN13: & $\MHp \approx\phantom{}1545\gev\quad$ & 
$\mstop2 + \mstop2 = \mh2 = \phantom{}1542\gev$ \\
\bottomrule
\end{tabular}
\vspace{1em}
\end{table}
%%%%%%%%%%%%%%%%%%%%% T A B L E %%%%%%%%%%%%%%%%%%%%%%%%%%%%%%%%%%%%%%%%%%%%%%

\clearpage
\newpage

%%%%%%%%%%%%%%%%%%%%%%%%%% F I G U R E %%%%%%%%%%%%%%%%%%%%%%%%%%%%%%%%%%%%%%%%%
%%%%%%%%%%%%%%%%%%%%%%%%%% F I G U R E %%%%%%%%%%%%%%%%%%%%%%%%%%%%%%%%%%%%%%%%%

\clearpage
\newpage

%%%%%%%%%%%%%%%%%%%%%%%%%%%%%%%%%%%%%%%%%%%%%%%%%%%%%%%%%%%%%%%%%%%%%%%%%%%%%%

\subsubsection*{\boldmath{$h_i$} decays into neutralinos}

\begin{figure}[t!]
\begin{center}
\begin{tabular}{c}
\includegraphics[width=0.49\textwidth,height=7.5cm]{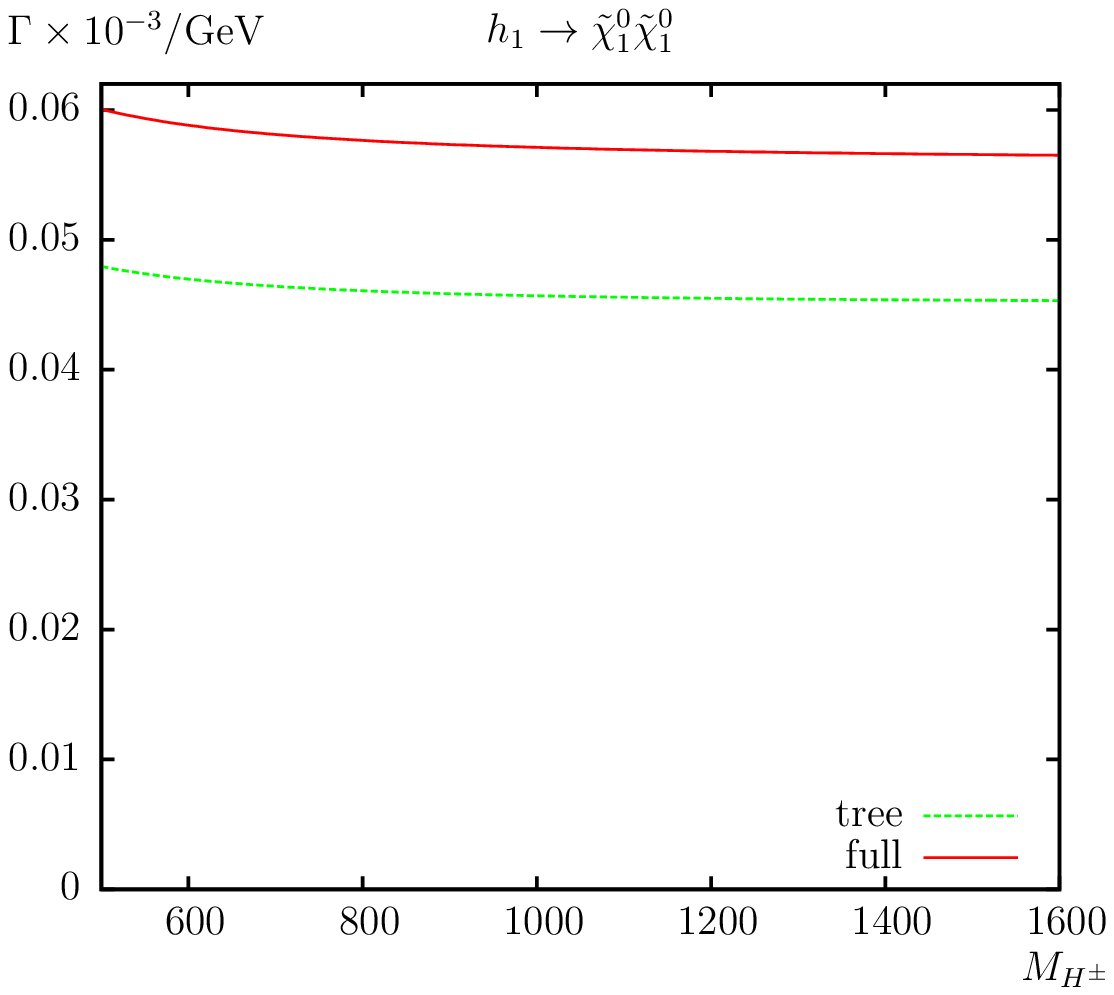}
\hspace{-4mm}
\includegraphics[width=0.49\textwidth,height=7.5cm]{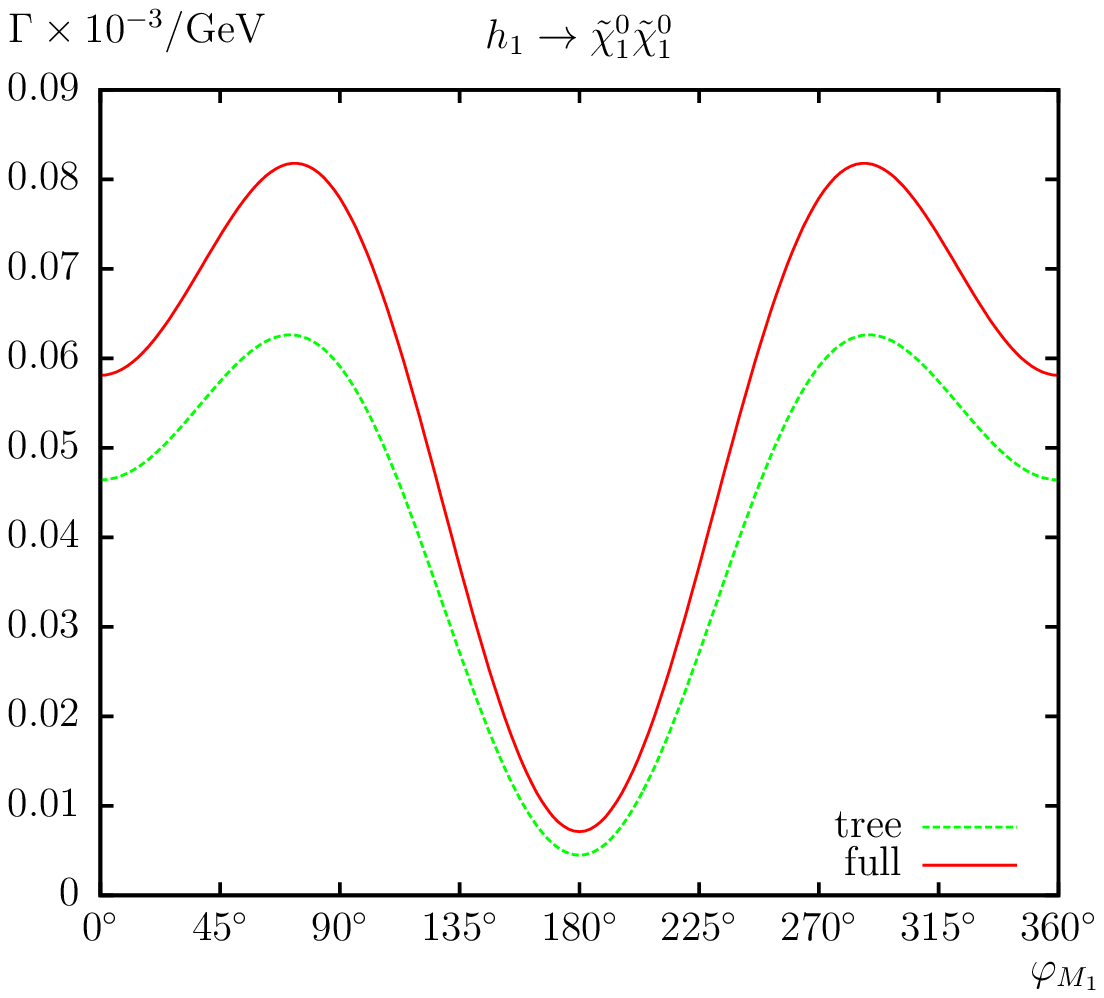} 
\\[2em]
\includegraphics[width=0.49\textwidth,height=7.5cm]{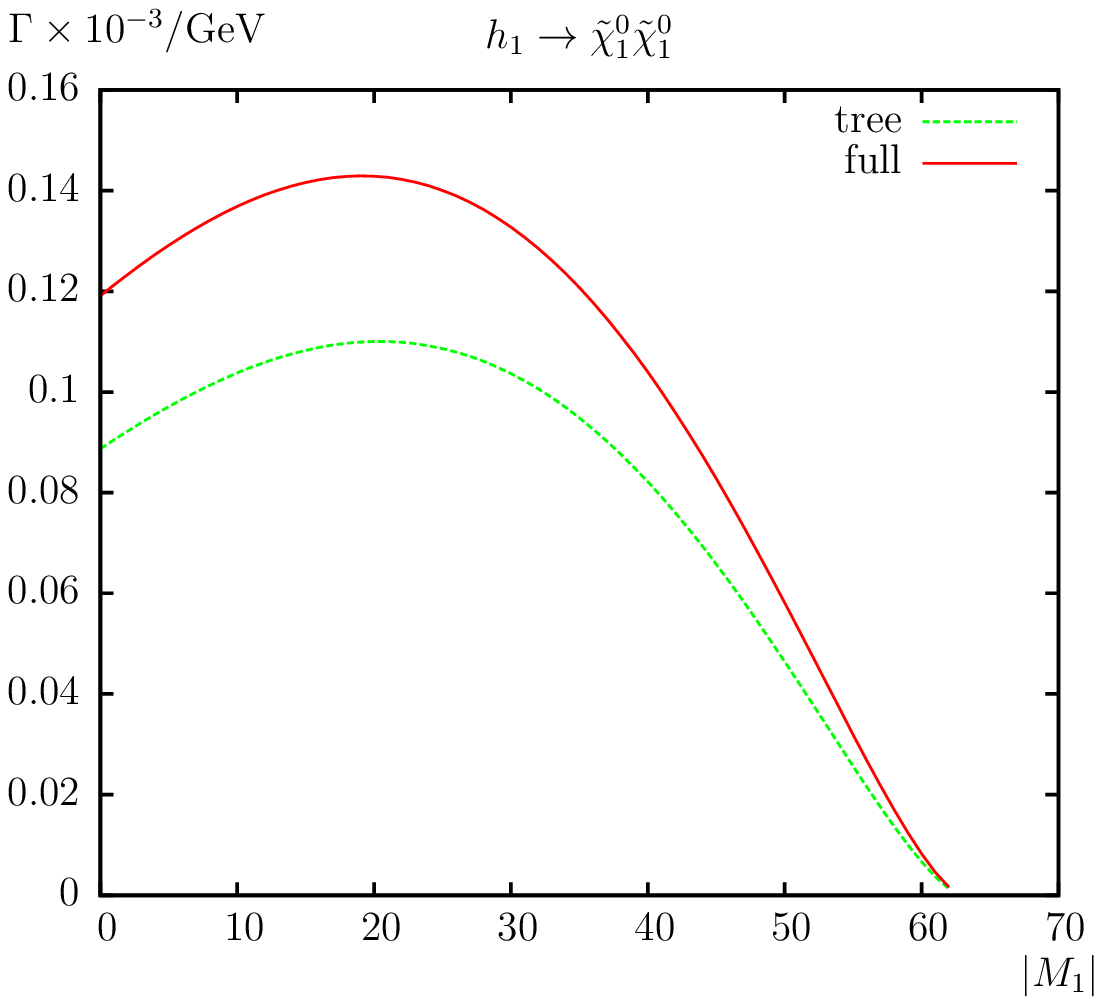}
\hspace{-4mm}
\includegraphics[width=0.49\textwidth,height=7.5cm]{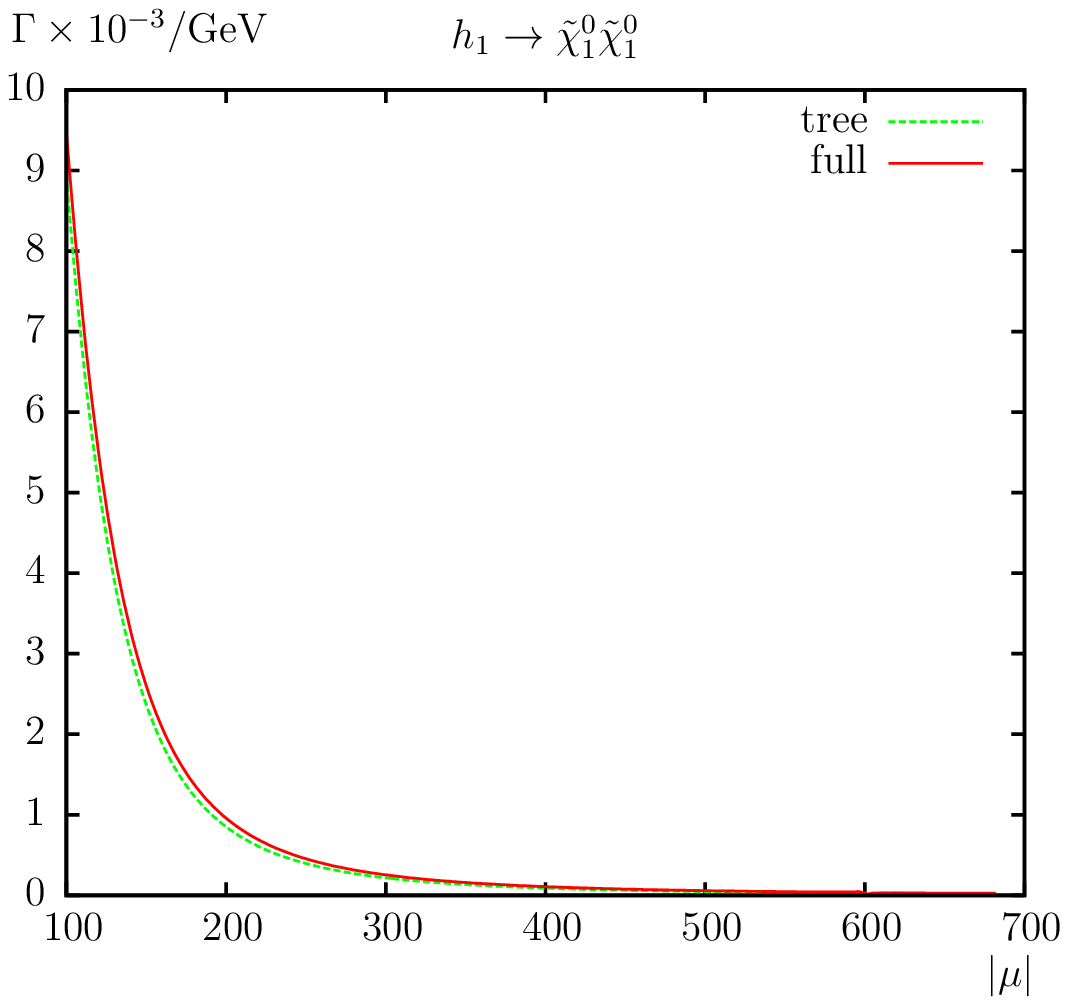}
\end{tabular}
\vspace{1em}
\caption{\label{fig:h1neu1neu1}
  $\Ga(h_1 \to \neu1 \neu1)$. 
  Tree-level and full one-loop corrected partial decay widths are shown
  with parameters chosen according to \Sce\ (see \refta{tab:para}) but 
  here with $|M_1| = 50\gev$. 
  The upper left plot shows the partial decay width with $\MHp$ varied. 
  The upper right plot shows the complex phase $\phiMe$ varied.
  The lower left (right) plot shows $M_1$ ($|\mu|$) varied.
}
\end{center}
\end{figure}

Next we consider $h_i$ decays into neutralinos with equal indices.
First, we present the decay $h_1 \to \neu1 \neu1$ in \reffi{fig:h1neu1neu1}.
Bounds on $\mneu1$ often assume an underlying GUT based on a simple Lie
group, leading to $M_1 = 5/3 (\SW/\CW)^2 M_2$. If the latter assumption is
dropped, hardly any direct bound on $\mneu1$ can be placed~\cite{masslessx}.
Therefore, we also treat $M_1$ as an independent parameter. 
The decay $h_1 \to \neu1\neu1$ constitutes an invisible decay of the light
Higgs boson, where bounds based on a combination of LHC and Tevatron data
yield an upper bound on an invisible branching ratio of 
$\sim 40\%$~\cite{BRinv}. Since the $\neu1$ constitutes a perfect Dark
Matter candidate in the MSSM~\cite{EHNOS}, in order
to determine the Dark Matter properties a precision measurement of this
process at the LHC or a future $e^+e^-$ collider will be necessary.
Consequently, a precise prediction of 
$\Ga(h_1 \to \neu1\neu1)$ is of particular interest.

\medskip

In the upper left plot of \reffi{fig:h1neu1neu1} we show the results for
$\MHp$ varied in \Sce, but with $|M_1| = 50\gev$ as the base scenario.
The full loop corrections are $\sim +25\%$ at $\MHp = 700\gev$ in the upper 
left plot.  $\phiMe$ is varied in the upper right plot.
One can observe a strong dependence of the decay width on $\phiMe$, which 
can change by a factor of 8.  The largest loop corrections are
found as $\sim +31\%$ for $\phiMe = 72^\circ, 288^\circ$ and $\sim +59\%$ at 
$\phiMe = 180^{\circ}$.  In the lower left plot of \reffi{fig:h1neu1neu1} 
we show the decay width with $M_1$ varied.  Close to $M_1 = 0$ the
lightest neutralino becomes massless.  For not too small values a decay
width of $\sim 10^{-4}\gev$ can be observed, going to zero at the
kinematic threshold. The one-loop corrections reach up to $\sim +30\%$ 
at $M_1 = 20\gev$. Finally, in the lower right plot $|\mu|$ is varied, 
and the decay width drops down to $\sim 10^{-5}\gev$ for $\mu > 600\gev$ 
and with corrections between $\sim +6\%$ and $\sim +28\%$.

\bigskip

\begin{figure}[ht!]
\begin{center}
\begin{tabular}{c}
\includegraphics[width=0.49\textwidth,height=7.5cm]{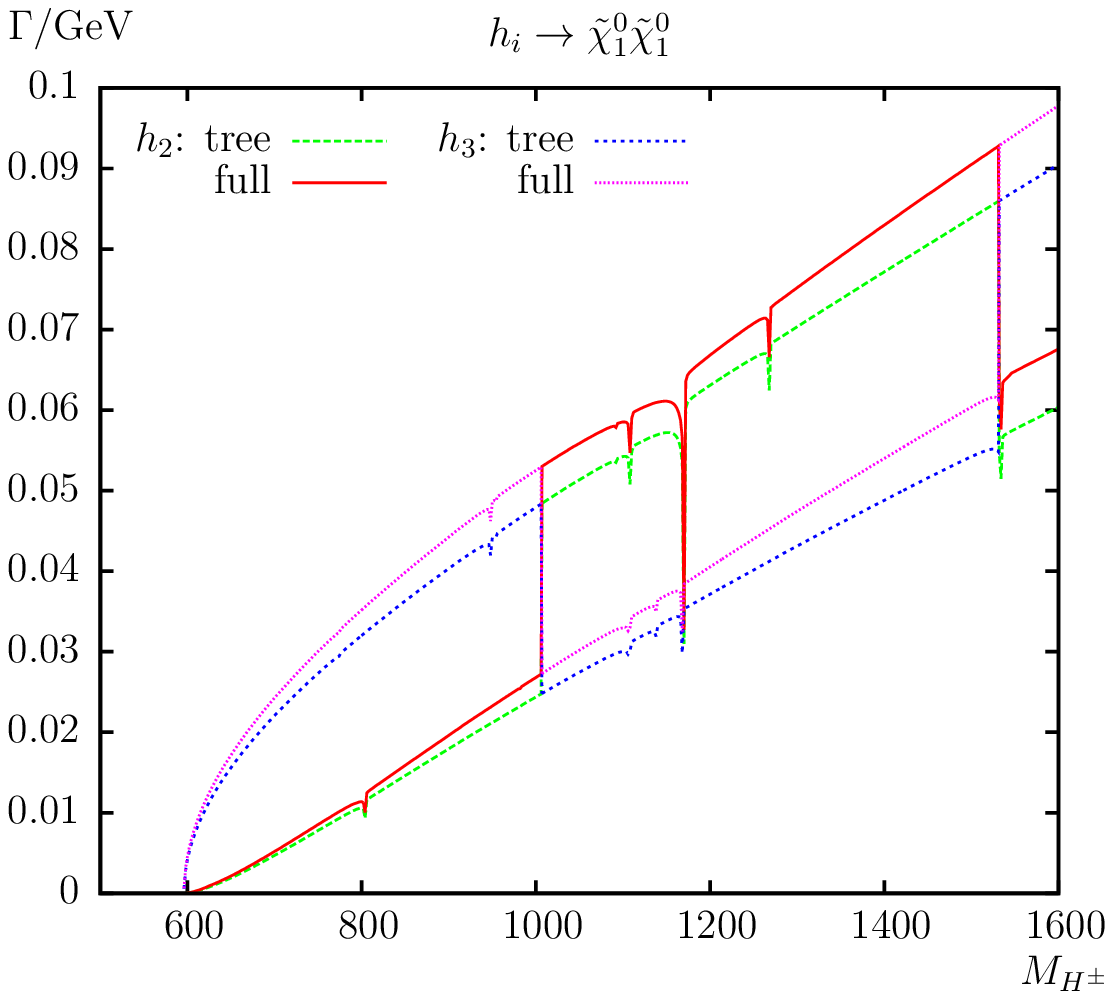}
\hspace{-4mm}
\includegraphics[width=0.49\textwidth,height=7.5cm]{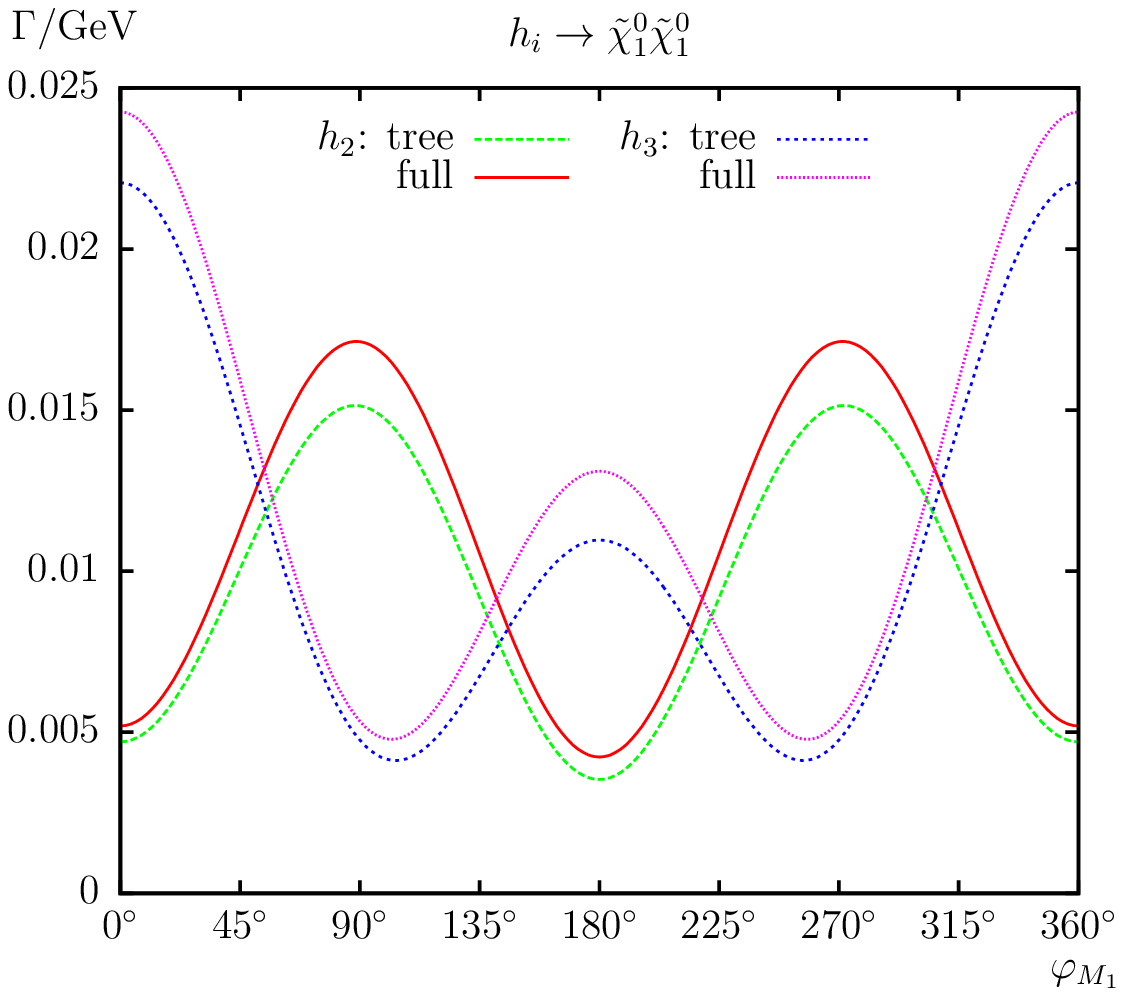}
\end{tabular}
\vspace{1em}
\caption{\label{fig:hneu1neu1}
  $\Ga(\hneueneue)$. 
  Tree-level and full one-loop corrected partial decay widths are shown. 
  The left plot shows the partial decay width with $\MHp$ varied. 
  The right plot shows the complex phase $\phiMe$ varied with parameters 
  chosen according to \Sce\ (see \refta{tab:para}).
}
\vspace{6em}
\begin{tabular}{c}
\includegraphics[width=0.49\textwidth,height=7.5cm]{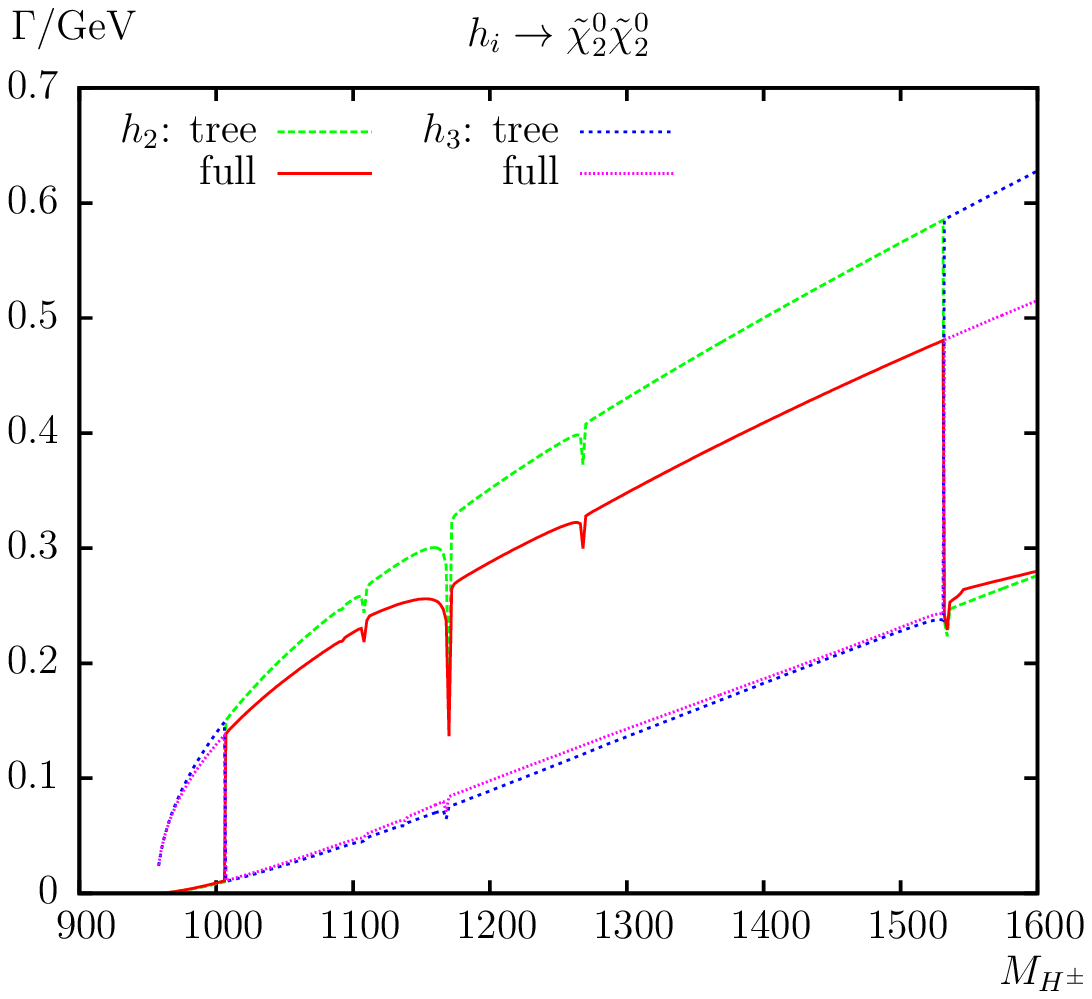}
\hspace{-4mm}
\includegraphics[width=0.49\textwidth,height=7.5cm]{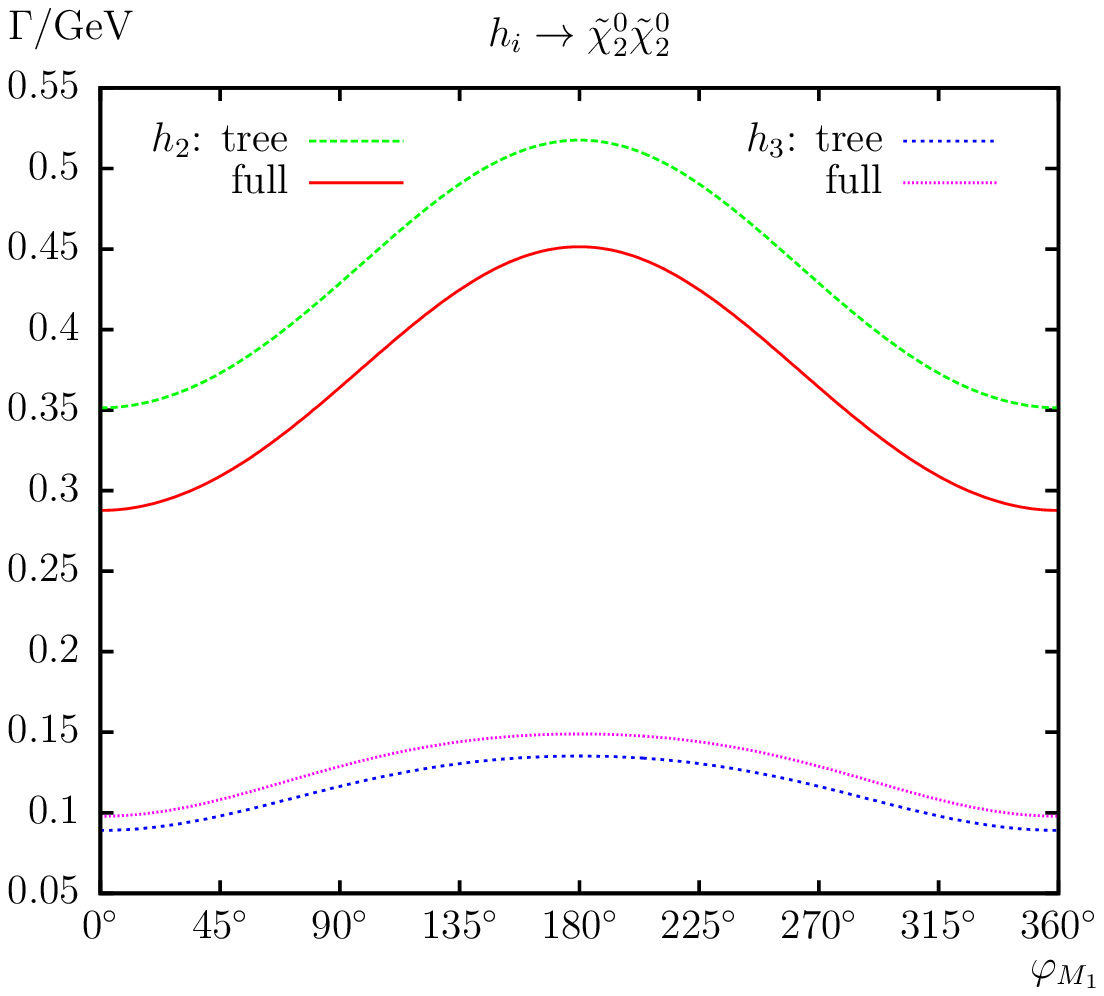}
\end{tabular}
\vspace{1em}
\caption{\label{fig:hneu2neu2}
  $\Ga(\hneuzneuz)$. 
  Tree-level and full one-loop corrected partial decay widths are shown. 
  The left plot shows the partial decay width with $\MHp$ varied. 
  The right plot shows the complex phase $\phiMe$ varied with parameters 
  chosen according to \Scv\ (see \refta{tab:para}).
}
\end{center}
\end{figure}

We now turn to the decays of the heavy neutral Higgs bosons.
In \reffi{fig:hneu1neu1} we present the results for the decays 
$\hneueneue$ ($i = 2,3$). The dependence on $\MHp$ is shown in the 
left plot, whereas the dependence on $\phiMe$ for $\MHp = 700\gev$ 
is given in the right plot. 
We start with $\Ga(\hneueneue)$ in the left plot. 
The first dip (lower lines) in the $h_2$ decay is the threshold 
TN1, see \refta{tab:nthreshold}.
The second dip (hardly visible in the upper lines) is the threshold 
TN4.%
\footnote{
  It should be noted that this threshold enter 
  \textit{into the tree-level} only via the 
  $\matr{\hat{Z}}$~matrix contribution. 
  These effects propagate also into the loop corrections via 
  $2 \Re \{\cMt^*\, \cMl^{}\}$.
  Furthermore $\matr{\hat{Z}}$ is calculated by \FH\ which uses 
  $\mb(\mb)$ and tree-level sfermion masses instead of the shifted 
  masses, causing a slight displacement in the threshold position.  
}
The third ``apparently single'' dip is (again) in reality coming from 
the thresholds TN6 and TN7.
The fourth (large) dip is the threshold TN10.
The last ``apparently single'' dip is in reality coming from the 
thresholds TN11 and TN12.
The full loop corrections are $\sim +11\%$ at $\MHp = 700\gev$.  
Also shown in this plot is the decay $h_3 \to \neu1\neu1$.
The first dip (upper lines) is in reality coming from the thresholds 
TN2 and TN3.  The second dip (lower lines) in the $h_3$ 
decay is the threshold TN5.  The third dip (lower lines) is 
the threshold TN8 and the last dip is the threshold TN9.  
The full relative corrections reach $\sim +10\%$ at $\MHp = 700\gev$.
The suppression of the $\CP$-even decay
(lower lines, going with \refeq{hEvenTree} at threshold, 
and then roughly linear with $\mh2$) vs.\ the $\CP$-odd decay 
(upper lines, going with \refeq{hOddTree} at threshold, 
and then roughly linear with $\mh3$) is again clearly visible.

In the right plot of \reffi{fig:hneu1neu1} we show the $h_2$
decay with the complex phase $\phiMe$ varied at $\MHp = 700\gev$.  
The variation with $\phiMe$ is found to be very large, 
changing the decay width by up to a factor of~5 where the
full relative 
corrections are up to $\sim +20\%$ at $\phiMe = 180^{\circ}$ for \Sce.
The $h_3$ decay with the complex phase $\phiMe$ shows also a very 
large variation at $\MHp = 700\gev$ and the loop corrections reach 
up to $\sim +19\%$ at $\phiMe = 180^\circ$.

%%%%%%%%%%%%%%%%%%%%%%%%%% F I G U R E %%%%%%%%%%%%%%%%%%%%%%%%%%%%%%%%%%%%%%%%%
%%%%%%%%%%%%%%%%%%%%%%%%%% F I G U R E %%%%%%%%%%%%%%%%%%%%%%%%%%%%%%%%%%%%%%%%%

\medskip

In \reffi{fig:hneu2neu2} we show the decays $\hneuzneuz$, in full 
analogy to \reffi{fig:hneu1neu1}. The same behavior of $h_2$ and 
$h_3$ concerning MC1 and MC2, as well as the $\CP$-even and 
$\CP$-odd decay can be observed. 
The dips (some are hardly visible) are the same as already described 
in \reffi{fig:hneu1neu1} beginning at $\MHp \approx 1092\gev$, 
see above.  The full relative corrections for the $h_2$ ($h_3$) decay 
are $\sim -18\%$ ($\sim +10\%$) at $\MHp = 1200\gev$, \ie \Scv.

In the right plot of \reffi{fig:hneu2neu2} we show the variation 
of $\Ga(h_2 \to \neu2\neu2)$ with $\phiMe$ at $\MHp = 1200\gev$. 
Here the loop corrections can vary between $\sim -18\%$ for 
$\phiMe = 0^\circ, 360^\circ$ and $\sim -13\%$ at 
$\phiMe = 180^{\circ}$. The $h_3$ decay with $\phiMe$ varied reach 
$\sim +11\%$ for $\phiMe \sim 90^\circ, 270^{\circ}$ in \Scv. 

\medskip

\begin{figure}[ht!]
\begin{center}
\begin{tabular}{c}
\includegraphics[width=0.49\textwidth,height=7.5cm]{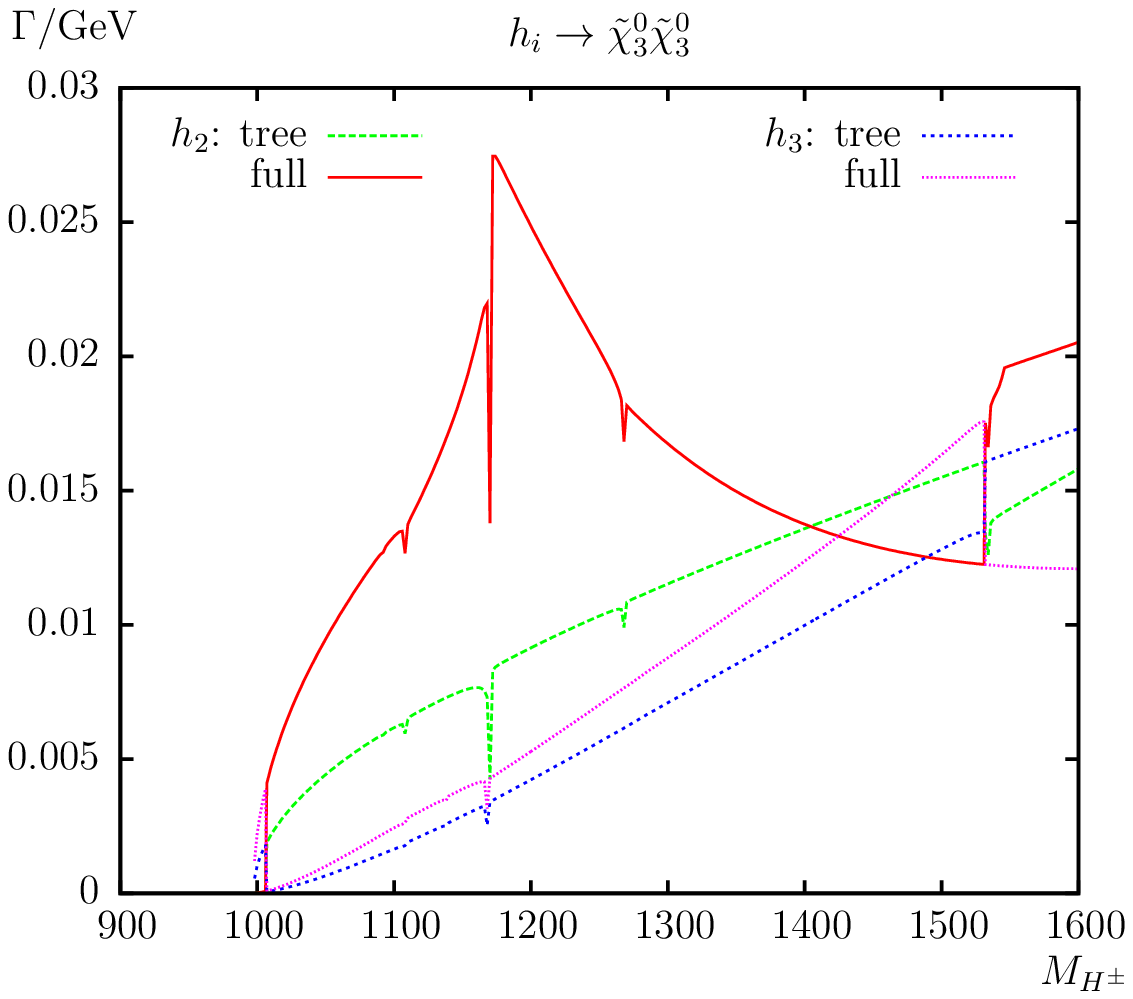}
\hspace{-4mm}
\includegraphics[width=0.49\textwidth,height=7.5cm]{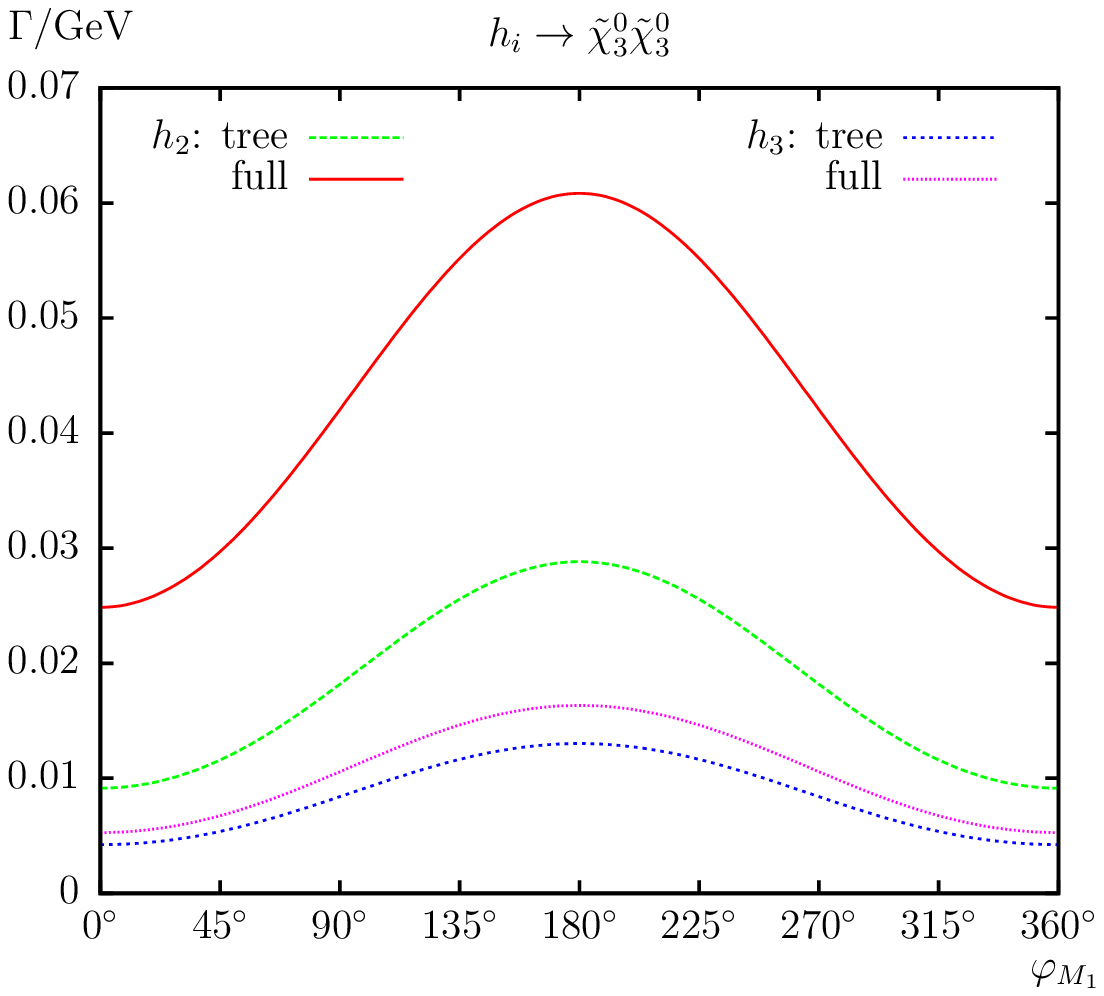}
\end{tabular}
\vspace{1em}
\caption{\label{fig:hneu3neu3}
  $\Ga(\hneudneud)$. 
  Tree-level and full one-loop corrected partial decay widths are shown. 
  The left plot shows the partial decay width with $\MHp$ varied. 
  The right plot shows the complex phase $\phiMe$ varied with parameters 
  chosen according to \Scv\ (see \refta{tab:para}).
}
\vspace{6em}
\begin{tabular}{c}
\includegraphics[width=0.49\textwidth,height=7.5cm]{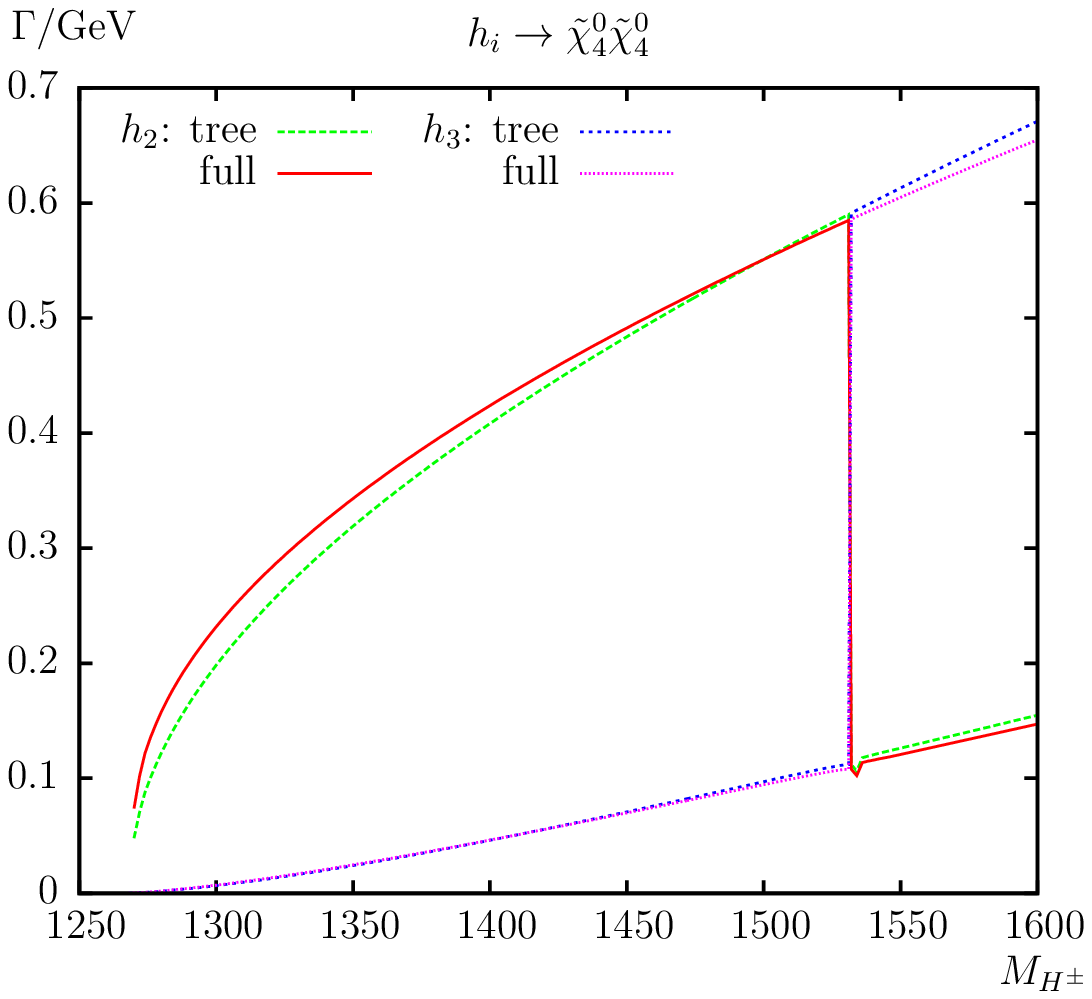}
\hspace{-4mm}
\includegraphics[width=0.49\textwidth,height=7.5cm]{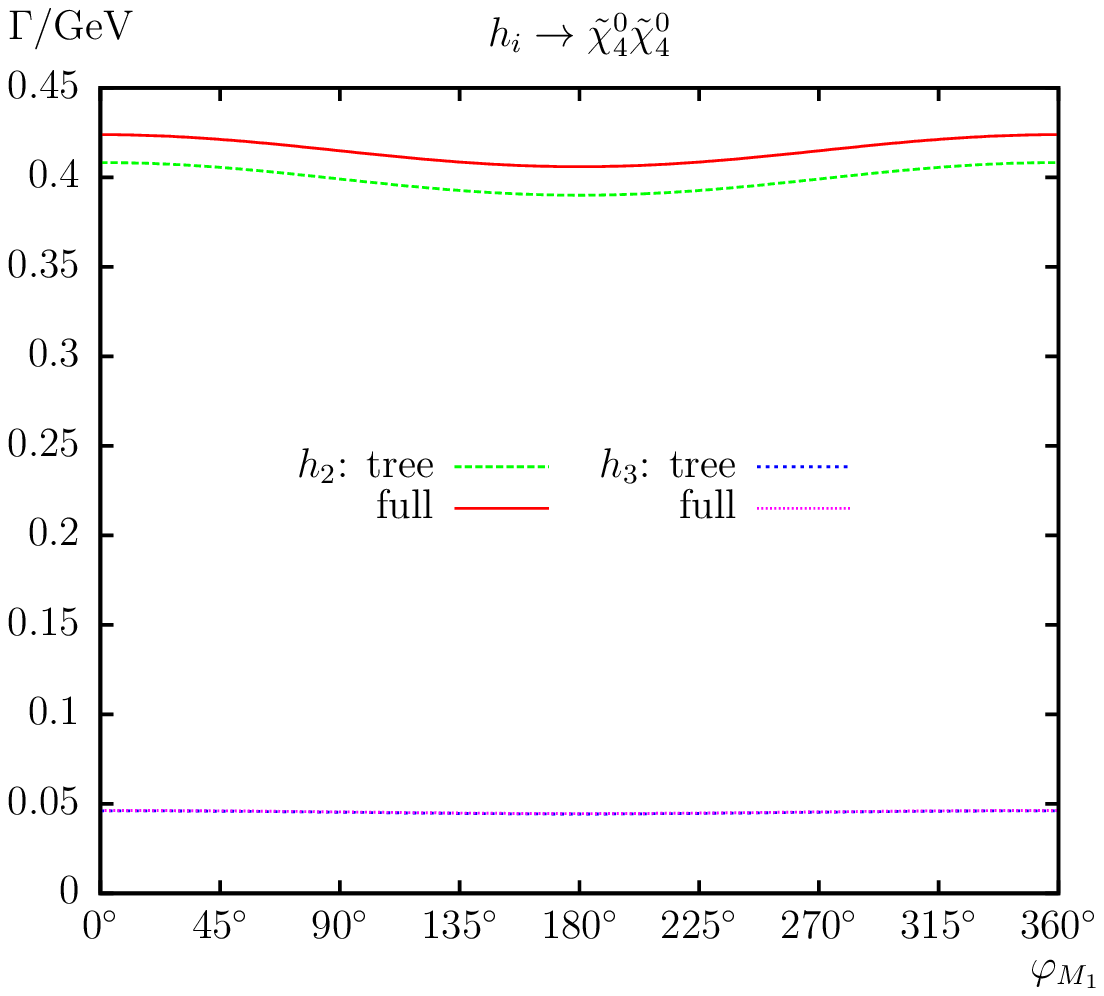}
\end{tabular}
\vspace{1em}
\caption{\label{fig:hneu4neu4}
  $\Ga(\hneuvneuv)$. 
  Tree-level and full one-loop corrected partial decay widths are shown. 
  The left plot shows the partial decay width with $\MHp$ varied. 
  The right plot shows the complex phase $\phiMe$ varied with parameters 
  chosen according to \Scf\ (see \refta{tab:para}).
}
\end{center}
\end{figure}

Next, in \reffi{fig:hneu3neu3} we present the decays $\hneudneud$, 
in full analogy to \reffi{fig:hneu1neu1}. The same behavior of 
$h_2$ and $h_3$ concerning MC1 and MC2 and the $\CP$-even/-odd
decay can be observed.  The dips (some are hardly visible) 
are again the same as described in \reffi{fig:hneu1neu1} beginning 
at $\MHp \approx 1092\gev$, see above.  
The ``knee'' at $\MHp \approx 1545\gev$ (red line) is the threshold 
TN13 (see \refta{tab:nthreshold}) in the $C$-functions
$C_{0,1,2}(\mneu3^2,\mh2^2\mneu3^2,m_t^2,\mstop2^2,\mstop2^2)$. 
The full one-loop corrections for the $h_2$ decay are $\sim +172\%$ 
at $\MHp = 1200\gev$.  This strange behavior is a numerical effect 
caused by an interplay of anomalous thresholds in
$C_{0,1,2}(\mneu3^2,\mh2^2,\mneu3^2,m_t^2,\mstop{\sind}^2,\mstop{\spri}^2)$
($\sind \ne \spri$) with the effects induced by the Higgs mixing matrix 
$\matr{\hat Z}$.
This effect is absent in the decay of the $h_3$, where we find the full 
relative corrections at the level of $\sim +25\%$ for $\MHp = 1200\gev$.

In the right plot of \reffi{fig:hneu3neu3} we show the variation of 
$\Ga(h_2 \to \neu3\neu3)$ with $\phiMe$ at $\MHp = 1200\gev$
(i.e.\ at an ``extreme'' point for the $h_2$ decay).
Here (for the same reasons as in the left plot) the loop corrections 
reach $\sim +111\%$ at $\phiMe = 180^{\circ}$.
Also in the right plot of \reffi{fig:hneu3neu3} we show 
$\Ga(h_3 \to \neu3\neu3)$ with $\phiMe$ varied in \Scv. 
Here the loop corrections can reach $\sim +26\%$ at 
$\phiMe = 90^{\circ}, 270^{\circ}$.

\medskip

In \reffi{fig:hneu4neu4} we present the decays $\hneuvneuv$, again 
in full analogy to \reffi{fig:hneu1neu1}.  The same behavior of $h_2$ 
and $h_3$ concerning MC2 and the $\CP$-even/-odd decay can 
be observed.  The full relative corrections for the $h_2$ decay are 
$\sim +4\%$ at $\MHp = 1400\gev$, \ie \Scf, while the $h_3$ decay 
shows relative corrections less than $+1\%$ at $\MHp = 1400\gev$.  

In the right plot of \reffi{fig:hneu4neu4} we show 
$\Ga(h_{2,3} \to \neu4\neu4)$ at $\MHp = 1400\gev$.
For both decays the variation of $\phiMe$ is very small, \ie the
loop corrections reach the same values (in \Scf) as in the left plot.

\bigskip

We now turn to the neutral Higgs decays to neutralinos with different 
indices.  In this case, contrary to the decay into identical charginos,
the $\CP$-asymmetries are also zero, due to the Majorana nature of the 
neutralinos.

\begin{figure}[htb!]
\begin{center}
\begin{tabular}{c}
\includegraphics[width=0.49\textwidth,height=7.5cm]{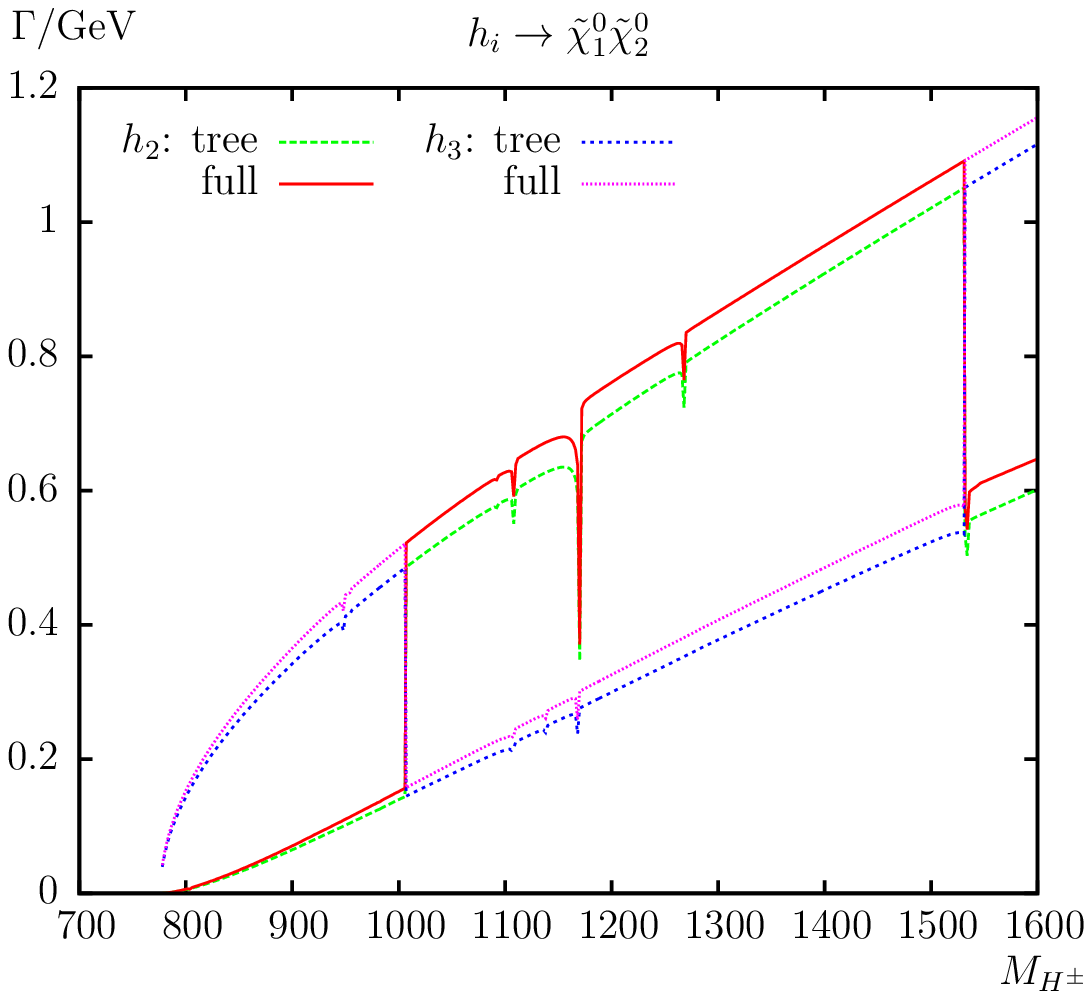}
\hspace{-4mm}
\includegraphics[width=0.49\textwidth,height=7.5cm]{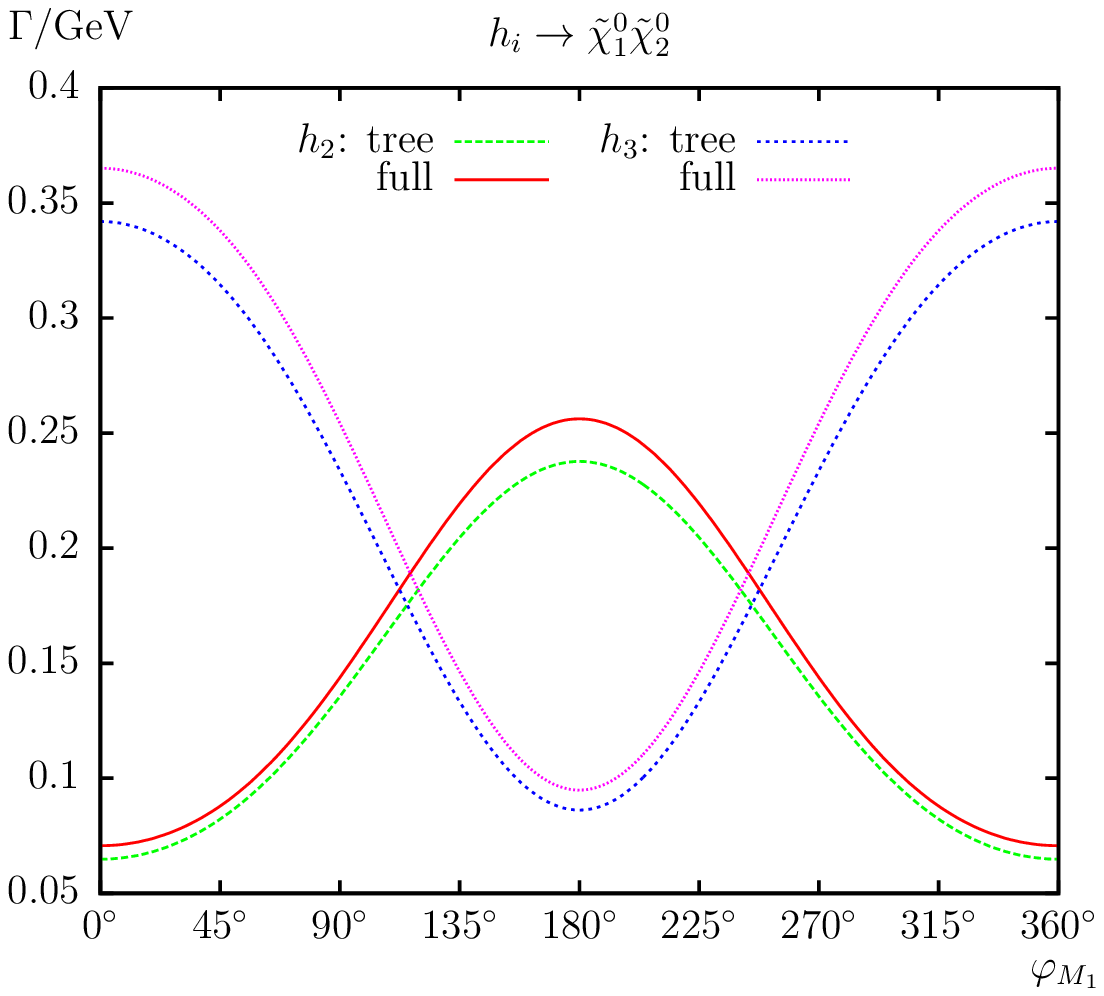}
\end{tabular}
\vspace{1em}
\caption{\label{fig:hneu1neu2}
  $\Ga(\hneueneuz)$. 
  Tree-level and full one-loop corrected partial decay widths are shown. 
  The left plot shows the partial decay width with $\MHp$ varied. 
  The right plot shows the complex phase $\phiMe$ varied with parameters 
  chosen according to \Scz\ (see \refta{tab:para}).
}
\vspace{6em}
\begin{tabular}{c}
\includegraphics[width=0.49\textwidth,height=7.5cm]{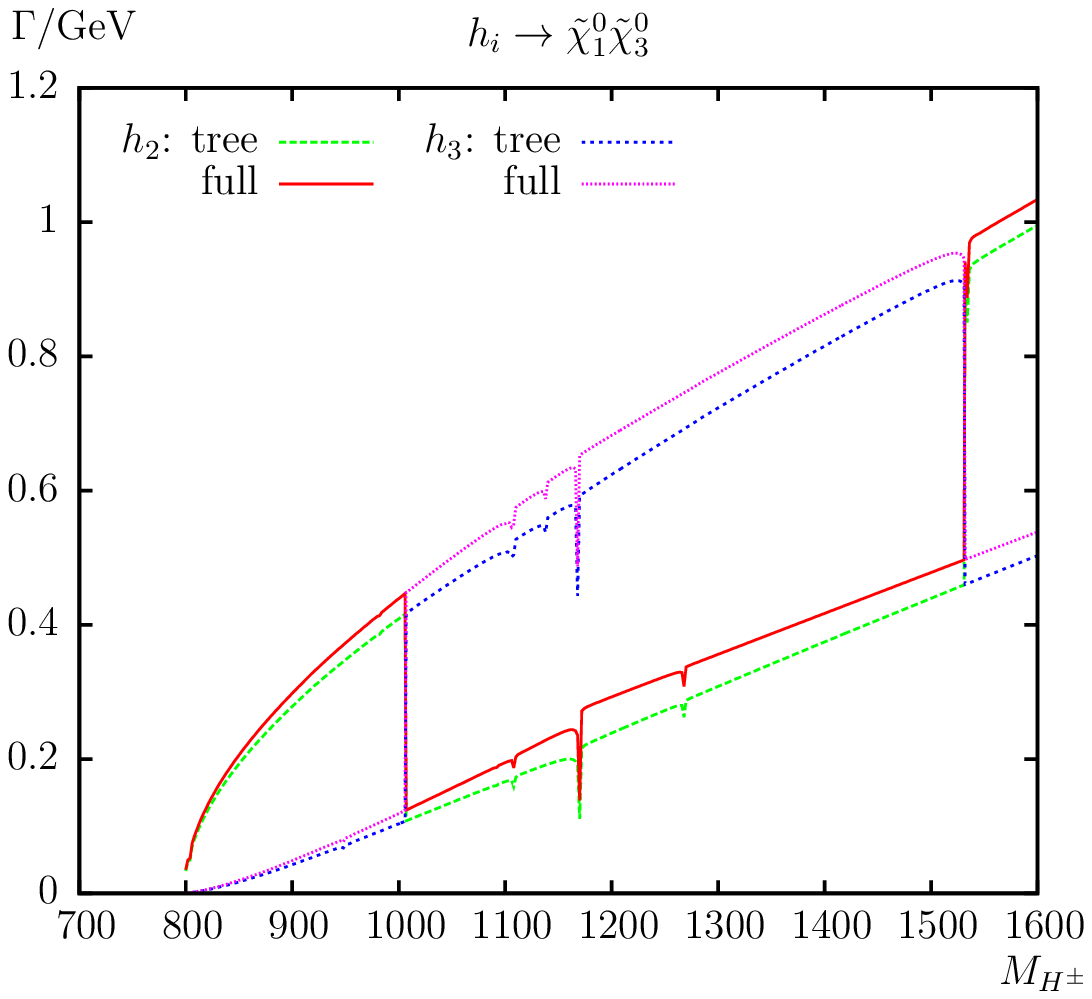}
\hspace{-4mm}
\includegraphics[width=0.49\textwidth,height=7.5cm]{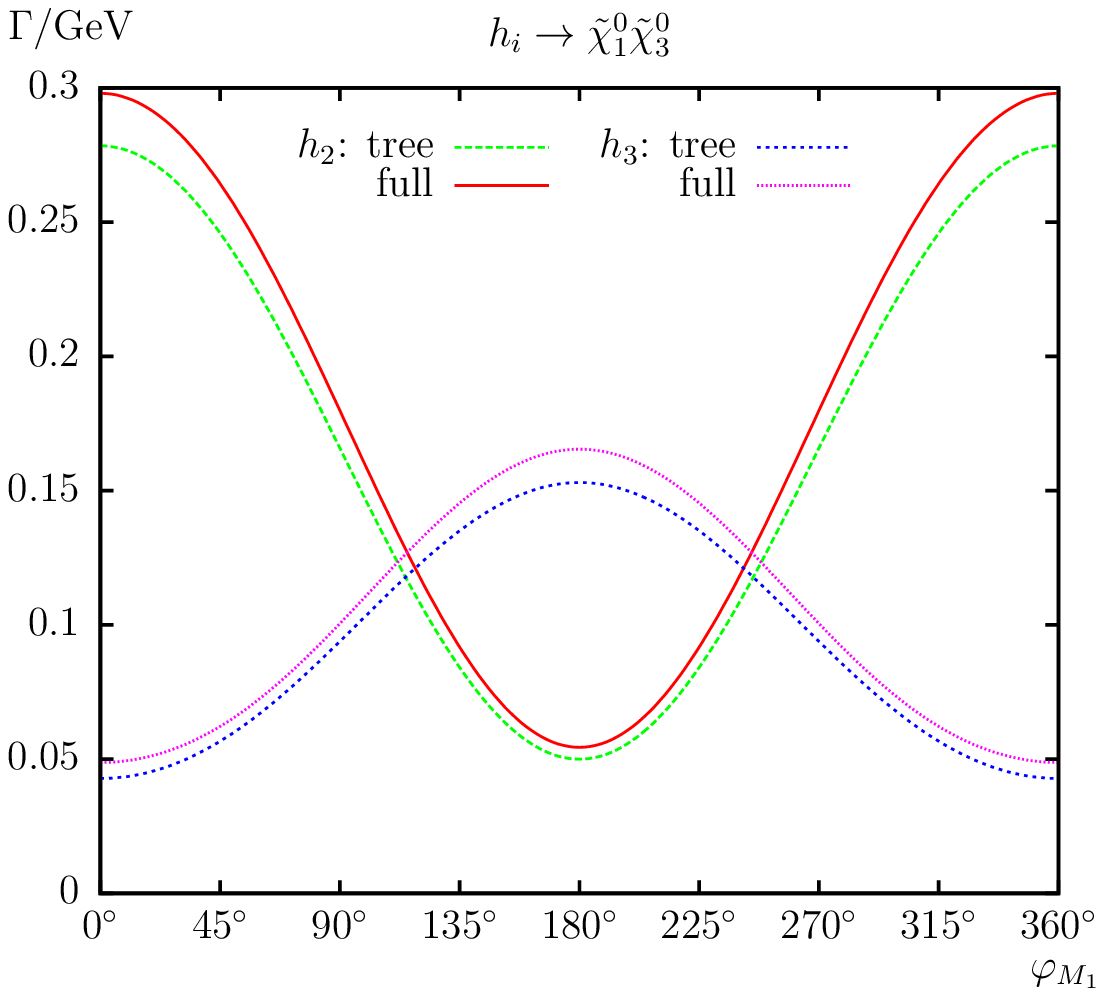}
\end{tabular}
\vspace{1em}
\caption{\label{fig:hneu1neu3}
  $\Ga(\hneueneud)$. 
  Tree-level and full one-loop corrected partial decay widths are shown. 
  The left plot shows the partial decay width with $\MHp$ varied. 
  The right plot shows the complex phase $\phiMe$ varied with parameters 
  chosen according to \Scz\ (see \refta{tab:para}).
}
\end{center}
\end{figure}

In \reffi{fig:hneu1neu2} we present the decay $\hneueneuz$.
In the left plot we show the results as a function of $\MHp$.
The dips are (again) the same as already described in 
\reffi{fig:hneu1neu1} beginning at $\MHp \approx 948\gev$, see above.
The full relative corrections for the $h_2$ decay are $\sim +9\%$ 
at $\MHp = 900\gev$ (\ie \Scz).  The full one-loop corrections for 
the $h_3$ decay at $\MHp = 900\gev$ reach $+7\%$.

In the right plot of \reffi{fig:hneu1neu2} we show the $h_i$ decay 
with the complex phase $\phiMe$ varied at $\MHp = 900\gev$.  
The variation with $\phiMe$ is found to be very large and the loop 
corrections vary between $\sim +9\%$ for $\phiMe \sim 0^\circ$ and 
$\sim +6\%$ at $\phiMe = 90^{\circ}$. 
We also show the $h_3$ decay in the right plot of \reffi{fig:hneu1neu2} 
with $\phiMe$ varied at $\MHp = 900\gev$.  The variation with $\phiMe$ 
is yet larger than in the $h_2$ case.  The full corrections can reach 
$\sim +10\%$ at $\phiMe = 180^{\circ}$.  
The general behavior can be understood as follows.  For $\phiMe = 0^\circ$
one finds $\CP(h_2 \neu1\neu2) = -\CP(h_3 \neu1\neu2) = +1$, leading
to the above discussed suppression of the $h_2$ decay, see 
\refeqs{hOddTree}, (\ref{hEvenTree}).  
Going to $\phiMe = 180^\circ$ changes the $\CP$-nature of the $\neu1$, 
leading to $\CP(h_2 \neu1\neu2) = -\CP(h_3 \neu1\neu2) = -1$ and the 
corresponding suppression of the $h_3$ decay.

\medskip

In \reffi{fig:hneu1neu3} we present the decay $\hneueneud$.
It should be noted that the decay $\hneueneud$ \textit{looks} quite 
similar to \reffi{fig:hneu1neu2} but with an interchange of $h_2$ 
with $h_3$.  In the left plot we show the results as a function of 
$\MHp$.  The dips are (again) the same as already described in 
\reffi{fig:hneu1neu1} beginning at $\MHp \approx 948\gev$, see above.
The full one-loop corrections for the $h_2$ decay reach $\sim +7\%$ 
at $\MHp = 900\gev$ (\ie \Scz).
The relative corrections for the $h_3$ decay at $\MHp = 900\gev$ 
are $+14\%$.
In comparison with \reffi{fig:hneu1neu2} one can observe an
``inversion'' of the relative size of the decays widths of the $h_2$ and
the $h_3$ (green/red lines vs.\ blue/purple lines).  This ``inversion'' 
is due to the fact that $\CP(\neu1\neu2) = -\CP(\neu1\neu3)$.

In the right plot of \reffi{fig:hneu1neu3} we show the $h_i$ decay 
with the complex phase $\phiMe$ varied at $\MHp = 900\gev$.  
This variation is (again) found to be very large, as can be seen in 
the right plot.  The loop corrections for the $h_2$ decay can reach 
$\sim +9\%$ at $\phiMe = 180^{\circ}$. 
In the right plot of \reffi{fig:hneu1neu3} we show also the $h_3$ 
decay with $\phiMe$ varied at $\MHp = 900\gev$.  There the loop 
corrections vary between $\sim +14\%$ for $\phiMe \sim 0^\circ$ and 
$\sim +7\%$ at $\phiMe = 90^{\circ}$.
Again the ``inversion'' (as in the left plot) can be observed.

\medskip

\begin{figure}[htb!]
\begin{center}
\begin{tabular}{c}
\includegraphics[width=0.49\textwidth,height=7.5cm]{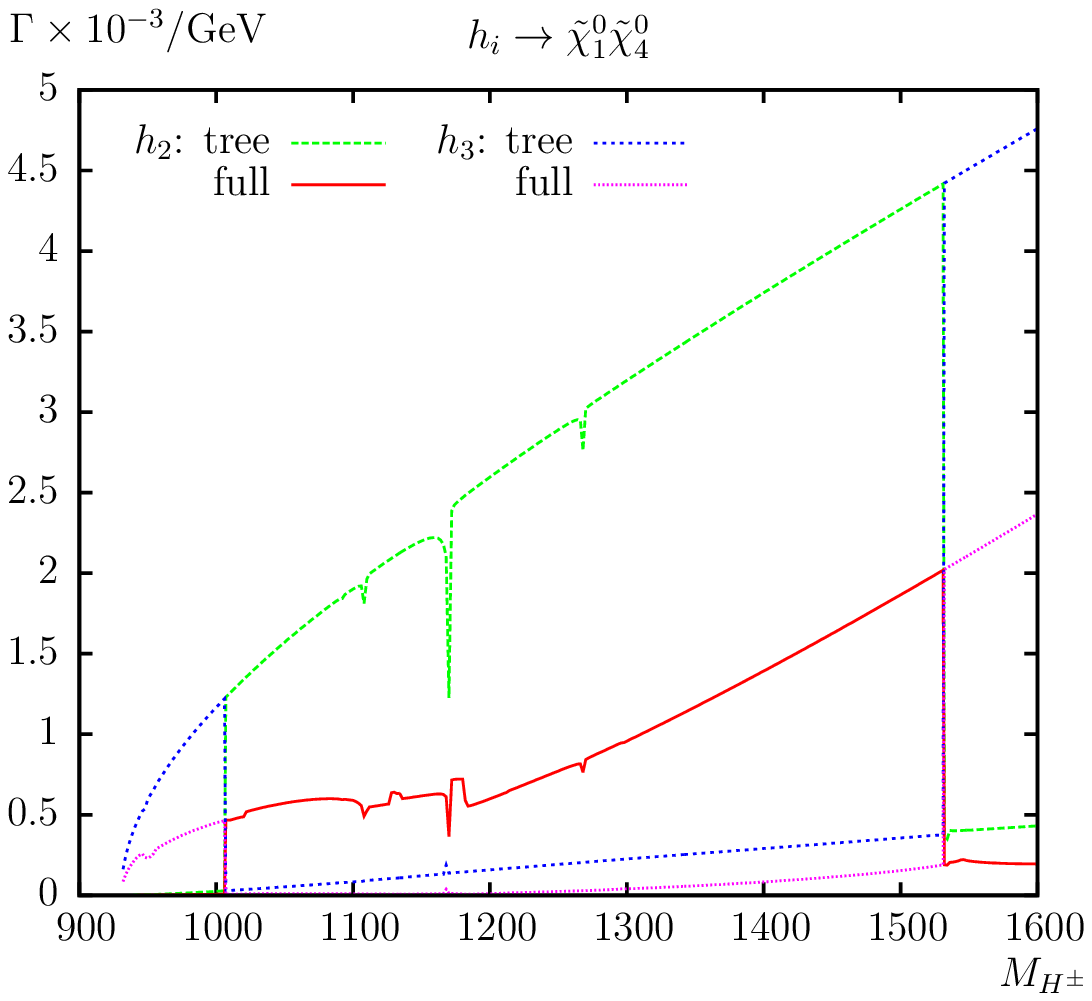}
\hspace{-4mm}
\includegraphics[width=0.49\textwidth,height=7.5cm]{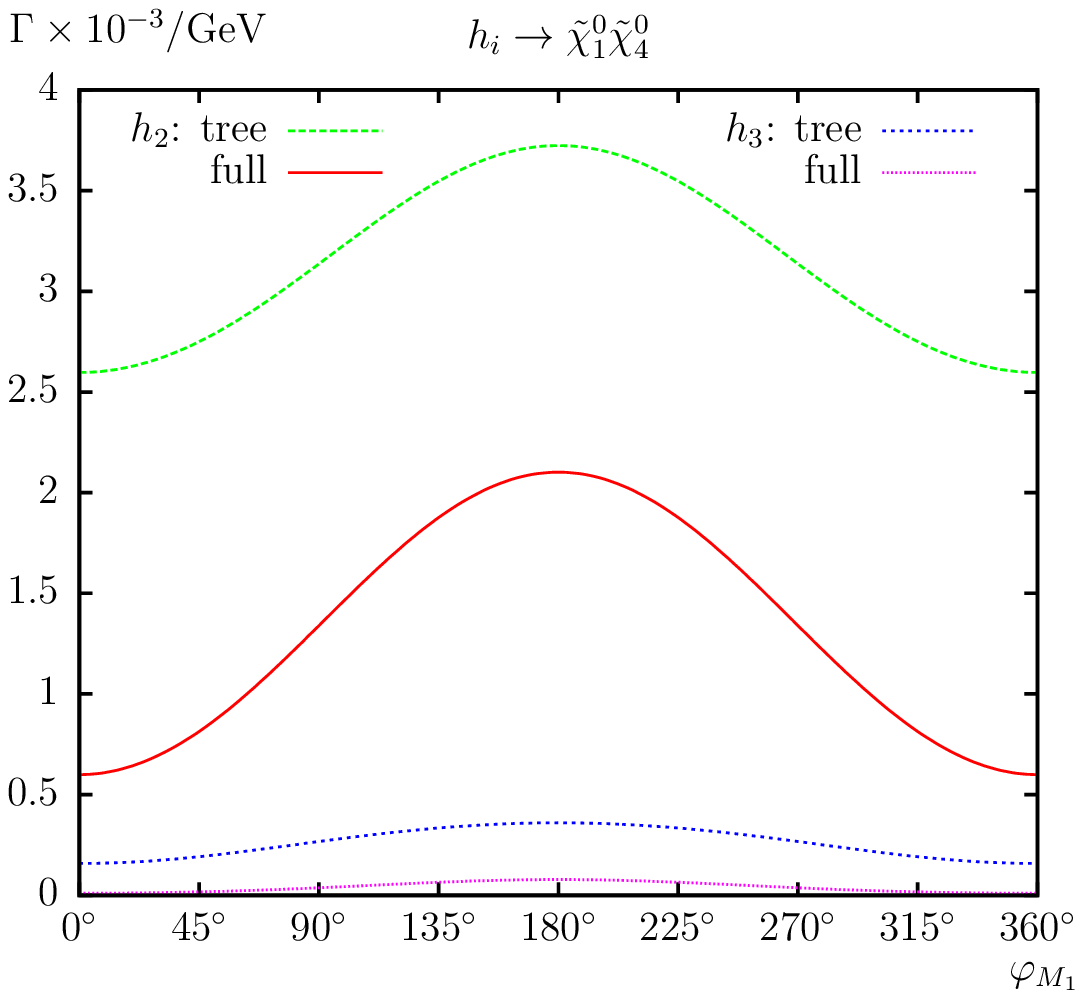}
\end{tabular}
\vspace{1em}
\caption{\label{fig:hneu1neu4}
  $\Ga(\hneueneuv)$. 
  Tree-level and full one-loop corrected partial decay widths are shown. 
  The left plot shows the partial decay width with $\MHp$ varied. 
  The right plot shows the complex phase $\phiMe$ varied with parameters 
  chosen according to \Scv\ (see \refta{tab:para}).
}
\vspace{6em}
\begin{tabular}{c}
\includegraphics[width=0.49\textwidth,height=7.5cm]{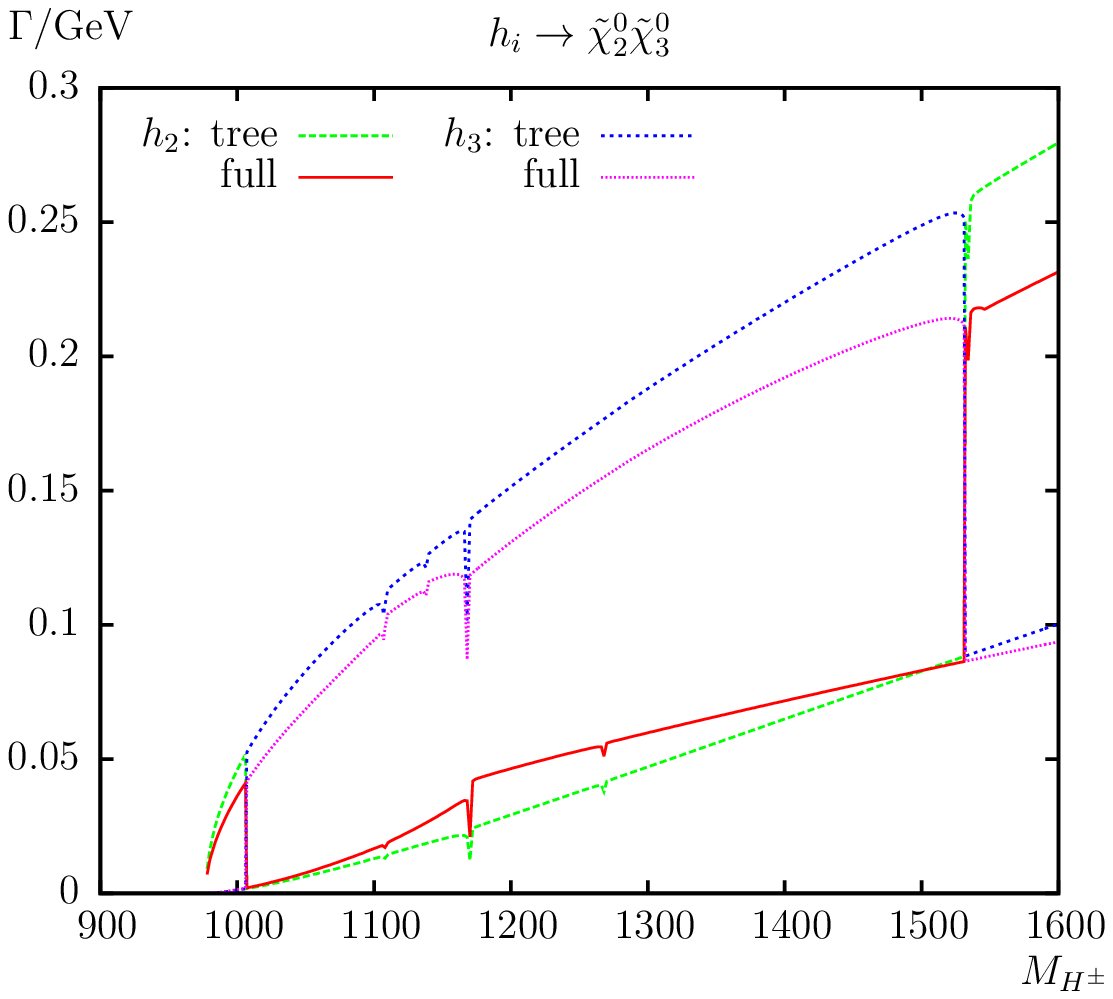}
\hspace{-4mm}
\includegraphics[width=0.49\textwidth,height=7.5cm]{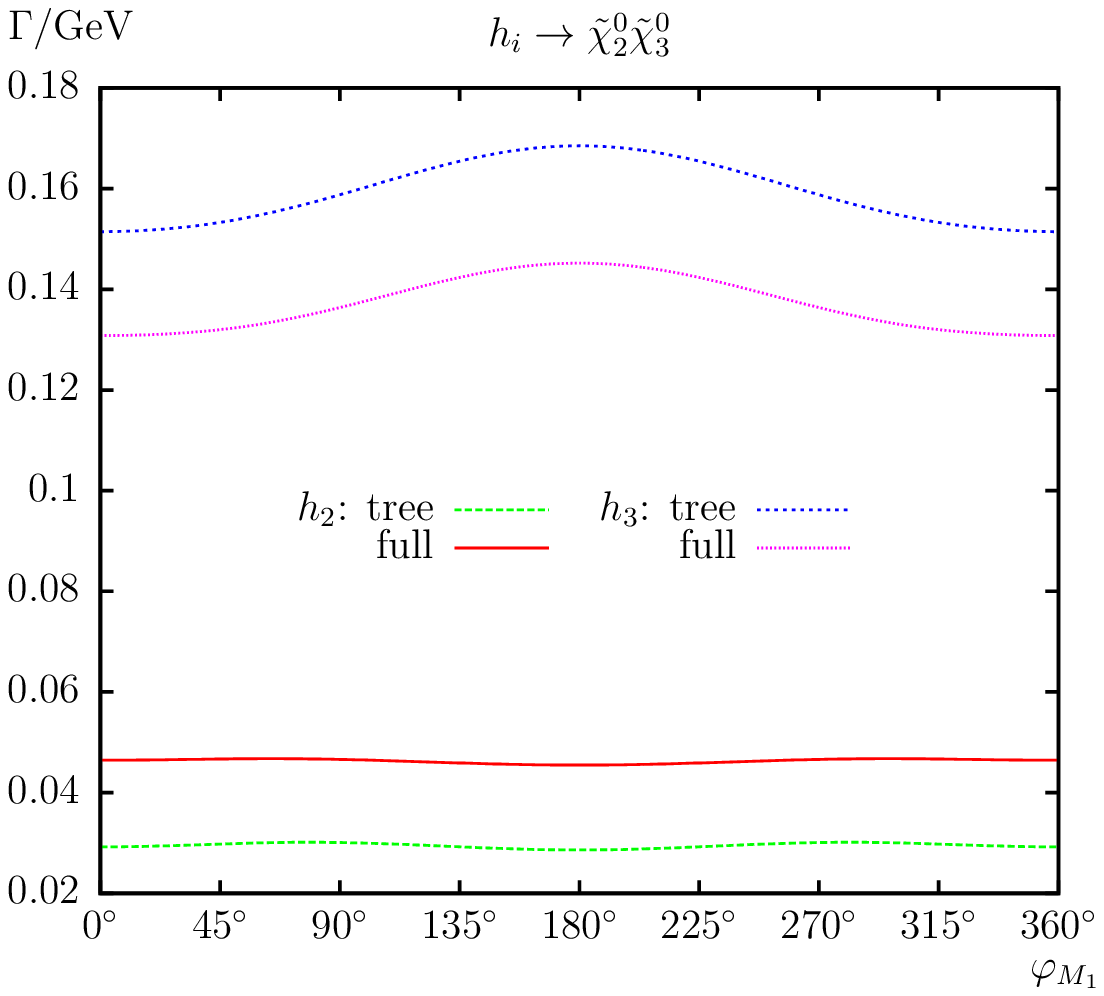}
\end{tabular}
\vspace{1em}
\caption{\label{fig:hneu2neu3}
  $\Ga(\hneuzneud)$. 
  Tree-level and full one-loop corrected partial decay widths are shown. 
  The left plot shows the partial decay width with $\MHp$ varied. 
  The right plot shows the complex phase $\phiMe$ varied with parameters 
  chosen according to \Scv\ (see \refta{tab:para}).
}
\end{center}
\end{figure}

In \reffi{fig:hneu1neu4} we present the results for $\Ga(\hneueneuv)$ 
as a function of $\MHp$ in the left plot.  
The tree-level decay width is accidently very small for the parameter 
set chosen, see \refta{tab:para}.  Because of this smallness, the 
relative size of the one-loop correction becomes larger then the 
tree-level, and can even turn negative.  Therefore, in this case we 
added $|\cMl|^2$ to the full one-loop result to obtain a positive 
decay width.  The dips are (again) the same as already described in 
\reffi{fig:hneu1neu1} beginning at $\MHp \approx 948\gev$, see above.
The anomalous thresholds (``steps'' in the red line)
could be traced back to the $C$-functions at
\begin{align*}
\MHp \approx 1020\gev: \qquad & 
C_{0,1,2}(\mh2^2,\mneu4^2,\mneu1^2,\mneu3^2,\mneu3^2,\MZ^2)\,, \\
\MHp \approx 1026\gev: \qquad & 
C_{0,1,2}(\mh2^2,\mneu4^2,\mneu1^2,\mcha2^2,\mcha1^2,\MW^2)\,, \\
\MHp \approx 1031\gev: \qquad &
C_{0,1,2}(\mneu1^2,\mh2^2,\mneu4^2,m_b^2,\msbot2^2,\msbot1^2)\,, \\
\MHp \approx 1035\gev: \qquad &
C_{0,1,2}(\mh2^2,\mneu4^2,\mneu1^2,\mneu4^2,\mneu2^2,\mh1^2)\,, \\
\MHp \approx 1182\gev: \qquad & 
C_{0,1,2}(\mneu1^2,\mh2^2,\mneu4^2,m_t^2,\mstop2^2,\mstop1^2)\,,
\end{align*}
(in this order). 
The full relative corrections for the $h_2$ ($h_3$) decay are 
$\sim -77\%$ ($\sim -93\%$) at $\MHp = 1200\gev$ (\ie \Scv).  

In the right plot of \reffi{fig:hneu1neu4} we show $\Ga(\hneueneuv)$ 
with the complex phase $\phiMe$ varied at $\MHp = 1200\gev$.
Here (for the same reasons as in the left plot) the loop corrections 
can be larger than the tree-level (and for consistency with the left 
plot we also add $|\cMl|^2$ here).
The loop corrections for the $h_2$ decay vary between $\sim -77\%$ 
at $\phiMe \sim 0^\circ, 360^\circ$ and $\sim -44\%$ at 
$\phiMe = 180^{\circ}$.
The loop corrections for the $h_3$ decay vary between $\sim -93\%$ 
at $\phiMe \sim 0^\circ, 360^\circ$ and $\sim -78\%$ at 
$\phiMe = 180^{\circ}$.

\medskip

In \reffi{fig:hneu2neu3} we present the decay $\hneuzneud$.
In the left plot we show the results as a function of $\MHp$.
The dips (some are hardly visible) are again the same as already 
described in \reffi{fig:hneu1neu1} beginning at $\MHp \approx 1092\gev$, 
see above.  
The full relative corrections for the $h_2$ decay reach up to 
$\sim +59\%$ at $\MHp = 1200\gev$ (\ie \Scv).
The loop corrections for the $h_3$ decay at $\MHp = 1200\gev$ 
are $\sim -14\%$.

In the right plot of \reffi{fig:hneu2neu3} we show the $h_i$ decay 
with the complex phase $\phiMe$ varied at $\MHp = 1200\gev$. 
For the $h_2$ decay the variation of $\phiMe$ is very small, \ie the
loop corrections reach (in \Scv) the same values ($\sim +59\%$) for 
all $\phiMe$.  The variation of $\phiMe$ in the $h_3$ decay is also 
small with corrections at the level of $\sim -14\%$.

\medskip

\begin{figure}[htb!]
\begin{center}
\begin{tabular}{c}
\includegraphics[width=0.49\textwidth,height=7.5cm]{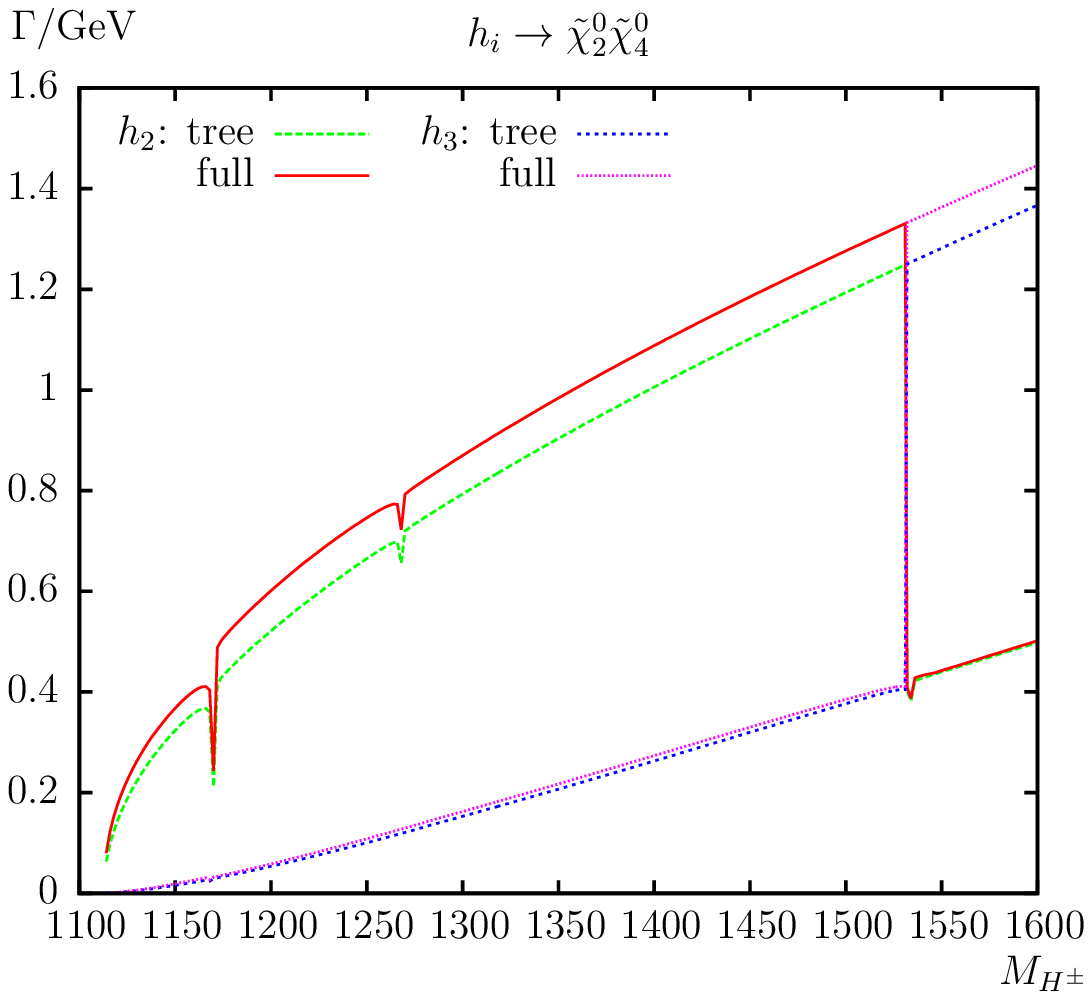}
\hspace{-4mm}
\includegraphics[width=0.49\textwidth,height=7.5cm]{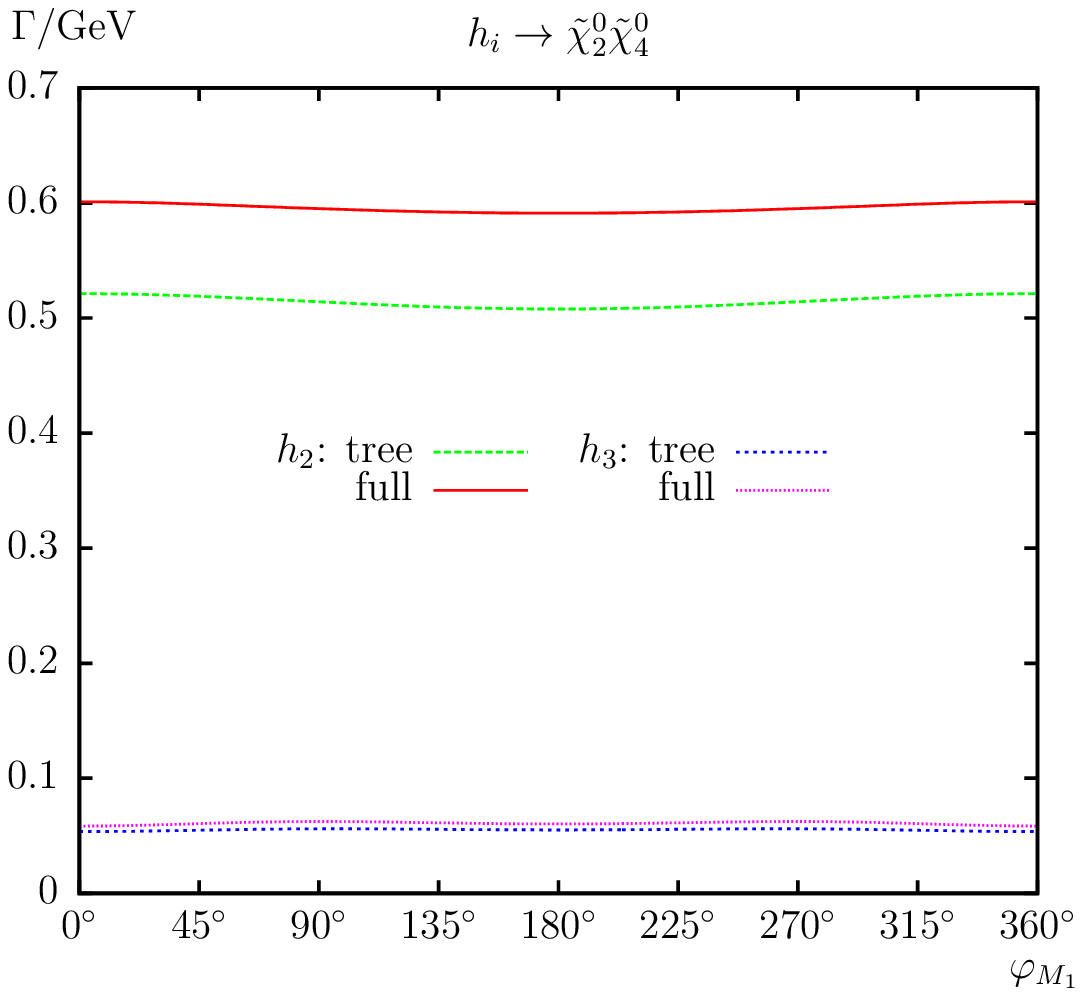}
\end{tabular}
\vspace{1em}
\caption{\label{fig:hneu2neu4}
  $\Ga(\hneuzneuv)$. 
  Tree-level and full one-loop corrected partial decay widths are shown. 
  The left plot shows the partial decay width with $\MHp$ varied. 
  The right plot shows the complex phase $\phiMe$ varied with parameters 
  chosen according to \Scv\ (see \refta{tab:para}).
}
\vspace{6em}
\begin{tabular}{c}
\includegraphics[width=0.49\textwidth,height=7.5cm]{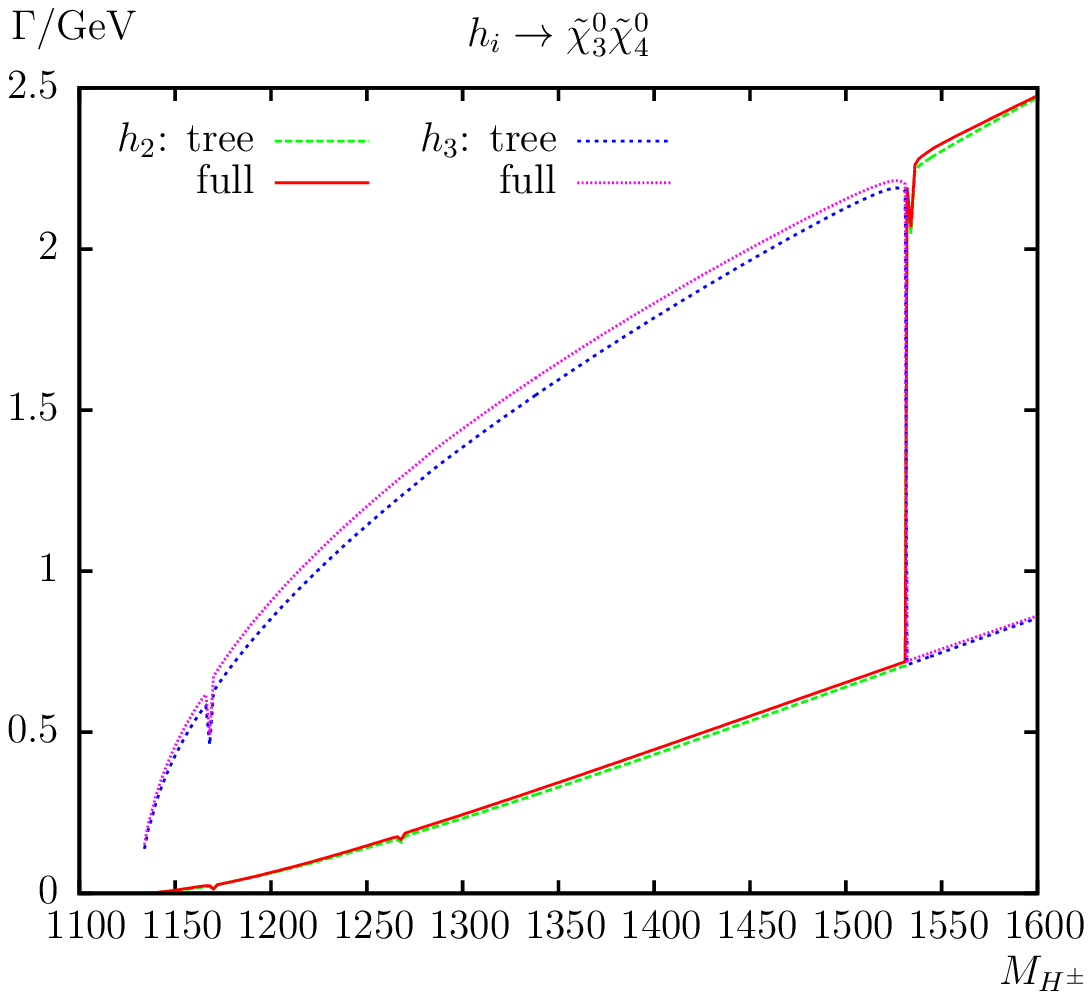}
\hspace{-4mm}
\includegraphics[width=0.49\textwidth,height=7.5cm]{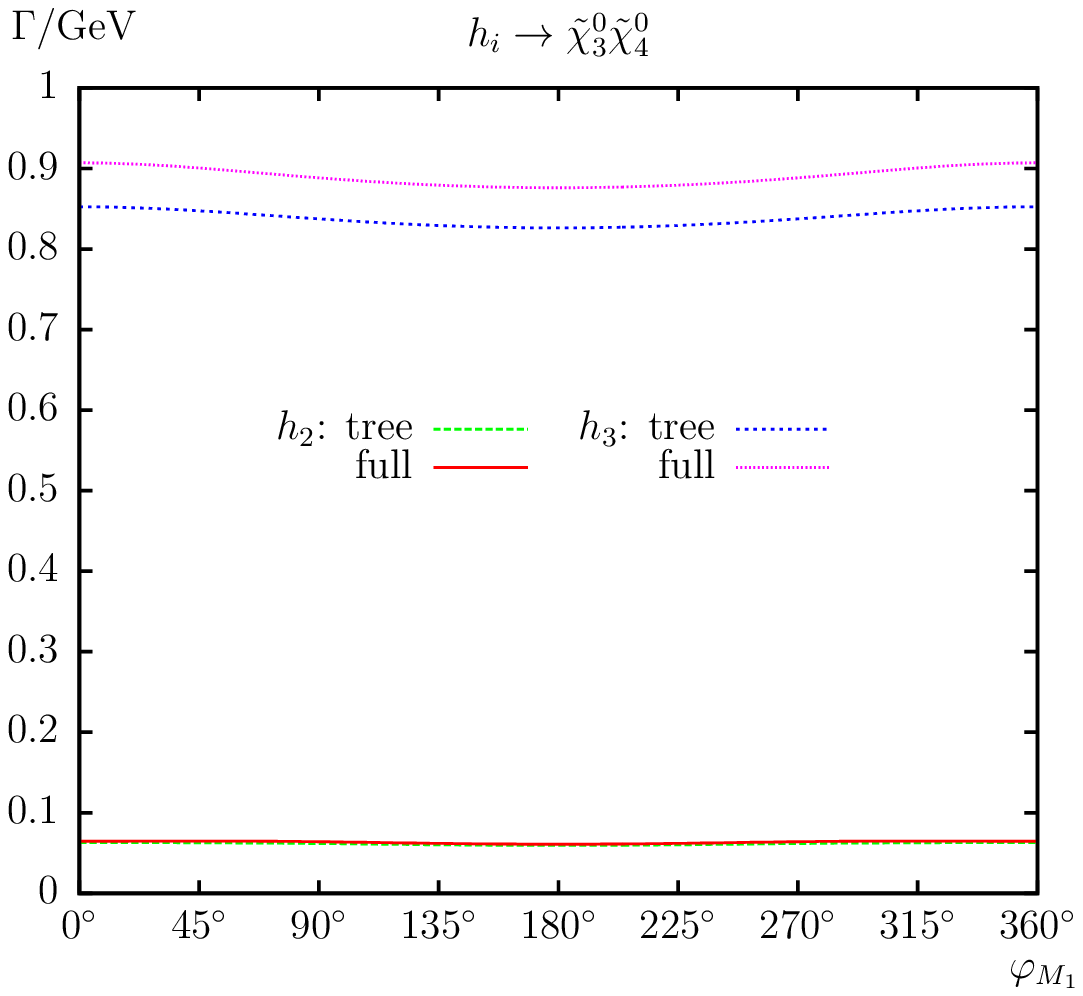}
\end{tabular}
\vspace{1em}
\caption{\label{fig:hneu3neu4}
  $\Ga(\hneudneuv)$. 
  Tree-level and full one-loop corrected partial decay widths are shown. 
  The left plot shows the partial decay width with $\MHp$ varied. 
  The right plot shows the complex phase $\phiMe$ varied with parameters 
  chosen according to \Scv\ (see \refta{tab:para}).
}
\end{center}
\end{figure}

In \reffi{fig:hneu2neu4} we show the decay $\hneuzneuv$.
In the left plot we show the results as a function of $\MHp$.
The dips are (again) the same as already described in 
\reffi{fig:hneu1neu1} beginning at $\MHp \approx 1171\gev$, 
see above. 
The full relative corrections for the $h_2$ decay reach up 
to $\sim +15\%$ at $\MHp = 1200\gev$, \ie \Scv.
The loop corrections for the $h_3$ decay at $\MHp = 1200\gev$ 
are $\sim +9\%$

In the right plot of \reffi{fig:hneu2neu4} the $h_i$ decay is shown
with the complex phase $\phiMe$ varied in \Scv. 
For both decays the variation of $\phiMe$ is very small, as expected, 
since $\neu2$ and $\neu4$ are determined largely by $M_2$ and $\mu$ 
in the neutralino mass matrix (for the parameters chosen as in 
\refta{tab:para}).  The loop corrections for the $h_2$ ($h_3$) decay 
reach $\sim +16\%$ ($\sim +9\%$) at $\phiMe = 180^{\circ}$.

\medskip

The final decays involving neutralinos are shown in \reffi{fig:hneu3neu4}.
The results as a function of $\MHp$ are given in the left plot.
The dips are (again) the same as already described in 
\reffi{fig:hneu1neu1} beginning at $\MHp \approx 1171\gev$, see above.
The full relative corrections are only $\sim +3\%$ at $\MHp = 1200\gev$ 
(\ie \Scv).  The full relative corrections at $\MHp = 1200\gev$ reach 
$+6\%$.
In comparison with \reffi{fig:hneu2neu4} one can observe (again) an
``inversion'' of the relative size of the decays widths of the $h_2$ and
the $h_3$, due to the fact that $\CP(\neu2\neu4) = -\CP(\neu3\neu4)$.

In the right plot of \reffi{fig:hneu3neu4} we show the $h_i$ decay 
with the complex phase $\phiMe$ varied in \Scv. 
For both decays the variation of $\phiMe$ is again very small,
since both neutralinos are largely determined by $\mu$ (for the
parameters chosen as in \refta{tab:para}).
The full one-loop corrections are the same as for the left plot.

\medskip

Overall, for the neutral Higgs decays to a neutralino pair we observed, 
again as expected, an increasing decay width $\propto \mh{i}$.%
\footnote{
  Exceptions are the $h_1 \to \neu1 \neu1$ decay 
  (see the upper left plot of \reffi{fig:h1neu1neu1}), 
  since $\mh1$ depends only very weakly on $\MHp$.  
  The next exception are the corrections to the $h_2$ decay 
  in the left plot of \reffi{fig:hneu3neu3} (red line), 
  due to an accidental interplay of anomalous thresholds with the 
  effects induced by the Higgs mixing matrix $\matr{\hat Z}$. 
  The final exception can be observed in the corrections to the 
  $h_2$ decay in the left plot of \reffi{fig:hneu1neu4} (red line), 
  because of the additional 2-loop corrections $|\cMl|^2$ 
  (see the discussion of \reffi{fig:hneu1neu4} above).
}
The full one-loop corrections reach a level of $10-20\%$ for decay widths 
of \order{1\gev}.  The variation with $\phiMe$ is found largest in cases 
where the $\CP$-nature of the decay depends strongly on the phase, 
there then changes by a factor of~5 or more can be observed.

\clearpage
\newpage

%%%%%%%%%%%%%%%%%%%%%%%%%% F I G U R E %%%%%%%%%%%%%%%%%%%%%%%%%%%%%%%%%%%%%%%%%
%%%%%%%%%%%%%%%%%%%%%%%%%% F I G U R E %%%%%%%%%%%%%%%%%%%%%%%%%%%%%%%%%%%%%%%%%

%\newpage

%%%%%%%%%%%%%%%%%%%%%%%%%% F I G U R E %%%%%%%%%%%%%%%%%%%%%%%%%%%%%%%%%%%%%%%%%
%%%%%%%%%%%%%%%%%%%%%%%%%% F I G U R E %%%%%%%%%%%%%%%%%%%%%%%%%%%%%%%%%%%%%%%%%

%\newpage

%%%%%%%%%%%%%%%%%%%%%%%%%% F I G U R E %%%%%%%%%%%%%%%%%%%%%%%%%%%%%%%%%%%%%%%%%
%%%%%%%%%%%%%%%%%%%%%%%%%% F I G U R E %%%%%%%%%%%%%%%%%%%%%%%%%%%%%%%%%%%%%%%%%

%\newpage

%%%%%%%%%%%%%%%%%%%%%%%%%% F I G U R E %%%%%%%%%%%%%%%%%%%%%%%%%%%%%%%%%%%%%%%%%
%%%%%%%%%%%%%%%%%%%%%%%%%% F I G U R E %%%%%%%%%%%%%%%%%%%%%%%%%%%%%%%%%%%%%%%%%

%\newpage

%%%%%%%%%%%%%%%%%%%%%%%%%% F I G U R E %%%%%%%%%%%%%%%%%%%%%%%%%%%%%%%%%%%%%%%%%
%%%%%%%%%%%%%%%%%%%%%%%%%% F I G U R E %%%%%%%%%%%%%%%%%%%%%%%%%%%%%%%%%%%%%%%%%

\clearpage
\newpage

%%%%%%%%%%%%%%%%%%%%%%%%%%%%%%%%%%%%%%%%%%%%%%%%%%%%%%%%%%%%%%%%%%%%%%%%%%%%%%%
%%%%%%%%%%%%%%%%%%%%%%%%%%%%%%%%%%%%%%%%%%%%%%%%%%%%%%%%%%%%%%%%%%%%%%%%%%%%%%%

\section{Conclusions}
\label{sec:conclusions}

We evaluated all partial decay widths corresponding to a two-body 
decay of the MSSM Higgs bosons to charginos and neutralinos,
allowing for complex parameters. 
In the case of a discovery of additional Higgs bosons a subsequent
precision measurement of their properties will be crucial determine
their nature and the underlying (SUSY) parameters. 
In order to yield a sufficient accuracy, one-loop corrections to the 
various Higgs-boson decay modes have to be considered. 
In this work we take another step in the direction of completion 
of the calculation of \textit{all} two-body decays at the one-loop level 
in the cMSSM in this stable and reliable renormalization scheme: we
calculated all two-body decay modes of the Higgs bosons to charginos and 
neutralinos in the cMSSM.

The decay modes are given in \refeqs{eq:hchacha} -- (\ref{eq:Hpmneucha}).
The evaluation is based on a full one-loop calculation of all decay 
channels, also including hard and soft QED radiation. 
We restricted ourselves to a version of our renormalization scheme which 
is valid for $|M_1| < |M_2|, |\mu|$ and $M_2 \neq \mu$ (where $M_1$ and 
$M_2$ denote the soft SUSY-breaking parameter of the $U(1)$ and $SU(2)$
gauginos, and $\mu$ is the Higgs mixing parameter) to simplify the
analysis, even though our set-up allows to switch to other parameter
regions, possibly implying a different renormalization, 
see the discussion in \citeres{LHCxC,LHCxN,LHCxNprod}.

We first reviewed the relevant sectors including some details on the
one-loop renormalization procedure of the cMSSM, 
which are relevant for our calculation. In most cases we follow
\citere{MSSMCT}. However, in the scalar fermion sector, where we differ
from \citere{MSSMCT} the relevant details are indicated.
We have discussed the calculation of the one-loop diagrams, the
treatment of UV and IR divergences that are canceled by the inclusion
of (hard and soft) QED radiation. 
We have checked our result against the literature, and in most
cases we have found good agreement, once our set-up was changed to the
one used in the existing analyses.

While the analytical calculation has been performed for \textit{all} 
decay modes to charginos and neutralinos, in the numerical analysis 
we mostly concentrated on the decays of the heavy Higgs bosons, with 
$h_1 \to \neu1\neu1$ being the only channel analyzed for the light
neutral Higgs boson.
For the analysis we have chosen a parameter set that allows
simultaneously a maximum number of two-body chargino/neutralino decay modes.
In the analysis either the charged Higgs boson mass or the phase of
$M_1$ has been varied.
For $\MHp$ we investigated an interval starting at $\MHp = 600\gev$ 
up to $\MHp = 1.6\tev$, which roughly coincides with the reach of 
the LHC for high-luminosity running as well as an $e^+e^-$ collider
with a center-of-mass energy up to $\sqrt{s} \sim 3\tev$.

In our numerical scenarios we compared the tree-level partial decay
widths with the full one-loop corrected  partial decay widths.  
We concentrated on the analysis of the decay widths themselves, since
the size of the corresponding branching ratios (and thus the size of 
their one-loop effects) is highly parameter dependent.

We found sizable corrections of $\sim 10\%$ in many channels,
sometimes going up to $\sim 20\%$.  Even larger corrections are only 
found in cases where the tree-level result is accidentally small and 
thus the decay likely not observable. Corrections at the $10-20\%$ level 
have also been found for the decay $h_1 \to \neu1\neu1$, which could
constitute an important channel for the determination of the Dark Matter
properties in the cMSSM. Consequently, the full one-loop 
corrections should be taken into account for the interpretation of the 
searches for charginos/neutralinos as well as for any future precision 
analyses of those decays.

The tree-level results, but also the size of the full one-loop
corrections often depend strongly on the complex phase analyzed, 
$\phiMe$. 
The one-loop contributions can vary by a factor of $\sim 2$ as a 
function of the complex phase.
Neglecting the phase dependence could lead to a wrong impression of 
the relative size of the various decay widths.

In the cases where a decay and its complex conjugate final state 
are possible, \ie the charged Higgs decays
we have evaluated both decay widths independently.  
The asymmetries, as a byproduct of our calculation, turn out to be 
relatively small, at the level of a few per-cent.

The numerical results we have shown are, of course, dependent on the choice 
of the SUSY parameters. Nevertheless, they give an idea of the relevance
of the full one-loop corrections. 
Decay channels (and their respective one-loop corrections) that may look 
unobservable due to the smallness of their decay width in our numerical
examples could become important if other channels are kinematically
forbidden. 
Following our analysis it is evident that the full one-loop corrections
are mandatory for a precise prediction of the various branching ratios.
The full one-loop corrections should be taken into account in any precise 
determination of (SUSY) parameters from the decay of (heavy) MSSM Higgs
bosons. 
It is planned to implement the evaluation of the branching ratios of the
(heavy) Higgs bosons into the Fortran code \FH, 
together with an automated choice of the renormalization scheme valid
for the full cMSSM parameter space.

%%%%%%%%%%%%%%%%%%%%%%%%%%%%%%%%%%%%%%%%%%%%%%%%%%%%%%%%%%%%%%%%%%%%%%%%%%%%%%

\subsection*{Acknowledgements}

We thank A.~Bharucha, T.~Hahn and F.~von~der~Pahlen for helpful discussions. 
The work of S.H.\ is supported in part by CICYT (grant FPA 2013-40715-P) 
and by the Spanish MICINN's Consolider-Ingenio 2010 Program under grant 
MultiDark CSD2009-00064.

%%%%%%%%%%%%%%%%%%%%%%%%%%%%%%%%%%%%%%%%%%%%%%%%%%%%%%%%%%%%%%%%%%%%%%%%%%%%%%%
%%%%%%%%%%%%%%%%%%%%%%%%%%%%%%%%%%%%%%%%%%%%%%%%%%%%%%%%%%%%%%%%%%%%%%%%%%%%%%%

%\newpage

%%%%%%%%%%%%%%%%%%%%%%%%%%%%%%%%%%%%%%%%%%%%%%%%%%%%%%%%%%%%%%%%%%%%%%%%%%%%%%%
%%%%%%%%%%%%%%%%%%%%%%%%%%%%%%%%%%%%%%%%%%%%%%%%%%%%%%%%%%%%%%%%%%%%%%%%%%%%%%%

\newcommand\jnl[1]{\textit{\frenchspacing #1}}
\newcommand\vol[1]{\textbf{#1}}


\begin{thebibliography}{99}
\bibitem{feynarts-mf}
The couplings can be found in \texttt{HMix.ps.gz} and 
\texttt{MSSM.ps.gz} as part of the \FA\ package~\cite{feynarts}.
\end{thebibliography}
\end{document}